\documentclass[manuscript]{acmart}

\usepackage{microtype}
\usepackage{array}
\usepackage{graphicx}
\usepackage{multirow}
\usepackage{amsmath,amsfonts}
\usepackage{amsthm}
\usepackage[mathcal]{eucal}
\usepackage[title]{appendix}
\usepackage{xcolor}
\usepackage{textcomp}
\usepackage{manyfoot}
\usepackage{booktabs}
\usepackage{algorithm}
\usepackage{algorithmicx}
\usepackage{algpseudocode}
\usepackage{listings}
\usepackage{tikz}
\usepackage{subfig}
\usepackage{bbm}
\usepackage{hyperref}
\usepackage{url}
\usepackage{float}
\usepackage{paralist}
\usepackage{fancyhdr}
\usepackage[acronym]{glossaries}
\usepackage[figuresright]{rotating}

\AtBeginDocument{%
  }

\newacronym{SLO}{SLO}{Service Level Objective}
\begin{document}

\title{Formal and Empirical Study of Metadata-Based Profiling for Resource Management in the Computing Continuum}

\author{Andrea Morichetta}
\orcid{https://orcid.org/0000-0003-3765-3067}
\affiliation{%
  \institution{Distributes Systems Group, Vienna University of Technology (\textit{TU Wien})}
  \city{Wien}
  \country{Austria}
}
\email{a.morichetta@dsg.tuwien.ac.at}
\authornote{Corresponding author}
\author{Stefan Nastic}
\orcid{https://orcid.org/0000-0003-0410-6315}
\affiliation{%
  \institution{Distributes Systems Group, Vienna University of Technology (\textit{TU Wien})}
  \city{Wien}
  \country{Austria}
}
\email{s.nastic@dsg.tuwien.ac.at}
\author{Victor Casamayor Pujol}
\orcid{https://orcid.org/0000-0003-2830-8368}
\affiliation{%
  \institution{Engineering Department, Universitat Pompeu Fabra (\textit{UPF})}
  \city{Barcelona}
  \country{Spain}
}
\email{victor.casamayor@upf.edu}
\author{Schahram Dustdar}
\orcid{https://orcid.org/0000-0001-6872-8821}
\affiliation{%
  \institution{Distributes Systems Group, Vienna University of Technology (\textit{TU Wien})}
  \city{Wien}
  \country{Austria}
}
\email{dustdar@dsg.tuwien.ac.at}

\renewcommand{\shortauthors}{Morichetta et al.}

\begin{abstract}
We present and formalize a general approach for profiling workload by leveraging only a priori available static metadata to supply
appropriate resource needs. Understanding the requirements and characteristics of a workload's runtime is essential. Profiles are essential for the platform (or infrastructure) provider because they want to ensure that Service Level Agreements and their objectives (SLOs) are fulfilled and, at the same time, avoid allocating too many resources to the workload. When the infrastructure to manage is the computing continuum (i.e., from IoT to Edge to Cloud nodes), there is a big problem of placement and tradeoff or distribution and performance. Still, existing techniques either rely on static predictions or runtime profiling, which are proven to deliver poor performance in runtime environments or require laborious mechanisms to produce fast and reliable evaluations. We want to propose a new approach for it. Our profile combines the information from past execution traces with the related workload metadata, equipping an infrastructure orchestrator with a fast and precise association of newly submitted workloads. We differentiate from previous works because we extract the profile group metadata saliency from the groups generated by grouping similar runtime behavior. We first formalize its functioning and its main components. Subsequently, we implement and empirically analyze our proposed technique on two public data sources: Alibaba cloud machine learning workloads and Google cluster data. Despite relying on partially anonymized or obscured information, the approach provides accurate estimates of workload runtime behavior in real-time.\end{abstract}



\maketitle



\newacronym{IoT}{IoT}{Internet of Things}
\newacronym{SLA}{SLA}{Service Level Agreement}
\newacronym{QoS}{QoS}{Quality of Service}
\newacronym{DAG}{DAG}{directed acyclic graph}
\newacronym{CRD}{CRD}{Custom Resource Definition}
\newacronym{ML}{ML}{Machine Learning}
\newacronym{AMQP}{AMQP}{Advanced Message Queuing Protocol}
\newacronym{MCDM}{MCDM}{multi-criteria decision making}
\newacronym{ILP}{ILP}{Integer Linear Programming}
\newacronym{GA}{GA}{Genetic Algorithm}
\newacronym{vCPU}{vCPU}{virtual CPU core}
\newacronym{XGBoost}{XGBoost}{eXtreem Gradient Boosting}
\newacronym{silhouette}{silhouette}{silhouette coefficient}
\newcommand{\SLO}[0]{\gls{SLO}}
\newcommand{\SLOs}[0]{\glspl{SLO}}
\newcommand{\QoS}[0]{\gls{QoS}}

\newcommand{\RMSEperc}[0]{$RMSE_{perc}$}

\newcommand{\userFeature}[0]{\textit{user}}
\newcommand{\jobNameFeature}[0]{\textit{job name}}
\newcommand{\workloadFeature}[0]{\textit{workload}}
\newcommand{\taskNameFeature}[0]{\textit{task name}}
\newcommand{\groupFeature}[0]{\textit{group}}

\newcommand{\xgboost}[0]{\gls{XGBoost}}
\newcommand{\euclideanDist}[0]{\textit{Euclidean}}
\newcommand{\cosineDist}[0]{\textit{Cosine}}
\newcommand{\manhattanDist}[0]{\textit{Manhattan}}

\newcommand{\StandardScalerTransform}[0]{\textit{StandardScaler}}
\newcommand{\MinMaxScalerTransform}[0]{\textit{MinMaxScaler}}
\newcommand{\RobustScalerTransform}[0]{\textit{RobustScaler}}
\newcommand{\PowerTransform}[0]{\textit{PowerTransform}}

\newcommand{\NetworkQosSLOsExperiment}[0]{Network \QoS{} \SLOs{} Compliance experiment}
\newcommand{\NetworkQosSLOsExperimentTitle}[0]{Network \QoS{} \SLOs{} Compliance}
\newcommand{\ScalabilityExperiment}[0]{Performance and Scalability experiment}
\newcommand{\ScalabilityExperimentTitle}[0]{Performance and Scalability}

\newcommand{\Polaris}[0]{\PolarisShortName{}}
\newcommand{\PolarisScheduler}[0]{Polaris Scheduler} 
\newcommand{\ServiceGraph}[0]{Service Graph}
\newcommand{\ClusterTopologyGraph}[0]{Cluster Topology Graph}

\newcommand{\Sort}[0]{\inlinecode{Sort}}
\newcommand{\PreFilter}[0]{\inlinecode{Pre\\-Filter}}
\newcommand{\Filter}[0]{\inlinecode{Filter}}
\newcommand{\PostFilter}[0]{\inlinecode{Post\\-Filter}}
\newcommand{\PreScore}[0]{\inlinecode{Pre\\-Score}}
\newcommand{\Score}[0]{\inlinecode{Score}}
\newcommand{\NormalizeScore}[0]{\inlinecode{Normalize\\-Score}}
\newcommand{\Reserve}[0]{\inlinecode{Reserve}}
\newcommand{\Permit}[0]{\inlinecode{Permit}}
\newcommand{\ServiceGraphPlugin}[0]{\inlinecode{Service\\-Graph}}
\newcommand{\ServiceGraphPluginTitle}[0]{ServiceGraph Plugin}
\newcommand{\NetworkQosPlugin}[0]{\inlinecode{Network\\-QoS}}
\newcommand{\NetworkQosPluginTitle}[0]{NetworkQoS Plugin}
\newcommand{\NodeCostPlugin}[0]{\inlinecode{Node\\-Cost}}
\newcommand{\NodeCostPluginTitle}[0]{NodeCost Plugin}
\newcommand{\PodsPerNodePlugin}[0]{\inlinecode{Pods\\-Per\\-Node}}
\newcommand{\PodsPerNodePluginTitle}[0]{PodsPerNode Plugin}
\newcommand{\AtomicDeploymentPlugin}[0]{\inlinecode{Atomic\\-Deployment}}
\newcommand{\AtomicDeploymentPluginTitle}[0]{AtomicDeployment Plugin}

\newcommand{\fakeCpu}[0]{\inlinecode{cpu}}
\newcommand{\fakeMemory}[0]{\inlinecode{memory}}
\newcommand{\baseNodePrefix}[1]{\inlinecode{base#1}}
\newcommand{\raspiNodePrefix}[1]{\inlinecode{raspi#1}}
\newcommand{\cloudNodePrefix}[1]{\inlinecode{cloud#1}}
\newcommand{\baseNode}[1]{\baseNodePrefix{-#1}}
\newcommand{\raspiNode}[1]{\raspiNodePrefix{-#1}}
\newcommand{\cloudNode}[1]{\cloudNodePrefix{-#1}}

\newcommand{\SchedulingStageSection}[1]{\textbf{#1.}}

\newcommand{\secref}[1]{\S\ref{#1}}
\section{Introduction}
\label{sec:intro}
Resource management in shared and virtualized systems across the Computing Continuum poses a major challenge for providers and operators~\cite{nastic2020sloc}. The key question is: \textit{how can we ensure Service Level Objectives (SLOs)~\cite{xiong2018extend} within this complex and interconnected infrastructure while optimizing its usage?}
Modern orchestration techniques address these challenges through scheduling workloads, resource placement, overcommitting and oversubscribing~\cite{householder2014cloud}, or handling resource bursts~\cite{noonan2016managing} and live migrations. In this context, profiling workloads plays a crucial role by helping to analyze and understand them. By reducing overprovisioning, profiling helps maintain efficiency while ensuring that stakeholders’ (e.g., application developers) goals are met.

\textbf{Workload profiling: state of the art and limitations.} The main approaches to workload profiling can be classified  into one of two general categories: 
(a)~they either attempt to exploit available information about past workload executions (historical data) to learn the workload's characteristics or 
(b)~they attempt to collect information about the workload's properties by actively observing it - usually by running the workload in a sandbox and probing it with synthetic traffic. Furthermore, they make at least one of the following assumptions:
\begin{inparaenum} [(i)]
    \item {\em Environment consistency} - that is, they assume that the sandboxed execution environment used for profiling faithfully resembles the production execution environment;
    \item {\em Performance consistency} - that is, the runtime performance of similar workloads will remain consistent over time.
    \item {\em Time consistency} - that is, there are no time constraints on how long it takes to make profiling decisions, i.e., the workload profile can be created ad hoc when needed;
    \item {\em Occurrence consistency} - that is, the same workload will run multiple times, and it will reoccur in the same shared computing environment in the (near) future.
\end{inparaenum}

Unfortunately, these consistency assumptions typically do not hold in practice. The reasons are multiple.
\begin{inparaenum} [(i)]
    \item The execution environment is typically inconsistent across multiple workload runs. 
    The primary reason is the infrastructure heterogeneity resulting from software and hardware updates, such as adding a new generation processor~\cite{jajoo2022case}. Additionally, due to the ``noisy neighbors'' phenomenon, the existing physical resources available in a host node can significantly vary. This scenario can cause significant variance in workload performance, rendering the profiles useless. 
    \item Further, several authors have pointed out that the runtime performance of similar workloads is not consistent~\cite{cui2023analysis} during their lifetime. It typically varies with time, even if the same preconditions are met, such as using the same input data~\cite{park20183sigma, jyothi2016morpheus}.
    \item The time allocated to the profiler to generate the workload's profile can significantly vary. It is use-case specific and typically inconsistent for different resource provisioning techniques. For example, time spent profiling a workload while it is pending to be scheduled must be orders of magnitude shorter than profiling a workload to prevent a bootstrapping problem when predicting SLO violations.
    \item Finally, previous work has shown that most general-purpose workloads are recurrent only to a limited degree, that is, only between 40\% and 60\% of workloads are reported to be recurrent~\cite{ferguson2012jockey, jyothi2016morpheus, jalaparti2015network}. By only looking at a single workload's history, approximately every other workload will fail to be successfully profiled. 
\end{inparaenum}

\textbf{Research challenges and requirements.}
Based on these limitations, we identify three main research challenges:
\begin{enumerate} [\bfseries (RC-1)]
    \item How can we derive accurate workload profiles in the face of a small sample size caused by non-recurrent workloads?
    \item How can we represent profiled characteristics so that they can capture the workloads' performance variance?
    \item How can we make the profiling process general, non-invasive, and transparent so that it can seamlessly facilitate various resource provisioning and management techniques? 
\end{enumerate}

\begin{figure}[ht!]
    \centering
    \includegraphics[width=0.95\columnwidth]{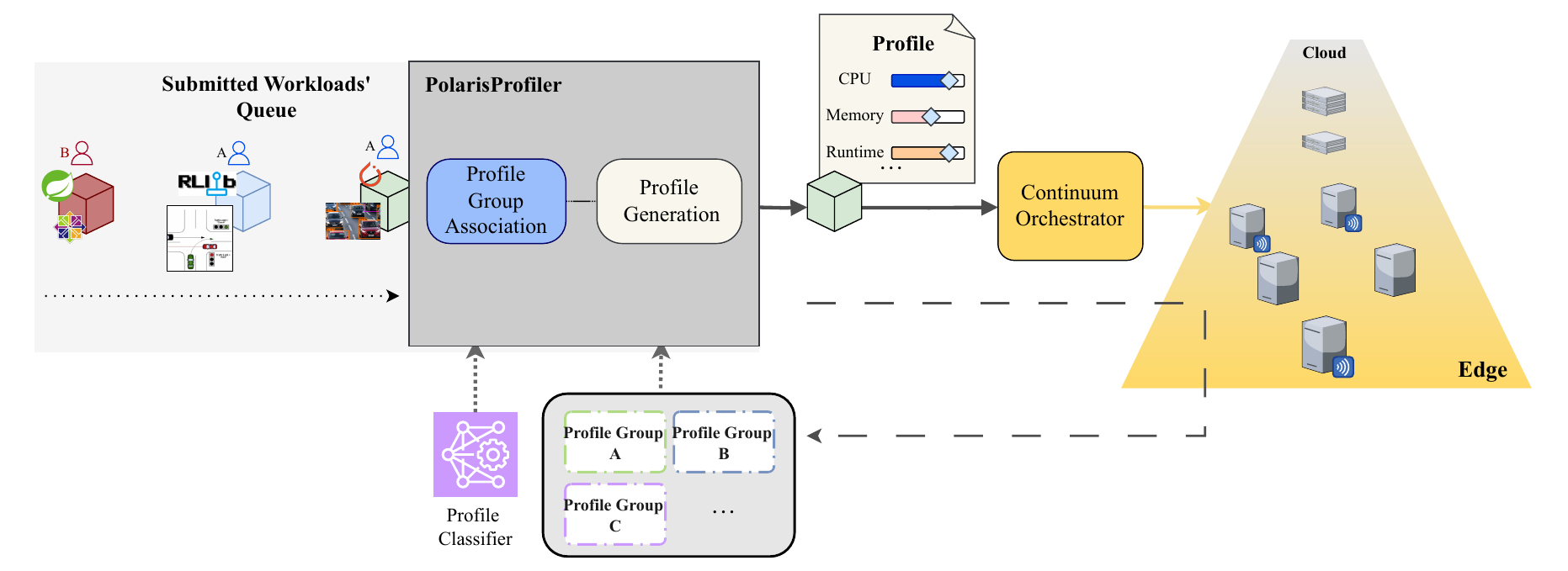}
    \caption{Overview of the PolarisProfiler's model.}
    \Description{Diagram illustrating PolarisProfiler workflow. On the left, new workloads is submitted to the computing continuum system. The first tool is the polaris profiler, which has two steps. The first is to associate a workload to a profile group and the second is to generate the profile, which is a list of expected runtime characteristics. The workload, with its profile, is therefore sent to the Computing Continuum orchestrator which will take the correct approaches to schedule it and scaling or migrating it.}
    \label{fig:profiling_model}
\end{figure}

To address the \textbf{RC-1}, we take a pragmatic approach by continuously analyzing \textit{all available workloads} from a shared infrastructure, grouping together the ones that show a \textit{comparable behavior} at runtime. Our approach only relies on de-facto standard telemetry data, which is typically readily available for any virtualized computing infrastructure. 
Furthermore, to make our approach generic, it must not rely on any particular assumption or precondition regarding the data and its preprocessing or preparation.   
Tackling \textbf{RC-2} requires a novel view of profile representation, moving away from traditional profiles, which attempt to represent the workload's runtime properties as static profile characteristics. 
The profile groups should encapsulate the runtime telemetry of various workload types that show homogeneous behavior. This characteristic allows to incorporate more detailed statistics, naturally \textit{reflecting workloads' performance variance}. 
Finally, we need to build agile profiling decisions to handle the \textbf{RC-3} and make our approach generally useful for various resource provisioning and management techniques. The profiling method should seamlessly associate workloads with profile groups upon arrival, ensuring \textit{non-invasiveness} and enabling \textit{transparent processing}. Moreover, the profile assignment should rely on widely accepted and well-known information, such as workload metadata.
The overall aim is to build a profiling method that doesn't just try to model the behavior of one type of workload either looking at its behavior at runtime or by modeling its previous runs.

\textit{PolarisProfiler} -- a novel profiling approach that leverages \textit{a priori} available, static metadata to enable generic and immediate workload profiling based on historic execution traces -- offers a solution to these challenges. We use the term \textit{a priori} to specify that we collect metadata information available at the submission (deployment or provisioning) phase. Examples are user data, application data, and OS parameters. This information does not change during the workload runtime; therefore, it is \textit{static} ( invariant, unchanging). Figure~\ref{fig:profiling_model} gives an overview of the approach. Once a workload is submitted to the managed platform, it is first associated with a profile group; subsequently, it is assigned a profile detailing its expected runtime characteristics. At this point, the orchestrator can make informed decisions on where to schedule and how to manage the workload. We achieve this through a {\em generic workload profile generator} component that automatically derives workload profiles based only on the readily available resource usage data. It does not rely on any specific assumptions or tailored feature engineering. 
The model represents dynamic profiles, which can capture the dynamic nature of the workload's runtime properties. Our dynamic profiles can be continuously updated, even after initial workload profiling, to reflect the workload's varying performance over time.  
The {\em metadata-based profile classifier} efficiently classifies new workloads and assigns runtime profiles through available, a priori, static metadata. 
This way, new workloads get nearly instantly assigned a profile that includes its expected runtime behavior. We achieve that thanks to profiles created based on similarity in resource usage, thus providing the ability to extract trends, anomalies, and seasonal patterns.
We show its potential through a {\em comprehensive case study} on two real-world, open-source traces. 
In this paper, we build on top of previous results, presented in~\cite{Morichetta2023PolarisProfilerAN} with the following contributions.

\begin{enumerate}
\item We introduce a formal representation of the PolarisProfiler model, extending and generalizing the previous definition of our solution. Through block schemas and flowcharts, we describe the components and their interactions to guarantee generalizability.
\item We introduce a quality metric, ACQUIRES, to evaluate profiles based on both profile-specific and general features and to manage the lifecycle of our system.
\item We introduce the functioning of the feedback loop. This aspect includes experimenting with new workloads that arrive in the system. It shows how this approach contributes to updating the profiles and keeping the system representative of the workload.
\item We expand our case study analysis by incorporating additional metrics (e.g., CPU, GPU, memory usage) and evaluating it on a dataset ten times larger than before.
\item We extend our analysis to estimate the capability of the proposed approach to work in different scenarios by leveraging the Google cluster data~\cite{verma2015large} traces.
\end{enumerate}

Despite only relying on \textit{static}, \textit{a priori} metadata, our methodology yields an overall error rate below 50\% for the 93\% of classified workloads for the Alibaba dataset. These results are competitive with the state-of-the-art approaches, with the difference that their specific focus is the estimation of AI workload duration. Similar performance is achieved for the Google cluster dataset, showing the generalizability of the approach.
We publicly release the code to allow transparency and reproducibility of our results\footnote{\url{https://github.com/polaris-slo-cloud/Profiling/edit/master/ml_data-profiling/README.md}}.

The rest of this paper is organized as follows: Section~\ref{sec:design} introduces the PolarisProfiler model and methodology, detailing the components and interactions within the profiling framework. Section~\ref{sec:case_study} presents the main case study, including the Alibaba dataset description, methodology, evaluation metrics, and results. In Section~\ref{sec:google} we explore how the PolarisProfiler can work with non-strictly machine learning workload. In Section~\ref{sec:related}, we review related work in the field of workload profiling and resource management. Section~\ref{sec:conclusion} concludes the paper, summarizing the key findings and contributions of the study.
\section{PolarisProfiler Model \& Methodology} 
~\label{sec:design}
\begin{figure}[ht!]
    \centering
    \includegraphics[width=0.5\columnwidth]{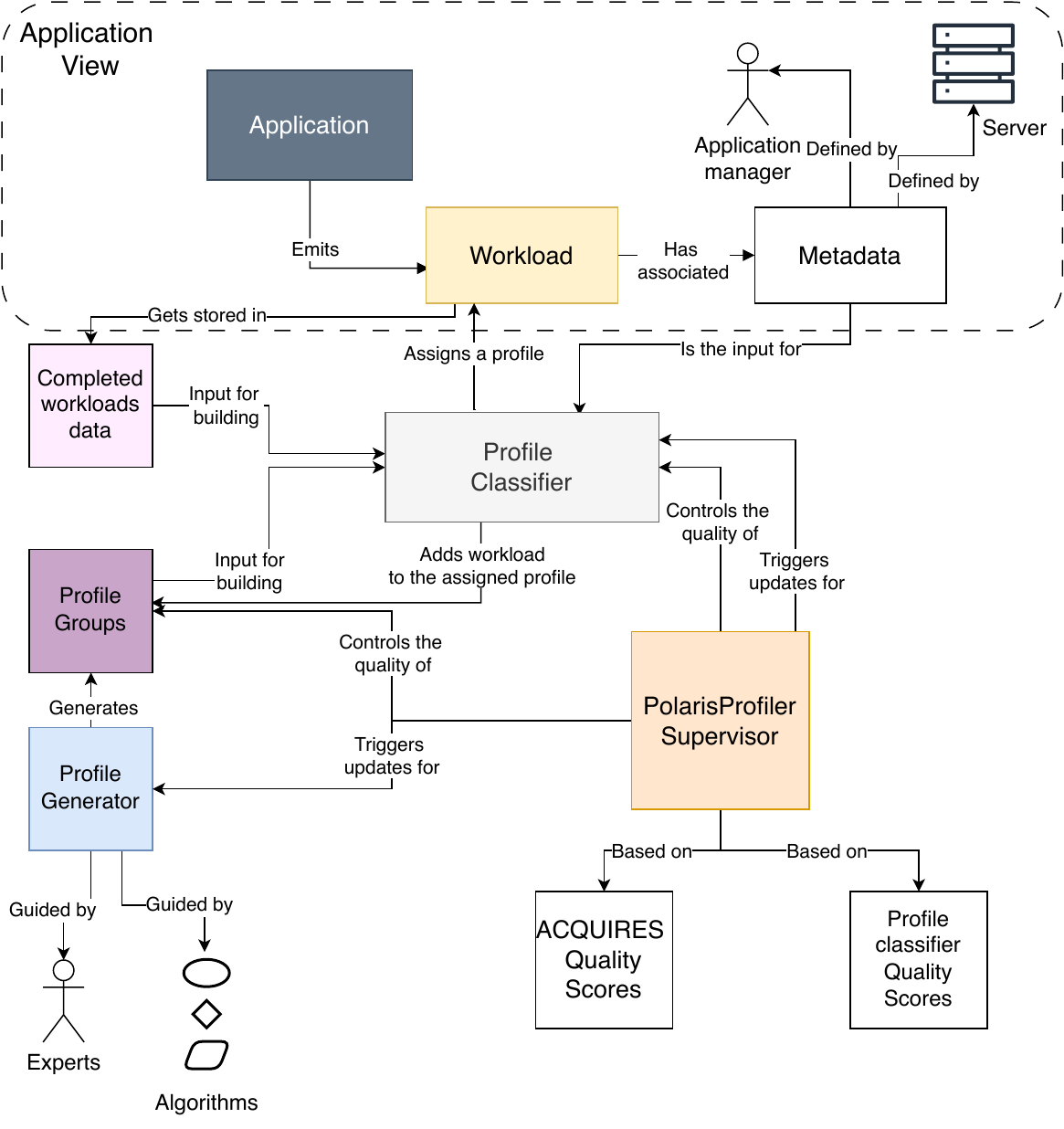}
    \caption{Visual representation of the PolarisProfiler components, actors and their interactions in the model lifecycle.}
    \Description{Diagram of the PolarisProfiler system showing the application view and interactions between components. At the top, the "Application" emits "Workload" which has associated "Metadata" defined by an application manager. The "Workload" gets stored in "Completed workloads data" and is input for the "Profile Classifier." The "Profile Classifier" assigns a profile to the workload, adds workload data to the assigned profile, and controls the quality of the profile, triggering updates for the "PolarisProfiler Supervisor.
    The "PolarisProfiler Supervisor" triggers updates for the "Profiles" and "Profiler." The "Profiles" and "Profiler" components are guided by experts and algorithms. Quality scores from the "PolarisProfiler Supervisor" are based on "ACQUIRES Quality Scores" and "Profile Classifier Quality Scores."}
    \label{fig:formal-representation}
\end{figure}

This section presents the core principles of our profiling methodology, PolarisProfiler. 
An overview of the model is summarized by Figure~\ref{fig:formal-representation}. When a stakeholder, such as an application owner, submits workloads to the Computing Continuum platform, standard metadata is attached. This metadata describes the workload's nature and requirements, such as its type (e.g., a machine learning task) and the associated resource characteristics (e.g., the virtual machine type and resource allocation).
The metadata accompanying the submitted workload serves as the input for the Profile Classifier module. The Profile Classifier uses this metadata to assign the workload to a specific profile group. This process involves labeling the workload as belonging to the group that exhibits the most similar metadata. 
The Profile Generator creates profile groups. 

\subsection{Profile Generator}

\begin{figure}[ht!]
    \centering
    \includegraphics[width=0.7\columnwidth]{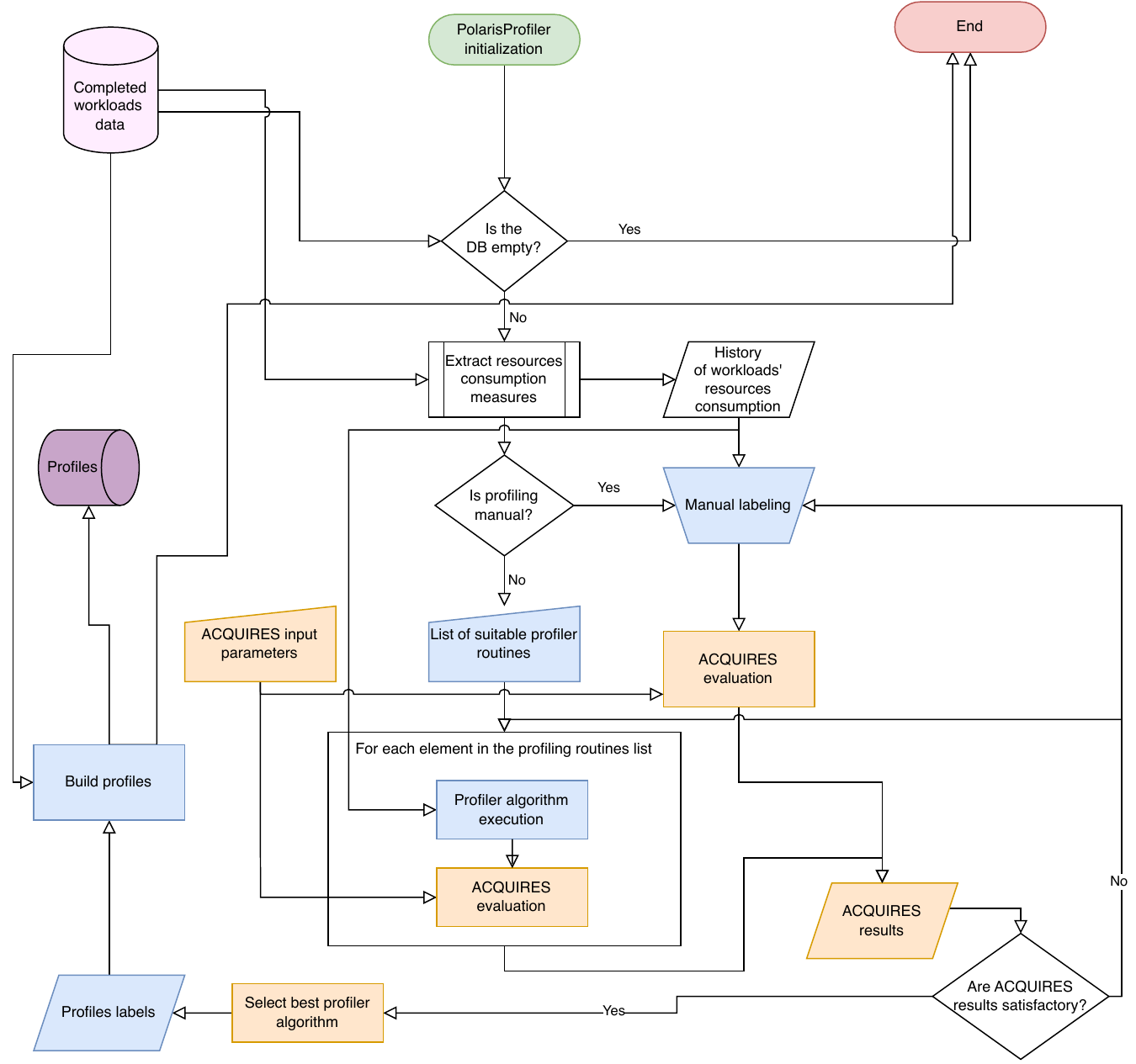}
    \caption{Flowchart diagram of the definition of the Profile Generator.}
    \label{fig:initialize-profiler}
\end{figure}

The Profile Generator is the first and the central element to develop in our approach. Its role is essential as it groups together various types of workload workload that showed comparable runtime behavior. The grouping is achieved through historical workloads usage traces to address \textbf{RC-2}, i.e., describe the workloads' runtime properties. The traces can include CPU, memory, GPU, disk usage, or execution duration measures. Figure~\ref{fig:initialize-profiler} shows the various steps of its generation.
Whatever the defined set of routines for profile groups generation, an essential step is to evaluate the categorization results; the Supervisor component takes care of this. 
At the end of the evaluation process, the selected Profile Generator mechanism produces groups to which new workload will be assigned. These groups contain relevant and specific runtime characteristics and metadata. To formalize the functioning of the Profile Generator, we define its components and objectives mathematically, ensuring a foundation for grouping workloads through a tool-independent design.
The goal of the Profile Generator is to cluster workloads $D=\{w_{1}, w_{2}, \dots, w_{n}\}$ into meaningful profile groups based on their runtime behaviors. Each workload $w_i$ is represented as $w_i = (r(w_i), m(w_i)),$ where $r(w_i) = \{r_1, r_2, \dots, r_l\}$ are the runtime features (e.g., CPU usage, memory usage, execution duration), and $m(w_i) = \{m_1, m_2, \dots, m_k\}$ are metadata features (e.g., application type, user). The Profile Generator utilizes only the runtime features $r(w_i)$ for clustering.
Clustering is based on a distance metric $d(w_i, w_j)$, which quantifies the dissimilarity between workloads based on their runtime features $d(w_i, w_j) = \text{dist}(r(w_i), r(w_j)),$ where $\text{dist}$ can be adjusted to the specific use case or picked among notorious ones, as Euclidean distance, Manhattan distance, or Cosine distance.
Alternatively, a similarity measure $s(w_i, w_j)$ can be derived as $s(w_i, w_j) = 1 - \frac{d(w_i, w_j)}{\max_{w_a, w_b \in D} d(w_a, w_b)}.$

\paragraph{Clustering objective}
The objective of the Profile Generator is to partition $D$ into an a priori or empirically defined number $k$ of disjoint profile groups $\{C_1, C_2, \dots, C_k\}$ such that $\bigcup_{i=1}^k C_i = D \quad \text{and} \quad C_i \cap C_j = \emptyset \; \text{for } i \neq j,$ and the intra-cluster distances are minimized, i.e., $\text{argmin}_{\{C_1, \dots, C_k\}} \sum_{i=1}^k \frac{1}{|C_i|} \sum_{w_a, w_b \in C_i} d(w_a, w_b).$
Workloads that do not meet a similarity threshold with any cluster are identified as outliers $\text{Outliers} = \{w \in D \mid \min_{i} d(w, \mu_i) > \tau \}$, where $\tau$ is the distance threshold.

\paragraph{Profile group representation}
Each profile group $C_i$ also includes various statistics of the workloads' runtime features (e.g., mean, variance). For example, the emerging runtime characteristics of a profile group can be represented as the mean of the runtime features of all workloads in the profile group (\textbf{centroid}) $\mu_i = \frac{1}{|C_i|} \sum_{w \in C_i} r(w)$, or the one with minimal average distance to all the other workload runtimes (\textbf{medoid}) $\mu_i = \text{argmin}_{w \in C_i} \sum_{w' \in C_i} d(w, w')$. These statistics may differ for each feature; for instance $S_{C_i}(r_j) = \{ \text{percentile}_p(r_j), \text{mean}(r_j), \text{median}(r_j) \mid p \in \mathcal{P} \},$ where $r_j$ is a runtime feature, $\mathcal{P}$ is the set of desired percentiles (e.g., 95th percentile for CPU usage, 20th percentile for memory usage), and $S_{C_i}(r_j)$ is the set of chosen statistics for $r_j$.
Furthermore, the profile group provides the metadata features $m(w)$ for all workloads in the cluster. These metadata features serve as input for the Profile Classifier.

\paragraph{General applicability}
This formalization does not prescribe specific clustering techniques, allowing flexibility. The first and more basic approach is to use manual labeling, i.e., by letting domain experts define a set of rules. Although formally feasible and appropriate for small, specific use cases, at-scale manual labeling is an impractical solution~\cite{abadi2016beckman, stonebraker2013data}, and rules updating can be cumbersome. 
When some labels or rules are available, a possible profiling routine can involve semi-supervised techniques~\cite{khan2022workload}. That way, we can learn underlying patterns in the data through a small set of labeled entries. Still, label definition is challenging and always relies on static rules.
Using unsupervised learning as well-known clustering algorithms as K-Means or DBSCAN~\cite{berral2020ai4dl,  kisous2022and, pang2022cloudcluster, park2022deepsketch, van2017automatic, wang2017lube} (eventually with the help of autoencoders~\cite{ng2011sparse}) represent for us the preferred solution as they can discover patterns without any previous knowledge. 

\paragraph{Summary}
In summary, using the introduced mechanisms and techniques, the Profile Generator addresses \textbf{RC-1} and \textbf{RC-2} and provides the input for \textbf{RC-3}. The appeal of this design is its adaptability. The Profile Generator leverages runtime features to generate profile groups that characterize workload behaviors, providing a flexible and generalizable framework adaptable to various clustering methods and tools. By focusing exclusively on runtime features, it ensures that profiles are directly reflective of execution properties, forming a solid basis for subsequent workload management. We provide the validation through empirical testing on two well-known datasets, Alibaba and Google cluster traces, showing how two different approaches can work on different data.

\subsection{Profile Classifier}

\begin{figure}[ht!]
    \centering
    \includegraphics[width=0.7\columnwidth]{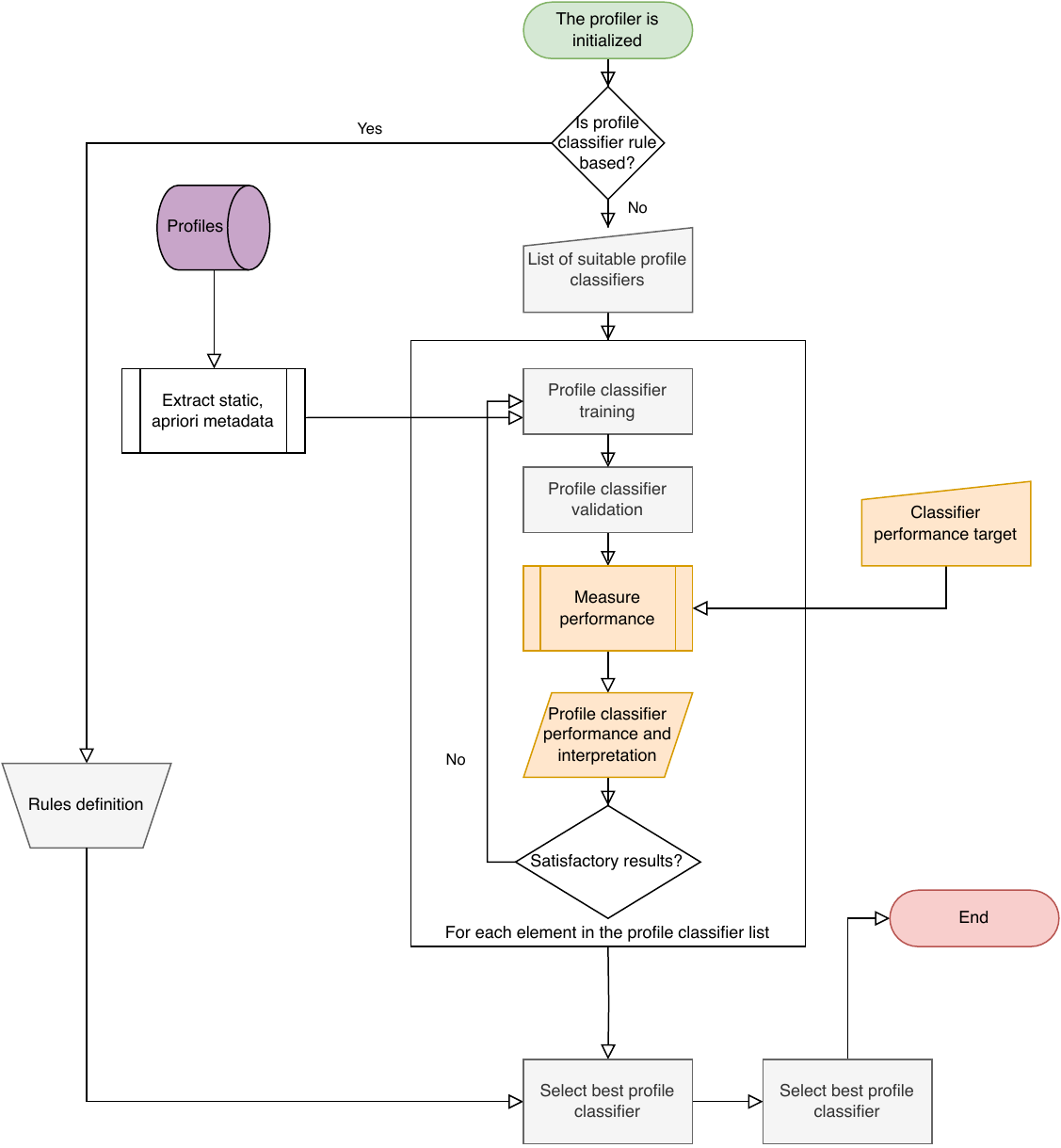}
    \caption{Flowchart diagram of the definition of the Profile Classifier.}
    \label{fig:initialize-profile-classifier}
\end{figure}


The Profile Classifier is responsible for assigning new workloads to existing profile groups based on metadata features. This process should be fast, scalable, and flexible. It builds upon the profile groups generated by the Profile Generator, leveraging the metadata of workloads within these groups for training. Figure~\ref{fig:initialize-profile-classifier} highlights the main steps. This approach highlights its novelty, as it trains exclusively on metadata features extracted from finalized profile groups.

\paragraph{Profile assignment objective}
Let $C = \{C_1, C_2, \dots, C_k\}$ be the set of profile groups generated by the Profile Generator, and let $m(w)$ represent the metadata features of a workload $w$. The Profile Classifier is a function $f: \mathcal{M} \to \mathcal{C},$ where $\mathcal{M}$ is the space of metadata features and $\mathcal{C}$ is the set of profile groups. For a new workload $w$, the classifier predicts the profile group $C_i$ with $f(m(w)) = C_i \quad \text{where} \quad C_i \in \mathcal{C}$.
Therefore, at runtime, the classifier assigns a new workload to a profile group based only on its metadata features $m(w)$, ensuring real-time efficiency and scalability.
The training dataset for the classifier consists of metadata features $m(w)$ and their associated profile group labels $C_i$: $\mathcal{D}_{\text{train}} = \{(m(w), C_i) \mid w \in C_i \text{ and } C_i \in C \}$.

\paragraph{Properties and representation of the Profile Classifier}
The Profile Classifier must satisfy key properties. First, the classifier must assign workloads to the correct profile group with high \textit{accuracy}. Validation involves computing the classification accuracy and other related metric, as for example F-Score on a test set $|\mathcal{D}_{\text{test}}|$.
Secondly, the classifier must handle large numbers of metadata features and workloads efficiently \textit{at scale}. Finally, the classifier must provide insights into how metadata features contribute to its decisions, ensuring transparency. This aspect can be achieved through \textit{interpretability} techniques like SHAP values.

\paragraph{General applicability}
The Profile Classifier is defined independently of specific implementation details. The two core elements are: (1) the inputs, i.e., the metadata features $m(w)$ of a workload, and (2) the outputs, i.e., the predicted profile group $C_i$ for a workload $w_k$.
This abstraction ensures the conceptual correctness of the model while allowing flexibility in implementation.
The Profile Classifier framework can indeed support various classification techniques, making it adaptable to different tools and technologies. It could be implemented using a rule-based classifiers, where it would leverage explicit rules derived from domain knowledge to assign workloads to profile groups. Alternatively, it could be implemented as a machine learning models, for example by employing decision trees, random forests, or gradient boosting models (e.g., XGBoost). Finally, it could also utilize deep learning for high-dimensional and complex metadata representations.
By separating the formal framework from the implementation, the Profile Classifier remains robust and versatile across diverse application scenarios.

\paragraph{Summary}
The Profile Classifier leverages the metadata features of profile groups generated by the Profile Generator, enabling accurate and interpretable assignment of new workloads. At runtime, it assigns new workloads to profile groups based solely on their metadata, ensuring scalability and real-time applicability. Its formalization ensures mathematical rigor, conceptual abstraction, and tool independence, making it adaptable to various classification techniques and real-world scenarios.

\subsection{Feedback loop}


\begin{figure}[ht!]
    \centering
    \includegraphics[width=0.7\columnwidth]{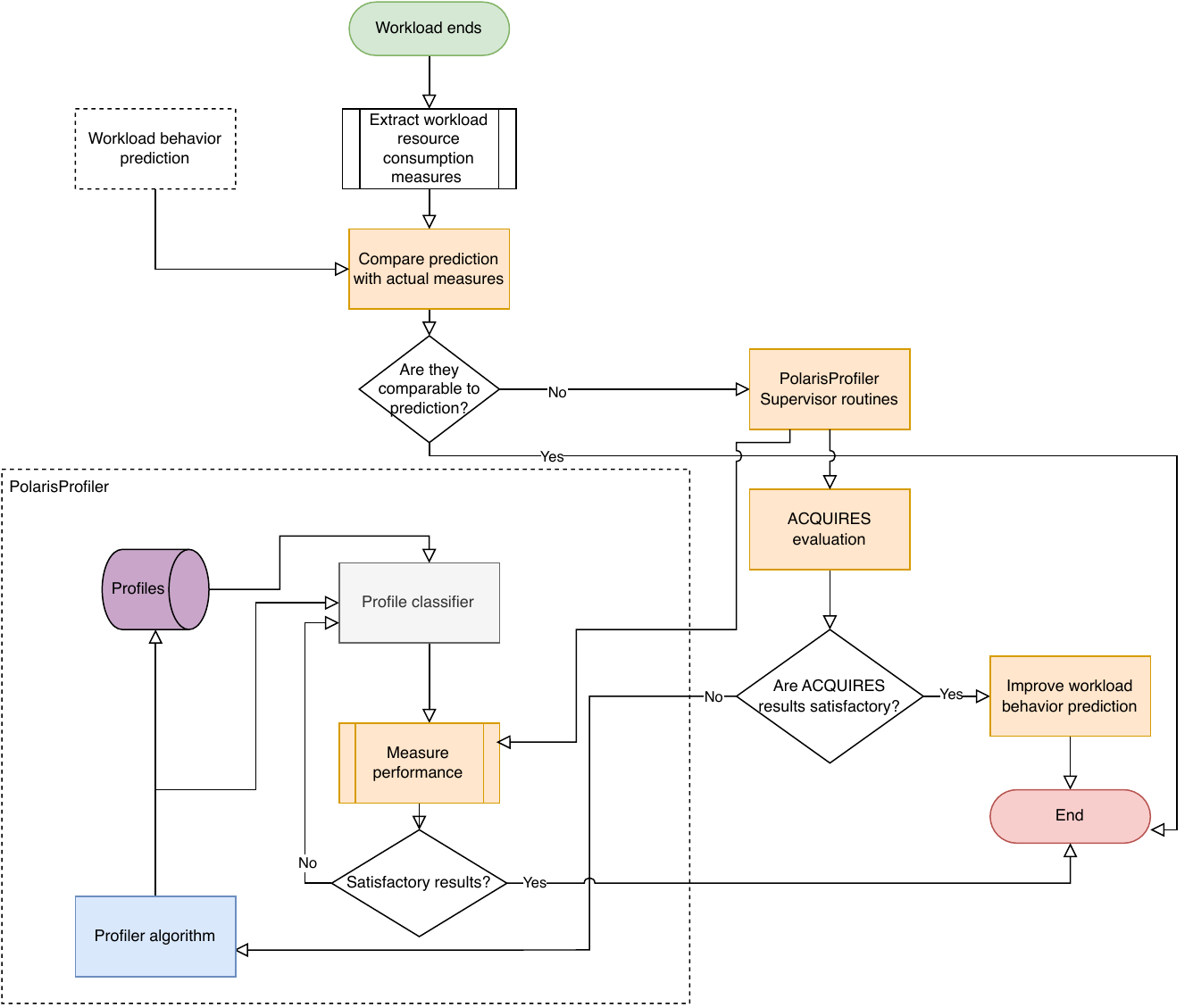}
    \caption{Flowchart diagram of the process of feedback loop for the PolarisProfiler.}
    \label{fig:PP-feedback-loop}
\end{figure}

The Feedback Loop is designed to maintain the accuracy and representativeness of the profile groups and the Profile Classifier over time. It dynamically adjusts the system based on the continuous influx of workloads and their runtime outcomes. This ensures that the profile groups remain representative and the classifier continues to assign workloads accurately. Key variables in the Feedback Loop include the frequency of updates to profile groups and the retraining of the Profile Classifier. These updates are triggered by specific conditions, such as the number of violations or the emergence of many outliers. Figure~\ref{fig:sys_logic} highlights this perspective.

\paragraph{Feedback loop objective}
Let $T$ be the set of discrete time points $\{t_1, t_2, ..., t_n\}$ where the system operates. At each time $t$, $C(t)$ represents the set of profile groups, $f(t)$ represents the classifier, $W(t)$ represents the set of active workloads. For a workload $w$ at time $t$, a violation occurs when a significant deviation between a workload's actual runtime feature values and the expected values derived from its assigned profile group. Specifically, for a runtime feature $r_j$. In detail, we express it as $V(w,t) = \bigvee_j (|r_j(w,t) - E[r_j(C_i)]| > \delta_j),$ where $r_j(w,t)$ is the actual resource usage $j$ at time $t$,$E[r_j(C_i)]$ is the expected usage from profile group $C_i$, and $\delta_j$ is a deviation threshold for the runtime feature $r_j$.

The Feedback Loop triggers updates when either of the following conditions is met: \textbf{(1)} the percentage of violations $V(w,t)$ exceeds a violation rate threshold $\tau_v$ during a time window $\Delta t$. The equation is $VR(t,\Delta t) > \tau_v$, where the violation rate is expressed as $VR(t,\Delta t) = \frac{|\{w \in W(t-\Delta t,t) : V(w,t)\}|}{|W(t-\Delta t,t)|}$. \textbf{(2)} the number of outliers, i.e., workloads not assigned to a profile group grows over a threshold: $|D_{outliers}(t)| / |D(t)| > \tau_o$. \textbf{(3)} the profile $C_i$ is not update for a while, i.e., its freshness decays after some time. We express it as $F(C_i,t) < tau_f$, where $F(C_i,t) = \exp(-\lambda(t - t_{last\_update}(C_i)))$, where $t_{last\_update}(C_i)$ is the last update time of profile $C_i$ and $\lambda$ is a decay parameter.
When one of these violations happen, i.e., when $UT(t)$ is true, according to: $UT(t) = \bigvee \begin{cases}
        VR(t,\Delta t) > \tau_v, & \text{Violation threshold} \\
        \min(CS(C_i,t)) < \tau_f, & \text{Freshness threshold} \\
        |D_{outliers}(t)| / |D(t)| > \tau_o, & \text{Outlier threshold}. \end{cases}$

Then, we trigger the update of the Profile Generator. That means When $UT(t)$ is true, we trigger a new reclustering for identifying new profile groups. Specifically, the new profile groups $C(t+1) = \text{Recluster}(D(t))$, where, $D(t)$ is the complete set of workloads up to time $t$ and $\text{Recluster}$ is the clustering algorithm (e.g., HDBSCAN). The final configuration $C(t+1)$ is persisted when $\text{ACQUIRES}(t) > \tau_{quality}$

\paragraph{Abstract representation of the feedback loop}
The Feedback Loop must satisfy the following properties. The first one is adaptivity; that means, the system must adapt to changes in workload characteristics over time.
The second characteristic is stability, which implies that the update frequency must be balanced, avoiding excessive retraining or clustering.
Finally, these updates must be accurate. In simple words, the update must improve the profile groups and, in general, reduce the number of violations.
These properties can be guaranteed through some empirical testing, e.g., by performing sensitivity analysis to determine optimal thresholds $\tau_{v}$ and $\tau_{o}$.

\paragraph{General applicability}
The Feedback Loop is defined abstractly as a monitoring and updating mechanism with three components. The input is represented by the workload outcomes, including runtime feature values and outlier detection. The outputs are updated profile groups and retrained Profile Classifier. The process happens based on triggers, i.e., conditions based on quality or freshness violations, or outliers.
This abstraction separates the monitoring logic from specific implementation details, ensuring flexibility.
The Feedback Loop framework supports various implementation techniques. For example, it can use rule-based thresholds or statistical models to detect deviations or more complex mathematical definitions. 

\paragraph{Summary}
The Feedback Loop ensures the ongoing accuracy and representativeness of profile groups and the Profile Classifier by dynamically monitoring workload outcomes and triggering updates based on violations and outliers. Its formalization supports mathematical rigor, abstract representation, and tool independence, making it robust and flexible for real-world deployment.

\subsubsection{Supervisor metrics}\label{subsubsec:supervisor}

\paragraph{ACQUIRES}\label{par:ACQUIRES}
To provide a unified measure of profile clustering performance, we develop the \textit{ACQUIRES (Algorithm's Cluster QualIty-REcall Score)} evaluation metric. 
This metric combines three complementary components, each of which carefully defined and normalized to measure distinct aspects of cluster quality: 

\begin{enumerate}
\item Outliers reduction: It measures the fraction of data points $|\mathcal O|$ classified as outliers, relative to the total number of points $|D|$. We aim at minimizing the number of outliers to ensure that the profile groups are representative of a larger fraction of the dataset. A lower outlier fraction results in a higher score. It is computed as: $\text{Outliers}_{\text{score}} = 1 - \frac{{|\mathcal O|}}{|D|}. \label{eq:outliers_score}$
\item Cluster count correctness: It evaluates how close the actual number of generated clusters ${\mathcal C}^{actual}$ matches the ideal number ${\mathcal C}^{optimal}$. The metric measures the relative deviation, ensuring that a perfect match yields a score of 1, while larger deviations lower the score. Accurate cluster counts is essential for density-based clustering methods (e.g., DBSCAN) as it can't predefined. Furthermore, we aim at penalizing over- or under-clustering, promoting balanced partitions. The equation is $|\mathcal C|_{\text{score}} = 1 - \frac{|{\mathcal C}^{optimal} - {\mathcal C}^{actual}|}{max({\mathcal C}^{optimal},{\mathcal C}^{actual})}. \label{eq:cl_size_score}$
\item Internal cohesion: We leverage the average silhouette score $\mathcal S$, a widely accepted measure~\cite{rousseeuw1987silhouettes, januzaj2023determining, morichetta2019lenta} of how well samples fit within their assigned clusters, to quantify internal cluster quality. Higher mean silhouette scores indicate more coherent clusters. In detail: ${SC}_{\text{score}} = mean(\mathcal S).  \label{eq:silh_score}$

\end{enumerate}

Each of these components focuses on a different, complementary aspect of clustering performance—outlier minimization, adherence to a desired cluster count, and internal coherence. To consolidate them into a single measure, we combine the sub-scores in a linear and equally weighted fashion. Assigning equal weights $w_1, w_2, w_3 = \frac{1}{3}$ reflects the initial assumption that all three dimensions—number of clusters, outliers, and cohesion—are of equal importance. This choice can later be refined or adjusted depending on the specific use case. For example, in the case of prototype-based clustering methods (e.g., K-Means), $w1$ and $w2$ should be set to $0$ as there are never outliers and the optimal ${\mathcal C}^{optimal}$ is defined in the initialization phase. The final equation is: $\text{ACQUIRES} =w_{1}|\mathcal C|_{\text{score}} + w_{2}\text{Outliers}_{\text{score}}  +  w_{3}{SC}_{\text{score}}. \label{eq:ACQUIRES}$

With ACQUIRES we offer a simple metric, easy to compute and interpret but at the same time that combines multiple clustering dimensions for robust evaluations.
This approach to metric design is in line with recent efforts to develop composite clustering metrics that balance multiple dimensions of quality simultaneously, and not only rely on the simple silhouette score~\cite{vardakas2024revisiting}. For instance, the Hybrid Clustering Score (HCS)~\cite{martino2022hybrid} similarly integrates well-known clustering indices (including the silhouette score) to provide a single metric that can be used for hyperparameter optimization. By adopting a comparable rationale, ACQUIRES is transparent, interpretable, and flexible, allowing it to be applied across a range of clustering tasks and domains. While this metric proves itself useful for the followed approach, future work may consider exploring alternative formulations—such as multiplicative combinations or logarithmic transformations.


\paragraph{Profile Classifier}
Concerning the Profile Classifier, the system can leverage the classic performance scores used for classification. In particular, the F-Score provides a good estimation of the label prediction distribution in a multi-label classification problem. In addition, we emphasize measuring other relevant parameters, such as execution time and resource consumption. These aspects are essential when dealing with real-time systems. Most importantly, it is crucial to have an interpretation of the results. This characteristic is of uttermost importance when dealing with the automation of complex decisions. Therefore, using model-interpretability tools aids the understanding of the decisions that the model has been taking.

\subsection{Impact for the infrastructure orchestration}

\begin{figure}[ht!]
    \centering
    \includegraphics[width=0.7\columnwidth]{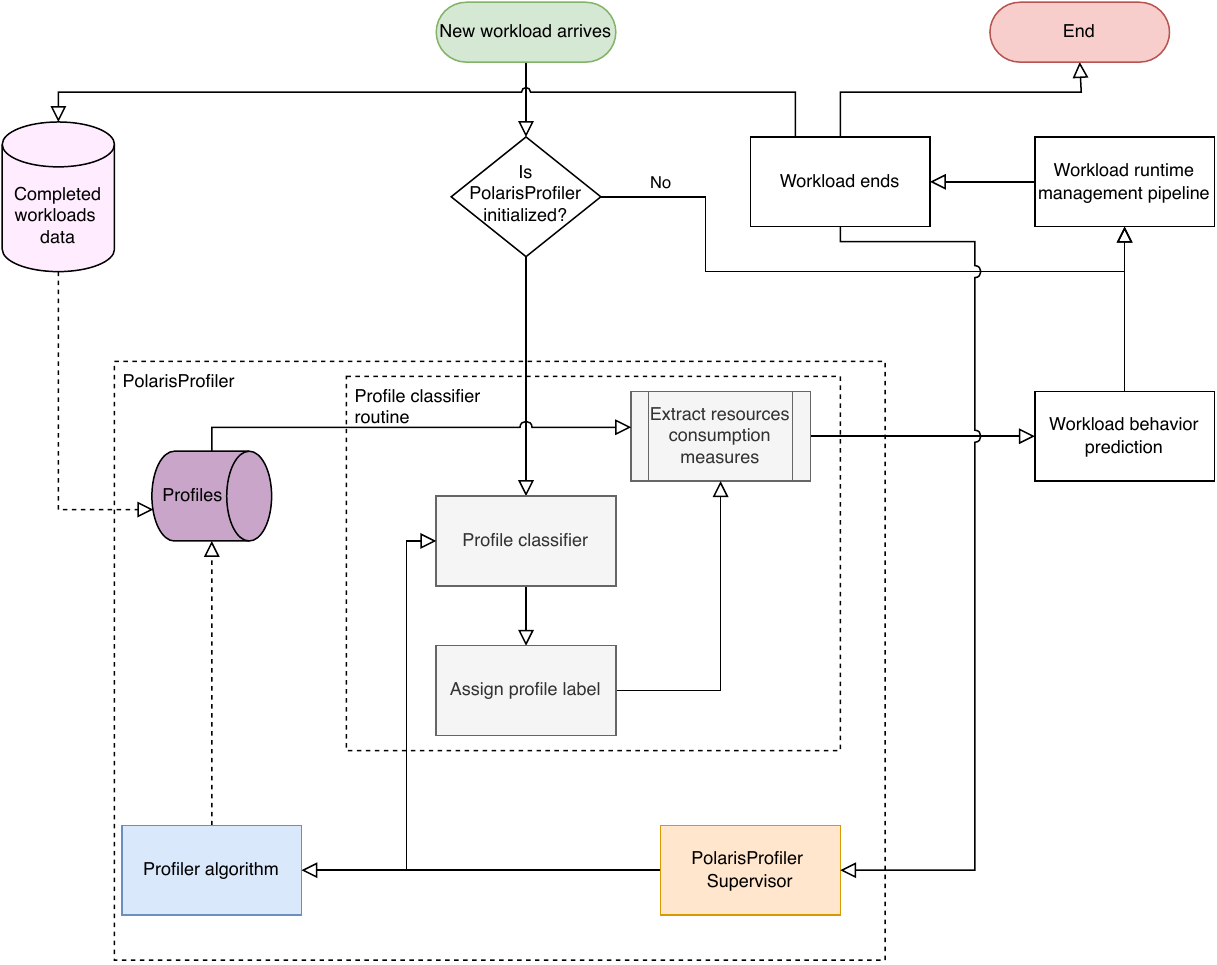}
    \caption{Flowchart diagram from the perspective of the application's workload.}
    \label{fig:formal-representation-appl}
\end{figure}

Figure~\ref{fig:formal-representation-appl} depicts in detail the information and action flow from the application's perspective. The workload, when submitted, gets into the PolarisProfiler routine. Here, it receives a label from the Profile Classifier, which pairs it with a specific profile. The system uses the information gathered from that profile to predict essential aspects of the workload execution, such as its duration and resource consumption. The predicted outlook feeds the runtime management pipeline with rich information that it can use for making better-informed decisions. 
For example, it can help the scheduling process by facilitating more informed decisions~\cite{nastic2021polaris}. Knowing the profile of a workload a priori can help in sampling more suitable machines~\cite{VPujol2023Intelligent} and filtering and scoring the ones best tailored to that model to serve the request.
Furthermore, *aaS solutions must satisfy users' SLOs~\cite{nastic2020sloc, PolarisMiddleware2021}. In this regard, achieving it in the bootstrapping phase takes work. There is a need to bring an application online and satisfy the defined SLOs by leveraging only a little information. In this context, the PolarisProfiler provides the information needed to assess the application behavior.
If we consider FaaS, there is a gap in how to tailor the correct resources from a heterogenous infrastructure~\cite{zhang2013harmony, nastic2022serverlessFabric, raith2023serverless} for specific functions.
Here, the PolarisProfiler aids in pairing the function characteristics with the most appropriate node configuration by highlighting patterns in node usage and application behavior.

\subsection{Scalability considerations}
\label{sec:methodology-scalability}
PolarisProfiler is built to profile workloads efficiently, sidestepping the intensive demands of runtime profiling by leveraging static metadata. This approach enables swift and accurate workload categorization. However, as workloads grow more diverse and accumulate, managing this complexity becomes challenging. To address this, PolarisProfiler can rely on incremental clustering, allowing new workloads to integrate seamlessly without disrupting the existing setup. For handling a surge in workloads that need to join the infrastructure, horizontal scalability offers an effective solution. Drawing on insights from Faroughi et al.~\cite{faroughi2018achieving}, horizontal scalability in density-based clustering can be achieved by splitting workload inputs into smaller pieces and distributing them across multiple nodes or parallel jobs, ensuring both efficiency and accuracy.

\section{Case Study}\label{sec:case_study}
We provide a reference implementation of PolarisProfiler and its main profiling processes. 
Specifically, we develop a profiling approach to optimize the scheduling of Machine Learning (ML) workloads. 
The rationale for targeting machine learning workload is that it represents a current challenge for large and distributed systems~\cite{verbraeken2020survey}. 
The need for a large amount of data and an increased necessity for the computing power of energy poses serious questions~\cite{bender2021dangers} and calls for optimization strategies both from the AI and systems communities.
Furthermore, the variety of algorithms and the specific behavior of ML models makes it not trivial to uncover utilization patterns~\cite{wan2021machine}. Therefore, guaranteeing SLOs while optimizing the infrastructure usage requires more elaborate strategies.

\subsection{Dataset}
Our study considers two months of ML workload traces (jobs) from the Alibaba Platform for Artificial Intelligence (PAI)~\cite{weng2022mlaas}. 
The platform's main target is businesses within the Alibaba group. It enables AI pipelines, offering different levels of abstraction, from a canvas UI where the users can drag and connect the elements for their pipeline to containers.
Once submitted, the supported frameworks~\footnote{PAI accepts frameworks like TensorFlow, PyTorch, Graph-Learn, and RLlib.} translate each workload into tasks with different roles, e.g., \textit{parameter servers (PS)} and \textit{workers} for a training workload and \textit{evaluator} for inference. 
Each task has one or more instances, deployed using Docker, and can run on multiple machines.
This dataset is relevant to our case study, showing several key characteristics. 
First of all, it contains real traces, reporting real machine usage. Furthermore, it discloses descriptive static and a priori metadata. The most suitable metadata contained in Alibaba's dataset is the user's name (\userFeature{}), the workload name (\jobNameFeature{}), the model used (\workloadFeature{}), and the type of the task, e.g., if it is training or inference and which architecture uses  (\taskNameFeature{}). Plus, the Alibaba trace comes with a \groupFeature{} tag, i.e., meta-information specified by tasks, such as entry scripts, command line parameters, data source, and sinks.

\begin{table}[]

\caption{Stratified sampling of 100\,001 elements based on the workload metadata feature.}
\label{tab:stratified_sampling}
\centering

    \begin{tabular}{l|c|c}
    \multicolumn{1}{r|}{\textbf{Workload}}                  & \multicolumn{1}{c|}{\textbf{Size}}    & \multicolumn{1}{c}{\textbf{Sampled size}} \\ \hline
bert                                                    & 10\,940\,142                          & 29\,818                                     \\
ctr                                                     & 9\,128\,957                           & 24\,881                                     \\
graphlearn                                              & 4\,888\,371                           & 13\,323                                     \\
inception                                               & 10\,781\,289                          & 29\,385                                     \\
nmt                                                     & 13\,537                               & 37                                        \\
resnet                                                  & 60\,863                               & 166                                       \\
rl                                                      & 849\,626                              & 2\,316                                      \\
vgg                                                     & 11\,768                               & 32                                        \\
xlnet                                                   & 15\,632                               & 43                                        \\ \hline \hline
\begin{tabular}[c]{@{}l@{}}\textbf{Total} \textbf{size}
\end{tabular} & 36\,690\,185                      & 100\,001                                 
\end{tabular}
\end{table}

We start with the assumption that we do not have insights about the Alibaba system.
First, we construct our case study filtering out all the workloads that are not \textit{terminated} since we do not have the resource usage information for them, obtaining circa 36 million instances. 
Then, we use stratified sampling to reduce the set to a manageable size. We base the stratification on the workload type, which, through the model names, gives us an explicit and more transparent understanding of the workloads and their instances. 
We extract a dataset $D$ with a cardinality $|D| = \text{100\,001 elements}$.  Table~\ref{tab:stratified_sampling} shows the categories and their sampled sizes.

We rely on 17 usage metrics to represent the workload runtime behavior.\footnote{Namely: the number of instances for that workload (\textit{inst num}), the starting and ending time (\textit{start time} and \textit{end time}), the planned resource usages (\textit{plan cpu}, \textit{plan mem}, and \textit{plan gpu}. Plus, the dynamic utilization metrics like \textit{CPU usage}, \textit{memory usage} (average and maximum), \textit{GPU usage}, \textit{GPU memory usage} (average and maximum), number of inputs and outputs (\textit{read count} and \textit{write count}), number of bytes	exchanged (\textit{read}, and \textit{write}) and the total workload \textit{duration}.} However, we need to verify that this information is capable of expressing relationships between workloads. We do so by relying on the Hopkins statistics~\cite{banerjee2004validating}. 
This test measures how well the data can be grouped, relying on the hypothesis that the data follows a Poisson point process. 
It outputs a score: if equal or above 0.3, the data have random distribution; the closer the values go to zero, the more the data could follow clusters.
We rely on the Python \texttt{pyclustertend} library for our analysis, that uses as default distance ``Minkowski,'' which results in the standard Euclidean distance.\footnote{\url{https://pyclustertend.readthedocs.io/en/master/}}${}^{,}$~\footnote{\url{https://scikit-learn.org/stable/modules/generated/sklearn.neighbors.BallTree.html}}
For the set $D$, the Hopkins score is \textbf{0.0033}, letting us believe in the possibility of obtaining meaningful profiles.

\subsection{Testing simple rule matching for profile generation}\label{subsec:dataset}
As we point out in the introduction, most profiling methods rely on \textit{occurrence consistency}~\cite{ferguson2012jockey, jyothi2016morpheus, jalaparti2015network}.
To test how this approach would work in the Alibaba case study, we assemble a \textit{baseline test} to evaluate the performance of single or combined static a priori metadata. The idea is to mimic the domain expert rule generation.
For this task, we rely on \workloadFeature{} and \taskNameFeature{}, who represent the most understandable metadata.
We analyze how well a single or a small group of metadata features can group workloads that behave similarly, i.e., that are close to our 17-dimensional problem (considering the 17 resource utilization metrics).
For the evaluation, we use the well-established unsupervised metric \Gls{silhouette}~\cite{rousseeuw1987silhouettes} ($SC_{score}$). It tells in a $[-1, 1]$ range how well each point lies within its group.\footnote{values closer to 1 representing a better fit} 
For a better assessment, we consider three distance measures: \euclideanDist{}, \cosineDist{}, and \manhattanDist{}.
For workload, the \manhattanDist{} achieved the highest score (0.21). Task name had the highest score with \euclideanDist{} (-0.07), though all values were negative. The combination of Workload and Task name also performed best with \manhattanDist{} (0.08).
This analysis suggests that a combination of metadata labels will be required to identify profiles; further, leveraging the workload's historical resource usage will also ensure that the obtained groups are cohesive. The goal is to have profiles that prove to be more cohesive than the groups obtained in this baseline test using just one metadata label.

\subsection{Developing the Profile Generator with unsupervised learning}\label{subsec:gen_profiles}
Here, we inspect which method can best implement the Profile Generator for the Alibaba dataset. 

\subsubsection{Candidate algorithms}
We focus our examination on density-based methods, as they generate an ``outliers' group,'' i.e., workloads not fitting any cluster, letting us explore irregular workloads and detect peculiar behaviors. Furthermore, they don't require input on the number of clusters. This aspect allows us to specify a desired target but lets the algorithm be free of movement.
In particular, here, we aim to have fine-grained clustering; therefore, we focus on methods that generate groups at different data densities.
The main algorithms are \textit{HDBSCAN}~\cite{campello2013density, mcinnes2017hdbscan} and \textit{OPTICS}~\cite{ankerst1999optics}.
HDBSCAN~\cite{campello2013density, mcinnes2017hdbscan} seeks to solve the single-density problem by generating a tree representation of all the possible clusters using a \textit{single-link} approach. Then, it extracts the best clusters by optimizing the overall cluster stability.
Similarly, OPTICS~\cite{ankerst1999optics}, given a fixed value for the minimal cluster size, draws out higher density clusters by looking at lower density ones. 

\subsubsection{Grid search for best Profile Generator model}
We aim at identify the best configuration for the Profile Generator, testing \textit{HDBSCAN} and \textit{OPTICS} on the case study dataset $D$ with combinations of four different parameters.
The first one is the data \textit{transformation} tool for the dynamic workload feature. 
We consider the \StandardScalerTransform{}, the \MinMaxScalerTransform{}, the \RobustScalerTransform{} -- particularly suitable for noisy datasets -- and the \PowerTransform{}, which produces a monotonic transformation.
An essential element in clustering is the \textit{distance metric}. For our scenario, we choose the \euclideanDist{} and the \manhattanDist{}.
Finally, both \textit{HDBSCAN} and \textit{OPTICS} need to input a parameter specifying the number of minimum points per cluster, \textit{MinPoints}. 
We choose a range from 50 to 1\,000~\footnote{Specifically, $\{50, 100, 200, 300, 400, 600, 1\,000\}$} to balance granularity and cluster representativeness.





\begin{table}[]
    \caption{Summary of the clustering results}
    \label{tab:clustering_summary}
    \centering
    
    \resizebox{0.6\columnwidth}{!}{
        \begin{tabular}{cc|l|l|l|l|l}
            \textbf{}                                         & \multicolumn{1}{r|}{\textbf{}} & \multicolumn{1}{r|}{\textbf{\begin{tabular}[c]{@{}r@{}}\#\\ outliers\end{tabular}}} & \multicolumn{1}{r|}{\textbf{\begin{tabular}[c]{@{}r@{}}\#\\ clusters\end{tabular}}} & \multicolumn{1}{r|}{\textbf{\begin{tabular}[c]{@{}r@{}}mean \\ $|$C$|$\end{tabular}}} & \multicolumn{1}{r|}{\textbf{\begin{tabular}[c]{@{}r@{}}avg\\ $SC_{score}$\end{tabular}}} & \multicolumn{1}{r}{\textbf{\begin{tabular}[c]{@{}r@{}}DB\\ score\end{tabular}}} \\ \hline
            
            \multicolumn{1}{c}{\multirow{4}{*}{\textit{HDBSCAN}}} & mean  & \textbf{40331.1}  & 68.6   & \textbf{1993.6} & 0.44 & \textbf{1.52} \\
            \multicolumn{1}{c}{}                                  & min & \textbf{18059} & \textbf{10} & \textbf{254.9} &\textbf{ 0.23} & \textbf{1.19} \\
            \multicolumn{1}{c}{}                                  & max & \textbf{64770} & \textbf{243} & \textbf{5859.5} & 0.65 & \textbf{2.01}\\
            \multicolumn{1}{c}{}                                  & std & 11124.0  & 70.3 & 1509.3 & 0.08 & 0.19 \\
            \hline
            \multicolumn{1}{c}{\multirow{4}{*}{\textit{OPTICS}}}  & mean  & 79578.3 & \textbf{66.2} & 889.5 & \textbf{0.60} & 1.19 \\
            \multicolumn{1}{c}{}                                  & min  & 66708 & 1 & 120.6 & -1.00 & 0.98      \\
            \multicolumn{1}{c}{}                                  & max & 98936 & 271 & 2979.5 & \textbf{0.81} & 1.40    \\ 
            \multicolumn{1}{c}{}                                  & std & 7643.2  & 79.3 & 752.1 & 0.20 & 0.10   \\
        \end{tabular}
   }
\end{table}

\begin{figure}[ht!]
  \centering
  \subfloat[a][\textbf{Minimum cluster size - ACQUIRES results} relation.]{\includegraphics[width=0.33\textwidth]{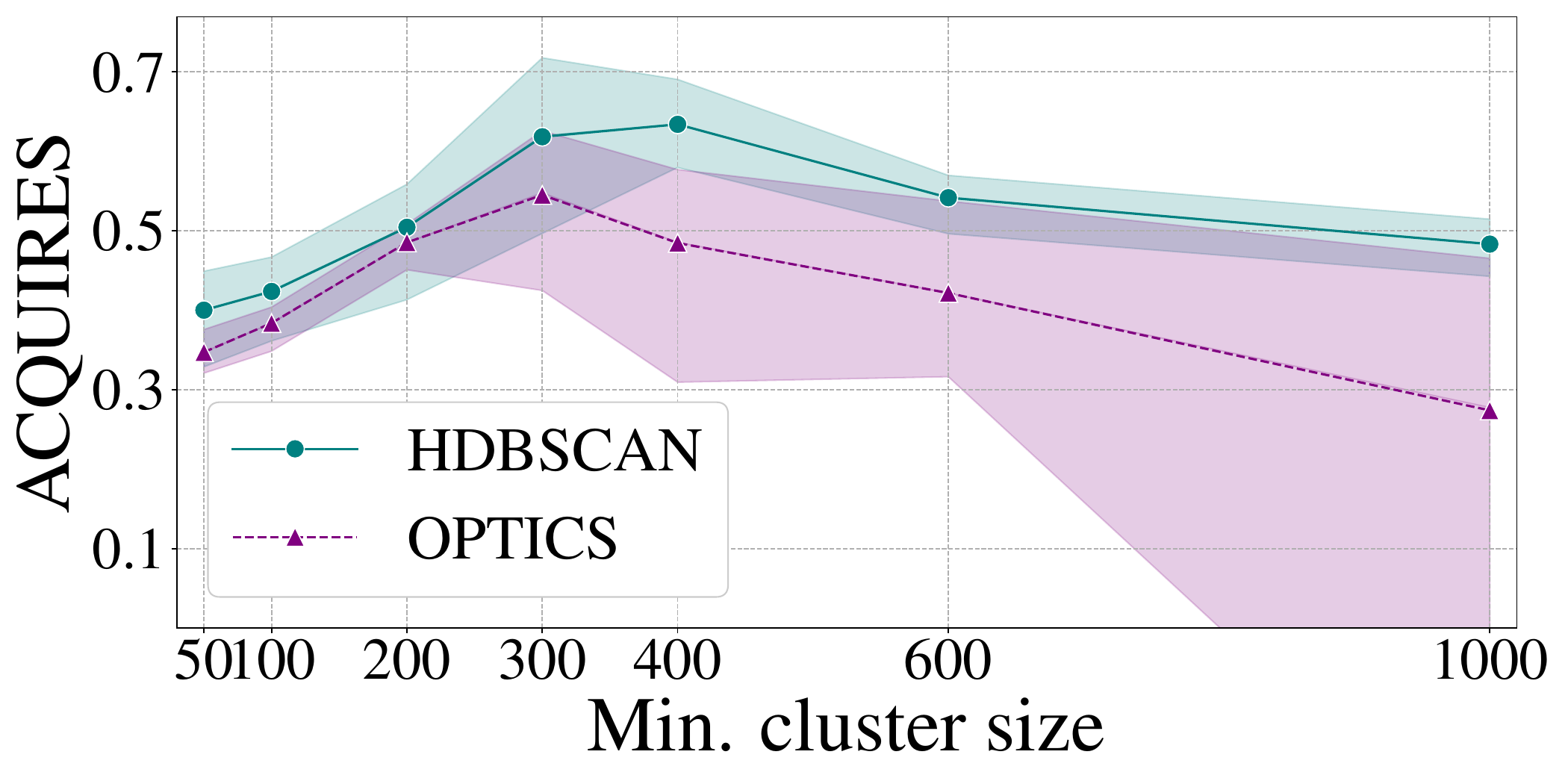} \label{fig:minsize-scorev1}}
  \subfloat[b][\textbf{Transform function - ACQUIRES results} relation.]{\includegraphics[width=0.33\textwidth]{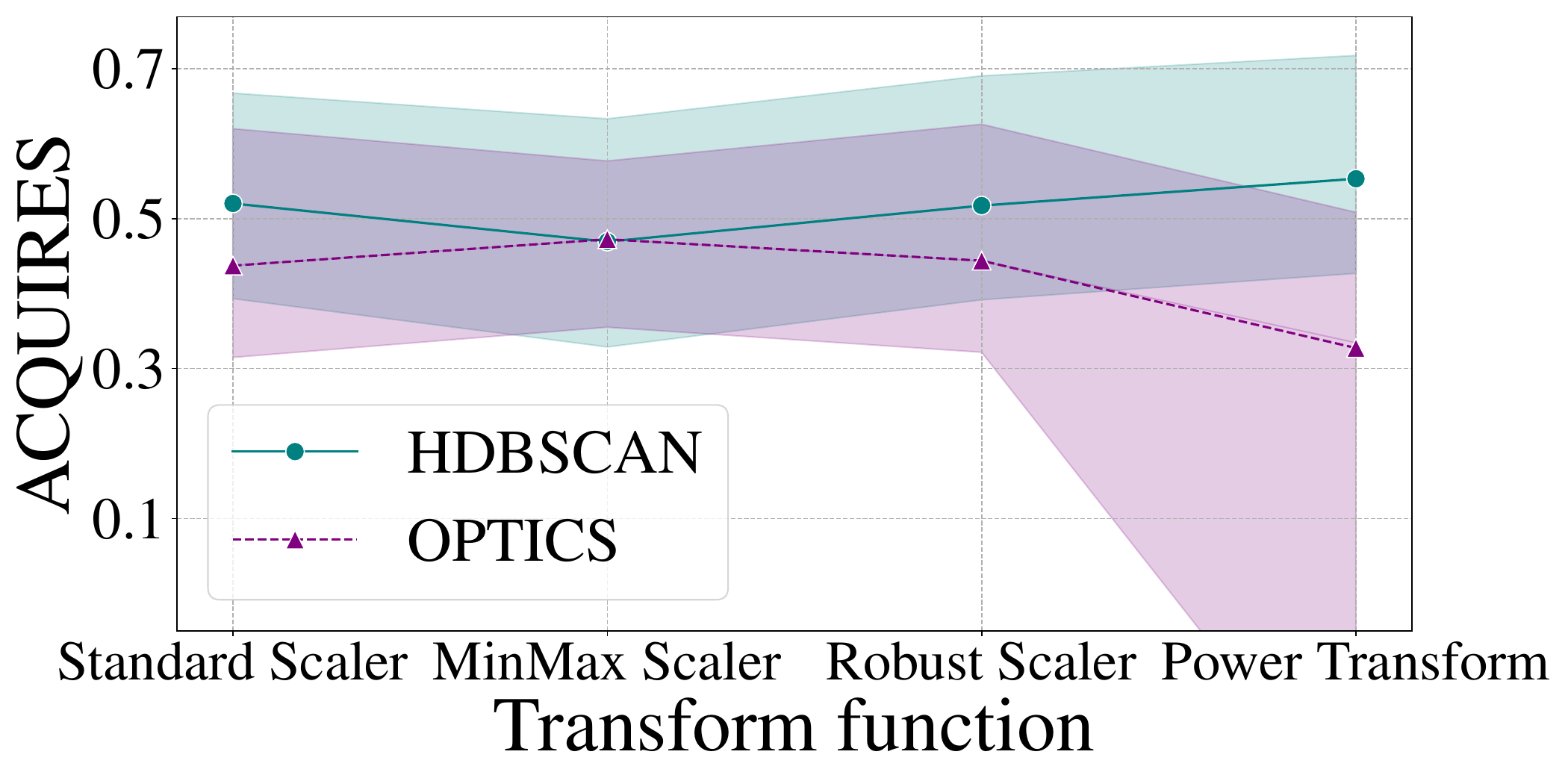} \label{fig:transform_funct-scorev1}}
  \subfloat[c][\textbf{Distance metric - ACQUIRES results} relation.]{\includegraphics[width=0.33\textwidth]{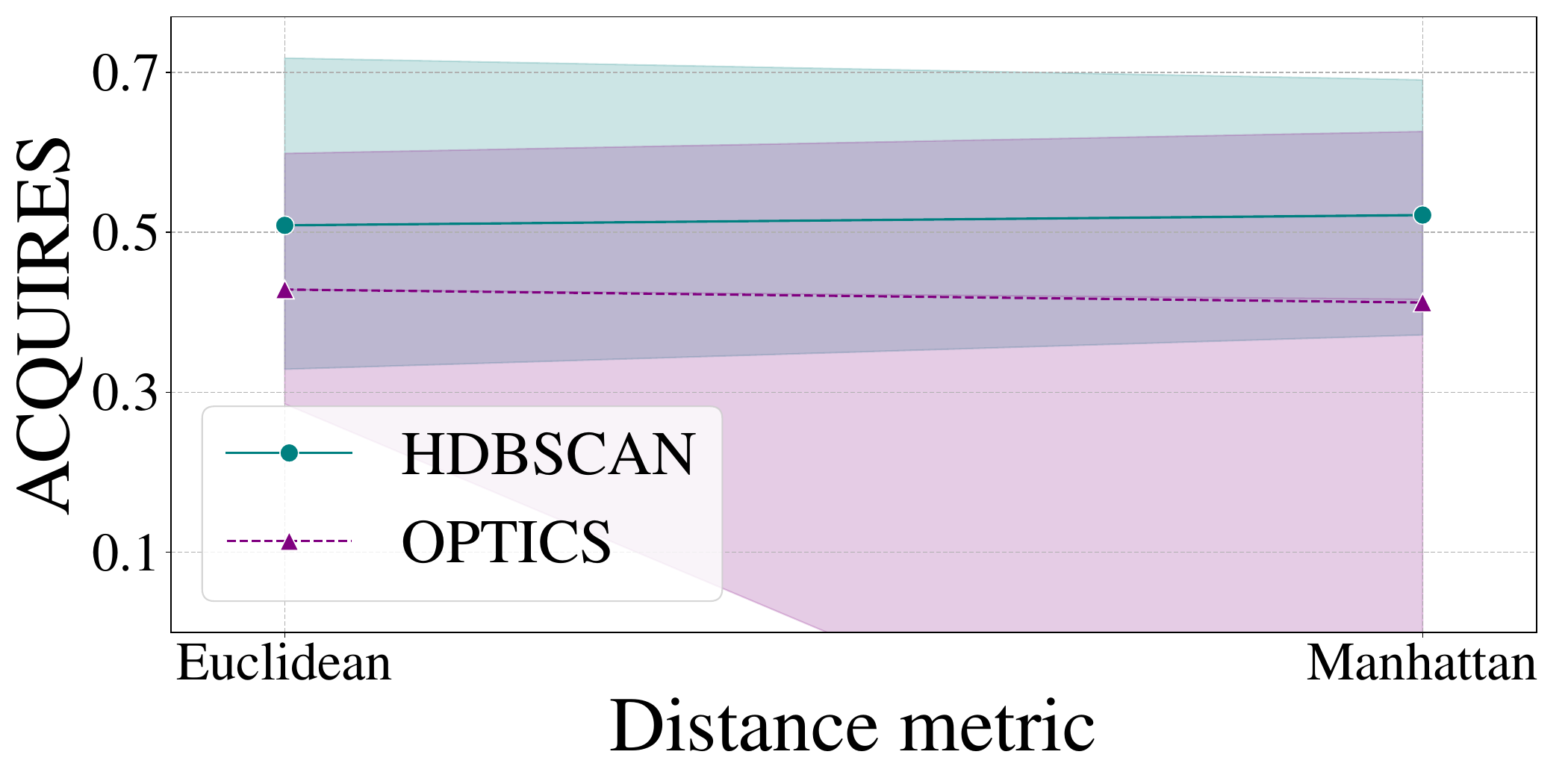} \label{fig:distance-scorev1}} 
  
  \caption{The plots depict the relationship between the main search parameters and the final score. The purple solid line and the blue dashed one represent the central tendency for HDBSCAN and OPTICS, respectively. The two colored areas show the confidence interval.} \label{fig:score1-results}
\end{figure}

We extract several statistics for each clustering result ($\mathcal C$). 
We consider the number of clusters generated, how many outliers $\mathcal O$ the clustering detects, and the average cluster size. 
Furthermore, we rely on unsupervised performance metrics, such as the overall $SC_{score}$ and the Davies Bouldin Score (DB Score). 
In addition, in our use case, we want to maximize the number of clustered points to have a significant representation in the profiles. 
Finally, we want to have an adequate number of clusters. We want to have more than one big group and avoid many small clusters. Therefore, we look at having a good balance between the number of clusters and their cardinality (mean $|$C$|$).
Table~\ref{tab:clustering_summary} summarizes the main statistics for HDBSCAN and OPTICS. 
We can see how HDBSCAN outperforms (highlighted in bold) OPTICS for most parameters. 
Furthermore, as we want to have a unique evaluation score, we show the ACQUIRES score (introduced in Section~\ref{par:ACQUIRES}) for the various HDBSCAN and OPTICS configurations. Figure~\ref{fig:score1-results} summarizes the results. The solid purple line and the blue dashed one represent the central tendency for HDBSCAN and OPTICS, respectively. The purple and blue areas show the value intervals. As we can notice, HDBSCAN generally guarantees a better ACQUIRES value for all the main search dimensions, i.e., the minimum cluster size, the transform function, and the distance metric. This behavior finds its ground because HDBSCAN produces far fewer outliers, a good number of clusters, and a solid $SC_{score}$. In particular, as summarized by the plots, we obtain the best results with HDBSCAN, using a minimum cluster size of 300 (Figure~\ref{fig:minsize-scorev1}), the \PowerTransform{} function (Figure~\ref{fig:transform_funct-scorev1}), and the \euclideanDist{} distance (Figure~\ref{fig:distance-scorev1}).

\subsubsection{HDBSCAN evaluation}\label{subsec:clusters_evaluation}

\begin{figure}[t]
    \centering
    \includegraphics[width=0.5\columnwidth]{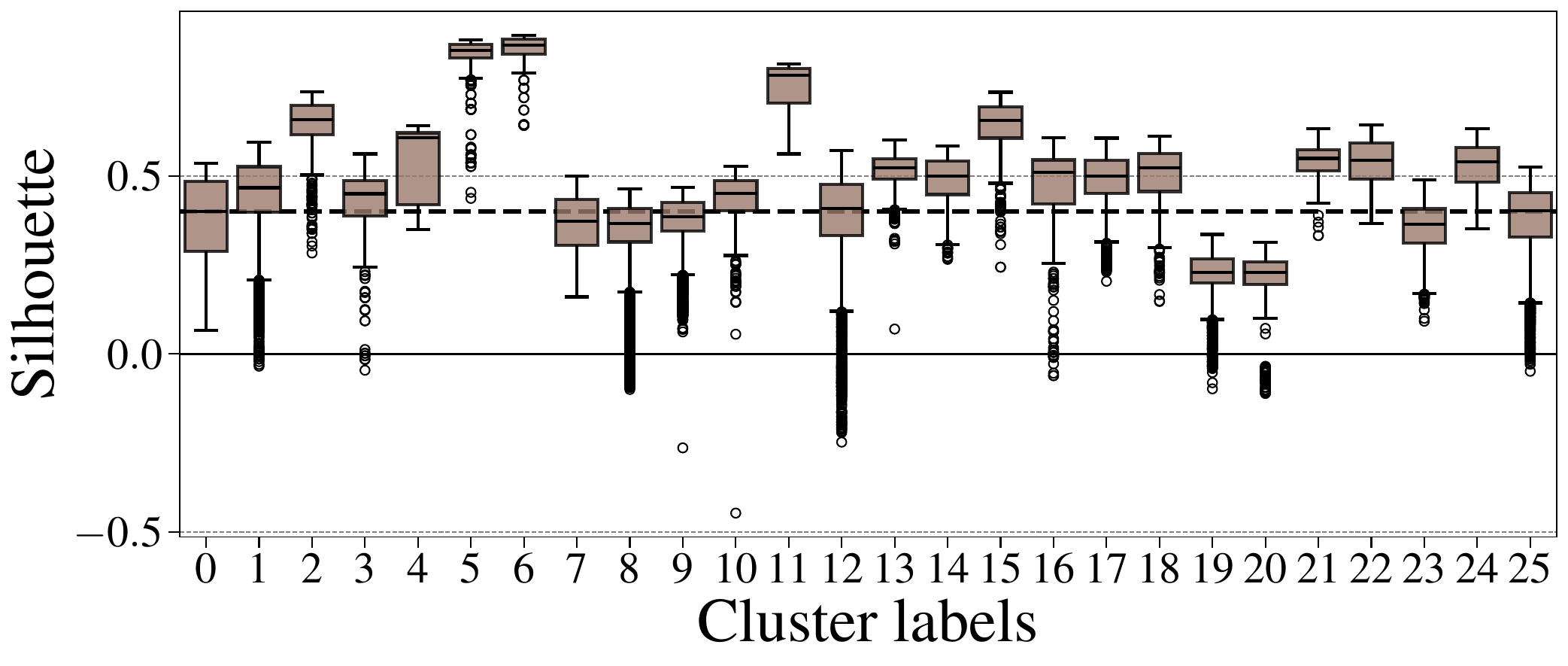}
    \caption{The silhouette coefficient scores for the points in the extracted clusters.}
    \label{fig:HDBSCAN_silh}
\end{figure}

Here, we inspect in detail the results of the selected HDBSCAN approach for generating dynamic profile models. 
First, we examine the performance in terms of cluster separation, relying again on the $SC_{score}$. 
We keep out from this analysis the ``outliers group.''
Figure~\ref{fig:HDBSCAN_silh} depicts the results.
The x-axis shows the profiles (each value, a profile). The y-axis reports the $SC_{score}$. 
The boxplots depict the variation of the $SC_{score}$ values for the points in each cluster.
The orange line represents the total average $SC_{score}$.
Figure~\ref{fig:HDBSCAN_silh} shows that overall, clusters 8 and 9 have a good $SC_{score}$, despite their large size.
Profiles 5, 6, and 11 are the ones that have the best $SC_{score}$. 
Their low cardinality and sample fit suggest that they represent particular and homogeneous workload instances. 
However, profile 10 has a significant amount of not well-fitted samples. 
Even if this last behavior is not negligible, it is unrealistic to expect perfect results with such cardinality.
Overall, the results are well grounded and show how, in the case study, HDBSCAN is a good candidate to implement our workload profile generator.

\begin{figure}[ht!]
  \centering
  \subfloat[a][CPU usage.]{\includegraphics[width=0.5\textwidth]{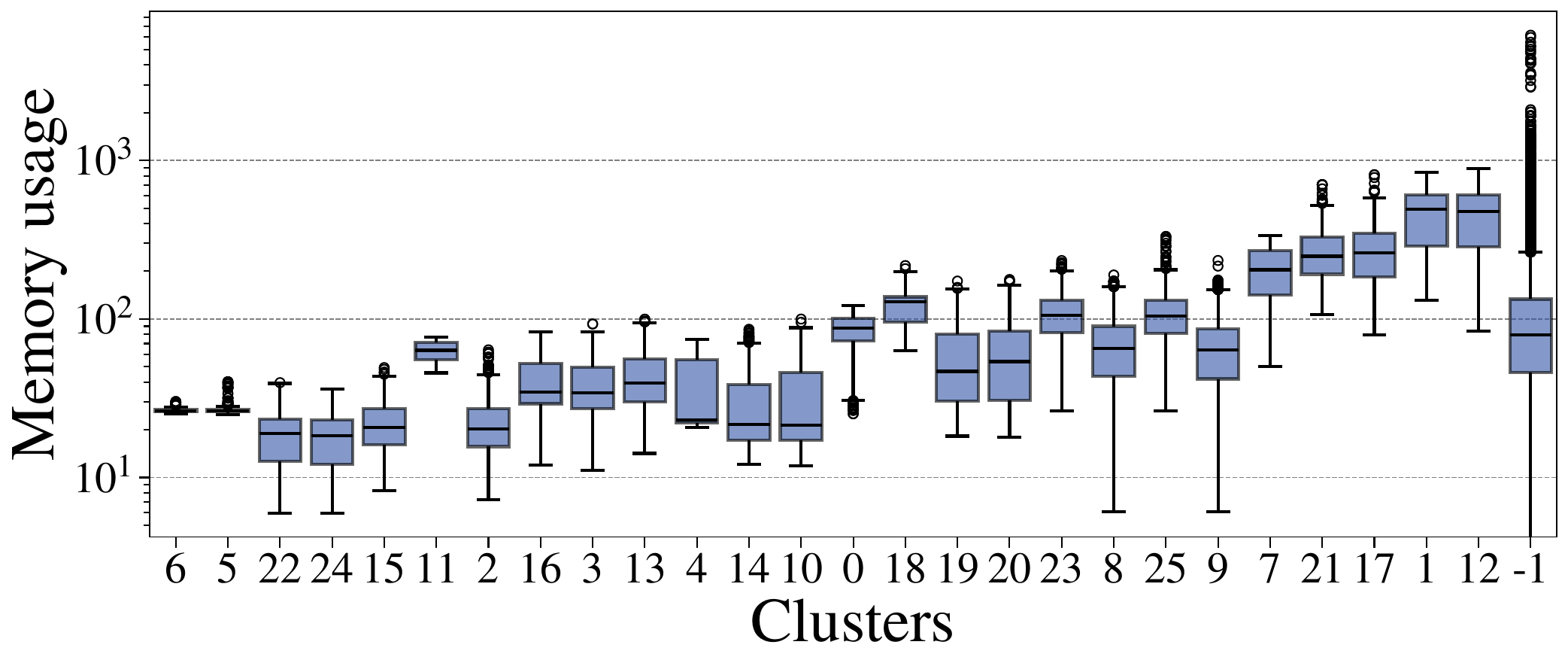} \label{fig:cpu_usage-boxplot}}
  \subfloat[b][GPU working utilization.]{\includegraphics[width=0.5\textwidth]{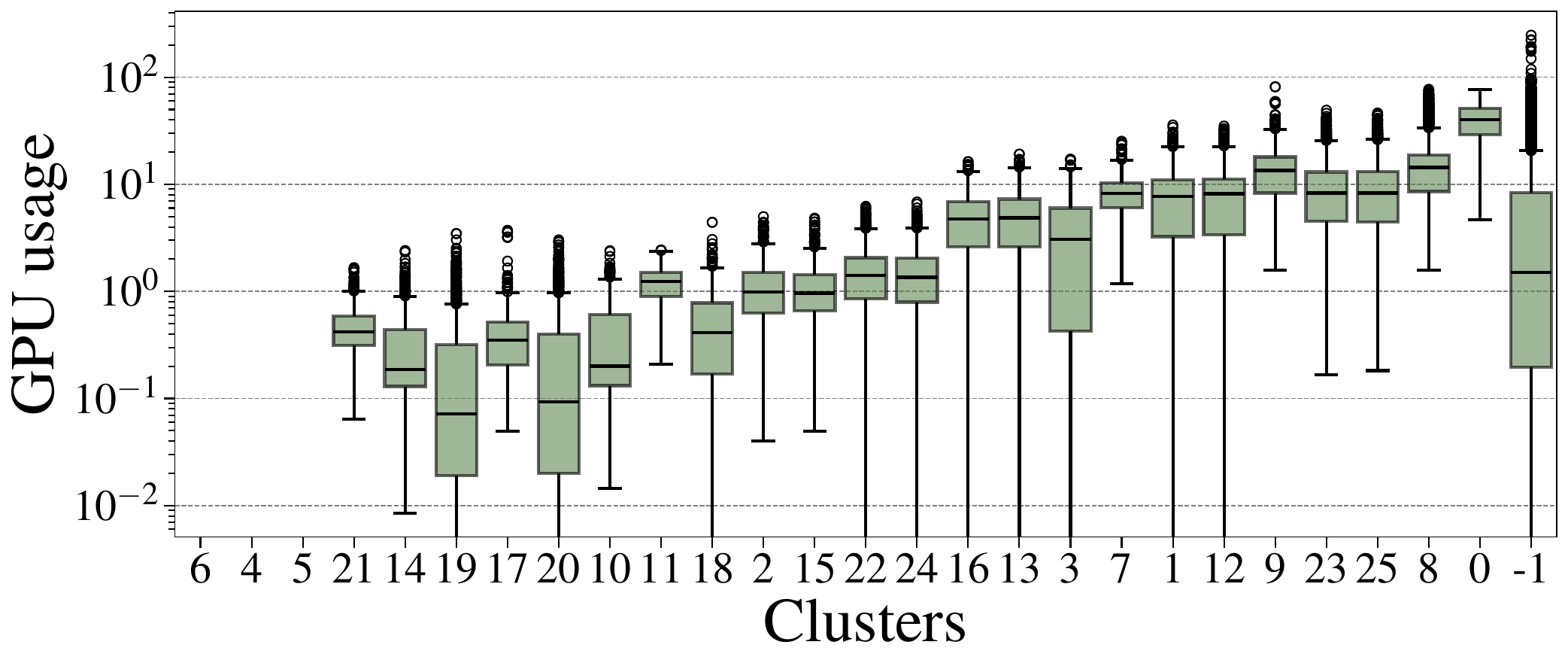} \label{fig:GPU-boxplot}} \\
  \subfloat[c][Memory usage.]{\includegraphics[width=0.5\textwidth]{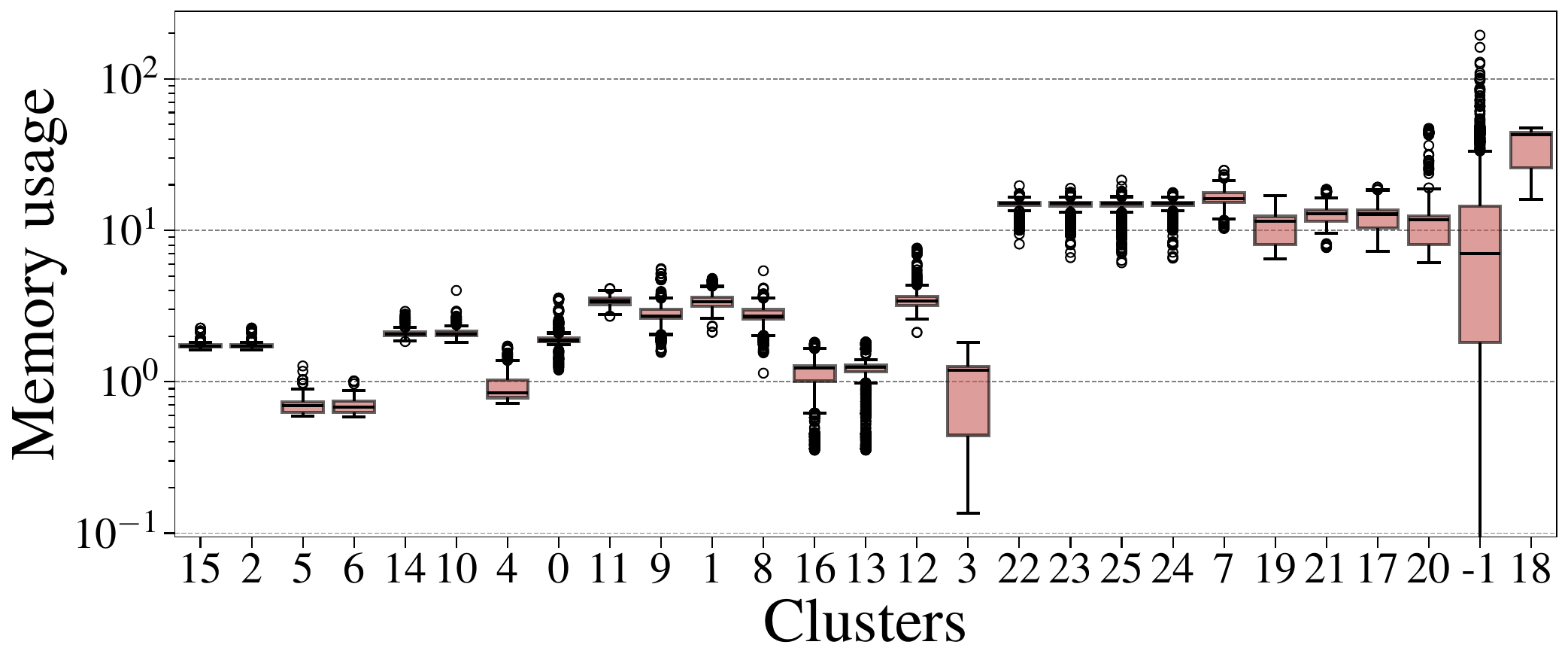} \label{fig:memory-boxplot}}  
  \subfloat[d][Workload duration.]{\includegraphics[width=0.5\textwidth]{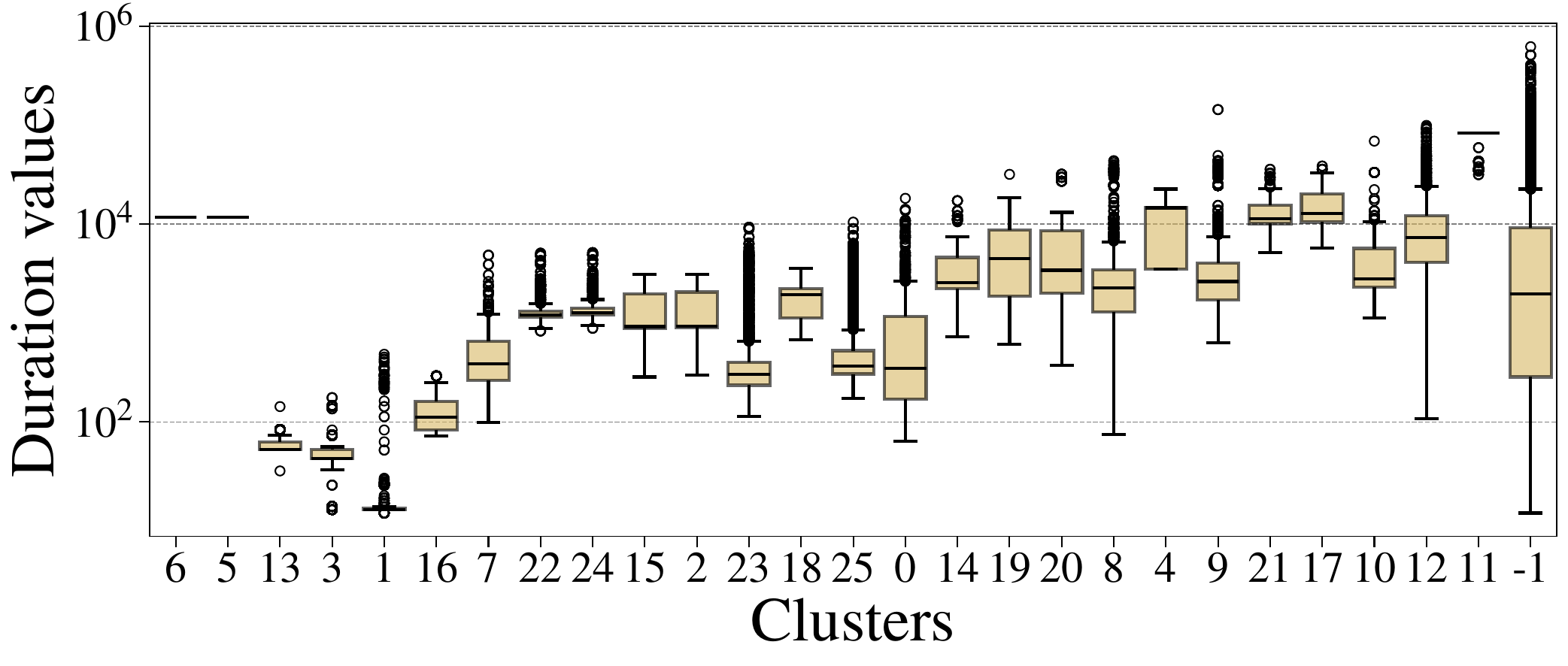} \label{fig:duration-boxplot}} 
  
  \caption{Box plots representing the distribution in the clustered profiles of CPU, GPU, Memory, and Duration. The y axis is in logarithmic scale.} \label{fig:clusters-dynamic-workload-data}
\end{figure}

\paragraph{Dynamic infrastructure usage data} 
We now represent the range of workload performance within the profiles to understand the core dynamic profile model. 
In detail, we examine the distribution of resource usage values across the profiles and their variability in each cluster. 
Figure~\ref{fig:clusters-dynamic-workload-data} depicts the results. 
Due to the page limit, we focus on the most representative features for the case study: CPU usage, memory usage, 
and GPU utilization, and workload duration. The color code is blue for CPU usage, green for GPU usage, red for memory usage, and yellow for workload duration; it is invariant and consistent from now on.
The figure shows the boxplots of the feature values grouped by the cluster labels. 
The plots sort the profiles, in the x-axis, by the considered feature standard deviation, in ascending order; higher values are at the right of the plot. 
We sort by the standard deviation to highlight the value of cohesiveness within each profile. 
The y-axis shows, for each feature, their values.\footnote{In Figure~\ref{fig:duration-boxplot}; we express values as multiples of $10^3$.}
The overall results show that most of the profiles have relatively low variation. The exception is the ``outliers group,'' labeled as ``-1,'' which naturally contains all the workloads that do not fit in the main profiles.
A particular case is maximum memory usage, where profile 18 has a broader value range than the outliers group. 
As a possible cause, this profile contains few workload samples and might include peculiar workloads. 
On the contrary, sizeable profiles, like 8 and 9, show a good homogeneity, with generally few noisy points present.
Looking back at the high-level representation of clustering results in Figure~\ref{fig:HDBSCAN_silh}, we can see that this cluster has low cardinality and might include peculiar workloads. On the contrary, large clusters, like 8 and 9, show a good homogeneity, with generally few noisy points present.
Overall, this first analysis suggests that the HDBSCAN clustering has managed to find homogeneous groups of workloads. 
Furthermore, such representation demonstrates the contribution of profiles to the estimation of the runtime characteristics of a workload.

\begin{figure}[ht!]
  \centering
  \subfloat[a][Top-10 Job names.]{
    \includegraphics[width=0.45\textwidth]{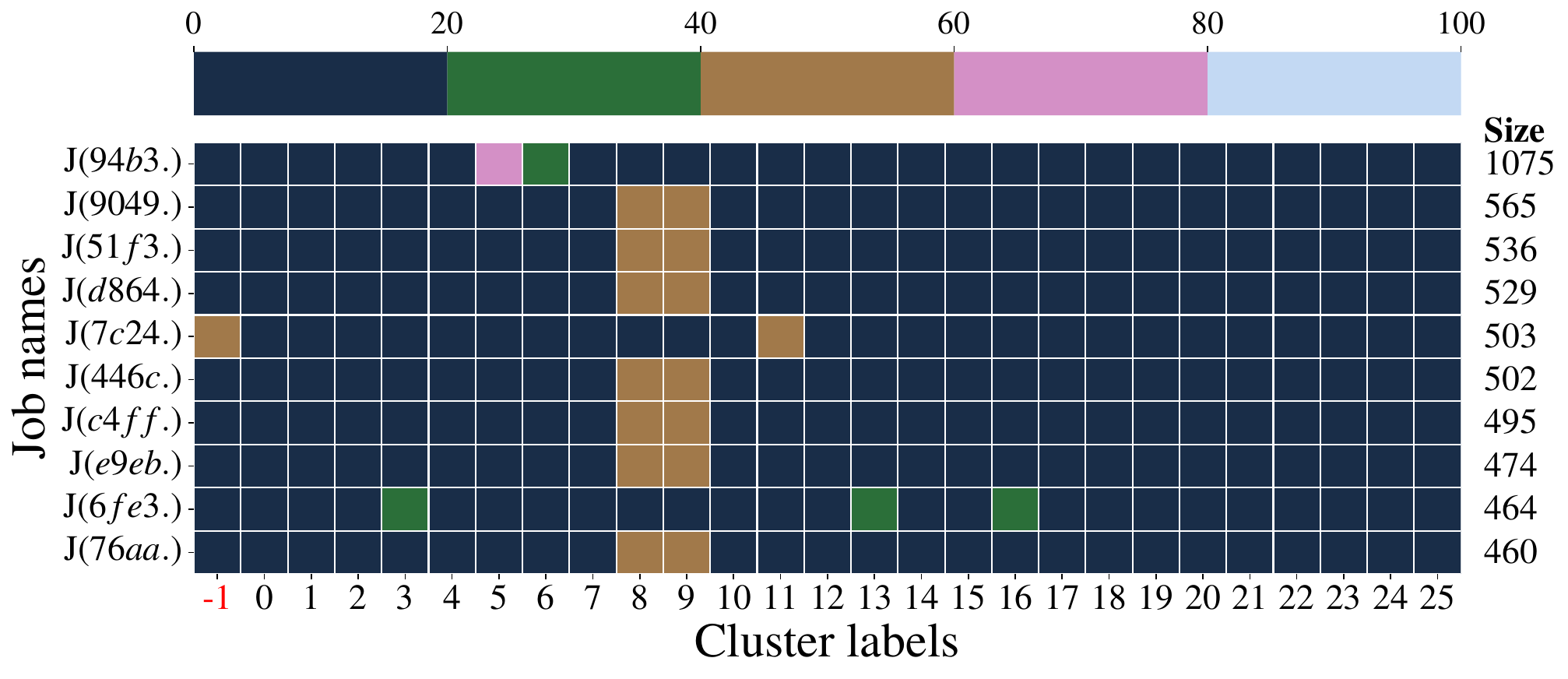}
    \label{fig:top10-job_names}
  }
  \subfloat[b][Workload type.]{
    \includegraphics[width=0.45\textwidth]{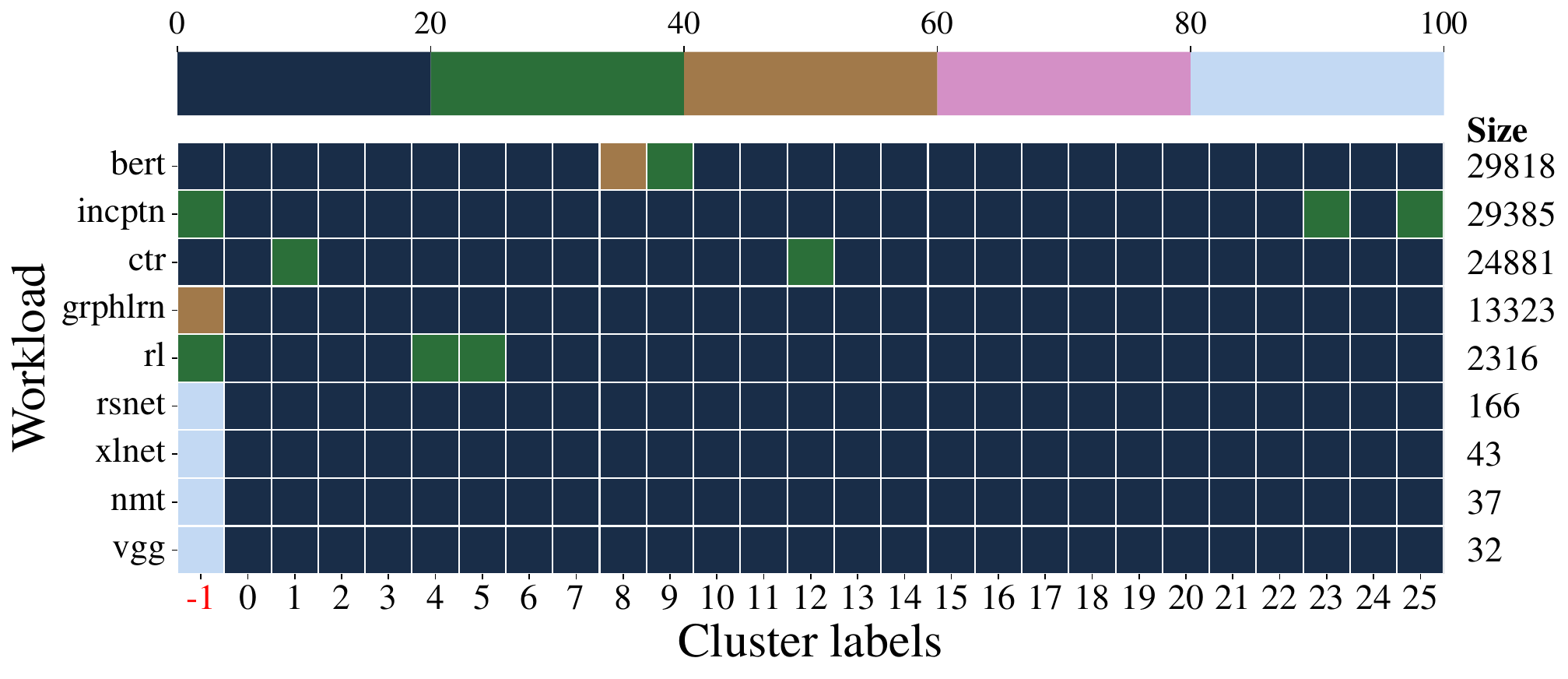}
    \label{fig:workload}
  } \\
  \subfloat[c][Task name.]{
    \includegraphics[width=0.45\textwidth]{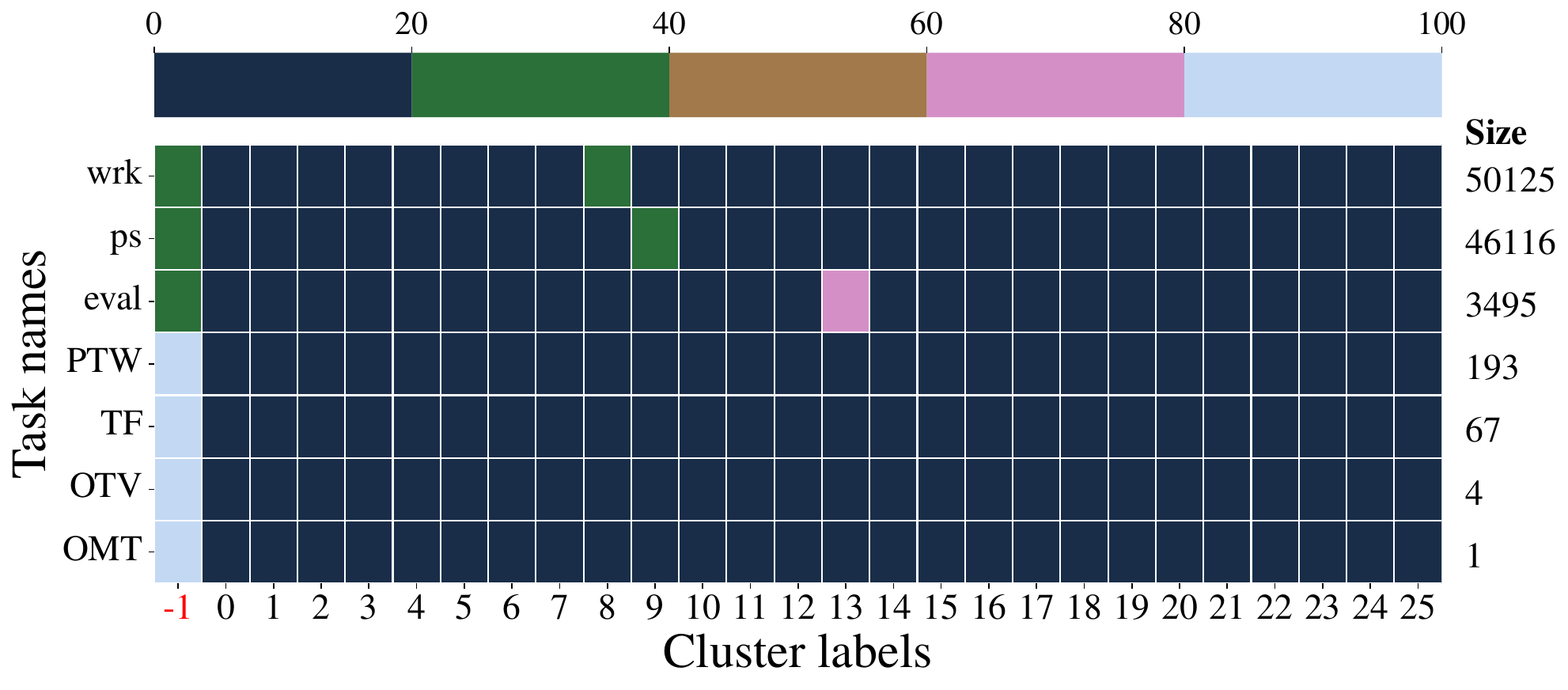}
    \label{fig:task_name}
  } 
  \subfloat[d][Top-10 Users.]{
    \includegraphics[width=0.45\textwidth]{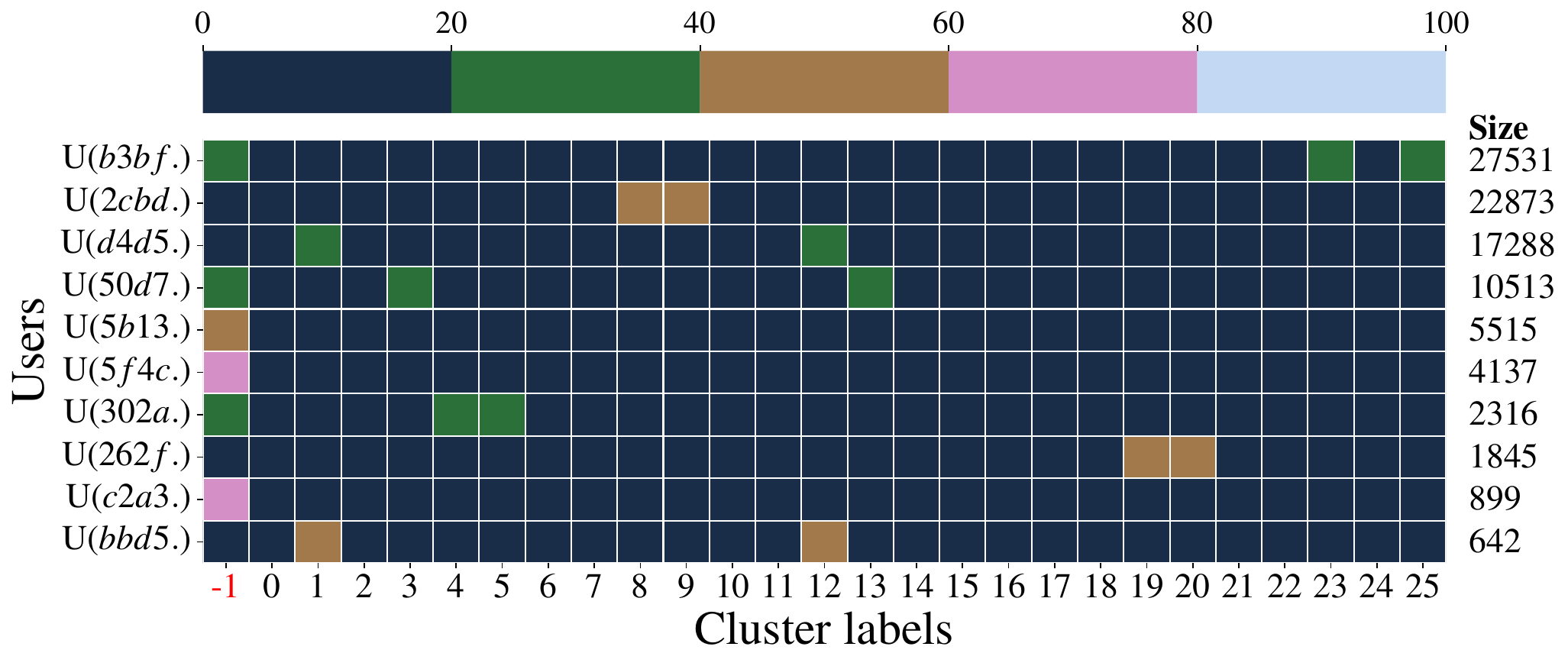}
    \label{fig:top10-users}
  } \\
  \subfloat[e][Top-10 Groups.]{
    \includegraphics[width=0.45\textwidth]{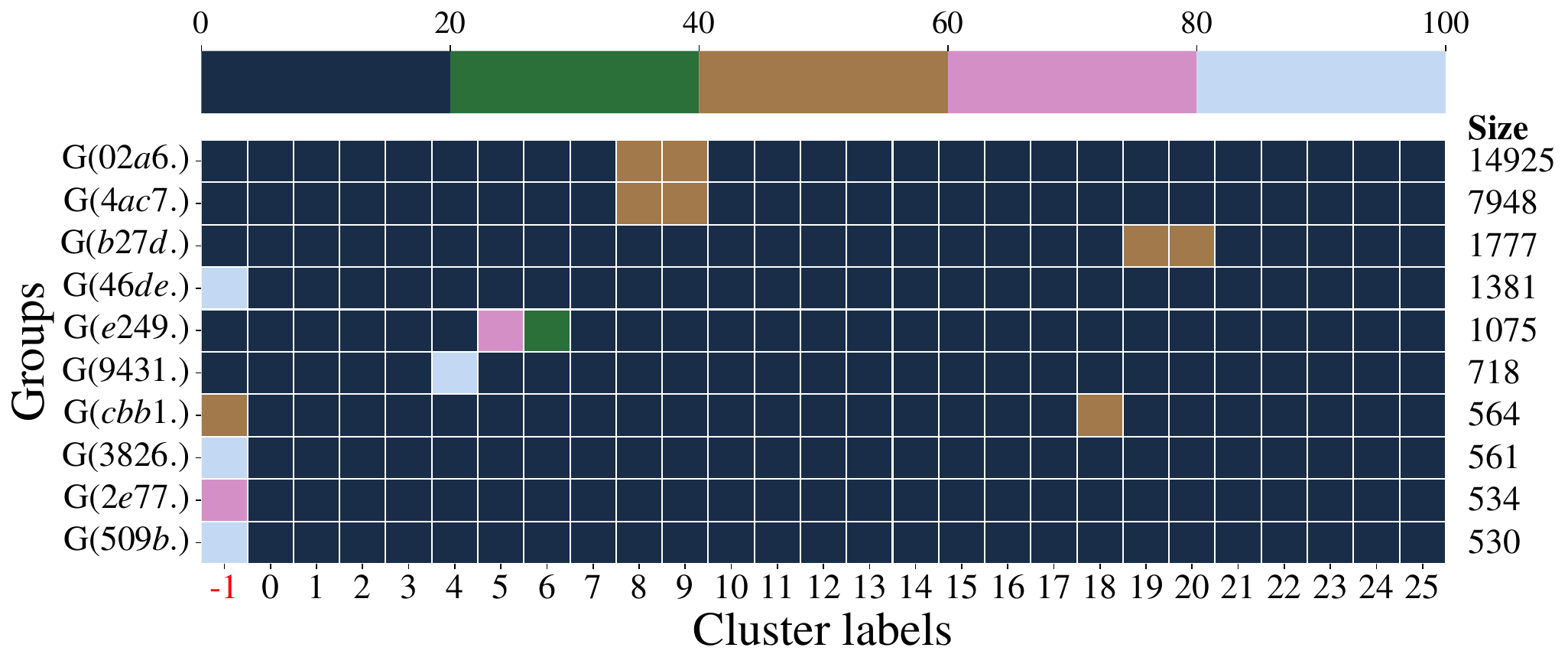}
    \label{fig:top10-groups}
  }
  \caption{Heatmaps reporting the distribution of the main values for the metadata features on the extracted profiles. Axis labels have been adjusted for better readability.}
  \label{fig:clusters-metadata}
\end{figure}

\paragraph{Metadata} 
Analyzing the metadata in the clusters is essential for \textbf{RC-3}, i.e., assigning profiles to new workloads.~\footnote{Related heatmap figures in the repository.}
Figure~\ref{fig:clusters-metadata} depicts the results obtained for the five static and a priori metadata features, i.e., job names, workload, task names, users, and groups. 
The heatmap shows the metadata feature value on the y-axis and, on the x-axis, the proportion in the extracted clusters of workloads with that metadata feature. 
For completeness, we add the outliers group marked red as ``-1'' on the x-axis. 
The cell colors depict proportional representation of each metadata feature value across the clusters. The proportion is divided in five quantiles. The dark blue shows the lower proportional representation (from 0 to 20\%) and the light blue the higher one (from 80 to 100\%). In detail, Figure~\ref{fig:top10-job_names} summarizes the pattern for the ten most recurring job names in the dataset. 
For seven out of ten job names, most of the values end in profiles 8 and 9, suggesting that these large groups contain various but similar workloads. 
These two profiles include, for the large part, ``bert'' workload. 
Furthermore, besides the ``rl'' workload, which characterizes profiles 5 and 6, the other workload feature values are scattered in the other clusters. 
Moreover, 
the clustering approach discarded the ``resnet,'' ``nmt,'' and ``vgg'' values. 
Looking at the cardinality of these values, which is lower than 500 -- our minimum cluster size -- we can understand why they are not in clusters. 
The task name distribution in Fig.~\ref{fig:task_name} confirms this outcome. 
Indeed, the last four values for task name distribution all have a cardinality below 200. 
These results show how the HDBSCAN-based profiling helps to distinguish workloads in the case study. 
This outcome is significant, considering that different workloads might show different patterns. 
Finally, looking at Figure~\ref{fig:top10-users} and Figure~\ref{fig:top10-groups}, representing the ten most recurring users and groups, we can see two patterns. 
Finally, some users and groups have a higher representation than others in the profiles pair 8 and 9 or the 19 and 20 pair. 
These two groups mainly refer to ``bert,'' as previously seen, and ``graphlearn.'' 
This outcome suggests that certain users focus on specific implementations, like ``bert'' and ``graphlearn'' and that these implementations have very specific meta-information embedded in the ``group'' metadata. 
Ultimately, this outline of the metadata distribution suggests that the clustering based on dynamic data can identify patterns in the metadata features and that combining these values in input can lead to accurately detecting profiles.

\subsection{Developing the metadata-based Profile Classifier}\label{subsec:profiling_new_jobs}
The final, essential step in the presented methodology is assigning a profile to newly submitted workloads. This task has to happen fast and by leveraging static, a priori metadata. 
We illustrate through the case study how to build such a classifier and discuss its performance. 
Furthermore, besides assigning new workloads to the profiles, we aim to understand the relevance of metadata features in the decision-making process through the model, which maps the input to the labels. 
Therefore,  we rely on the interpretable \gls{XGBoost} classification model due to its performance in classifying and its \textit{white box} characteristics.


\begin{table*}[ht!]
\centering
\caption{Class-level classification score reports.}
\label{tab:xgboost_class_report}
\resizebox{\textwidth}{!}{
\begin{tabular}{|c|c|c|c|c|c|c|c|c|c|c|c|c|c|c|c|c|c|c|c|c|c|c|c|c|c|c|c|c|c|}
\hline
\textbf{Profile} & \textbf{0} & \textbf{1} & \textbf{2} & \textbf{3} & \textbf{4} & \textbf{5} & \textbf{6} & \textbf{7} & \textbf{8} & \textbf{9} & \textbf{10} & \textbf{11} & \textbf{12} & \textbf{13} & \textbf{14} & \textbf{15} & \textbf{16} & \textbf{17} & \textbf{18} & \textbf{19} & \textbf{20} & \textbf{21} & \textbf{22} & \textbf{23} & \textbf{24} & \textbf{25} & \textbf{Macro Avg} & \textbf{Weighted Avg} \\ \hline
\textbf{Precision} & 0.90 & 0.99 & 1.00 & 1.00 & 1.00 & 0.66 & 0.00 & 0.94 & 1.00 & 1.00 & 1.00 & 1.00 & 1.00 & 1.00 & 1.00 & 1.00 & 0.99 & 1.00 & 1.00 & 1.00 & 1.00 & 1.00 & 0.97 & 0.84 & 0.99 & 0.84 & 0.93 & 0.95 \\ \hline
\textbf{Recall} & 1.00 & 1.00 & 1.00 & 0.99 & 1.00 & 1.00 & 0.00 & 0.99 & 0.98 & 1.00 & 0.98 & 1.00 & 1.00 & 1.00 & 1.00 & 0.99 & 1.00 & 1.00 & 1.00 & 1.00 & 1.00 & 1.00 & 0.26 & 0.99 & 0.30 & 1.00 & 0.90 & 0.95 \\ \hline
\textbf{F1-Score} & 0.94 & 1.00 & 1.00 & 0.99 & 1.00 & 0.80 & 0.00 & 0.96 & 0.99 & 1.00 & 0.99 & 1.00 & 1.00 & 1.00 & 1.00 & 0.99 & 1.00 & 1.00 & 1.00 & 1.00 & 1.00 & 1.00 & 0.41 & 0.91 & 0.46 & 0.91 & 0.90 & 0.94 \\ \hline
\textbf{Support} & 415 & 1172 & 177 & 607 & 143 & 134 & 68 & 134 & 2509 & 2316 & 128 & 64 & 1104 & 468 & 135 & 158 & 370 & 116 & 99 & 435 & 437 & 57 & 383 & 1489 & 426 & 1536 & 15080 & 15080 \\ \hline
\end{tabular}}
\end{table*}

\subsubsection{Training the Profile Classifier}\label{subsec:training_classifier}
We use the dataset of clustered elements $D^{\mathcal C}$, leaving out the outliers group. 
From each clustered workload, we extract their static, a priori metadata features, namely: \textit{job name}, \textit{user}, \textit{task name}, \textit{group}, and \textit{workload}. 
Overall, we obtain a set with a cardinality of 75\,398 and a dimension of 5, i.e., the metadata features.  
For the model generation, we subdivide the collection in \textit{training} and \textit{validation} sets, with an 80-20 ratio. 
The \gls{XGBoost} algorithm has limited support for categorical data. So, we must transform the input features into numerical ones. Valid approaches are \textit{one-hot encoding} or recurring to the \textit{embedding} networks. The latter requires a long training time; therefore, we use the former approach.
After this transformation, the set dimension grows to 21\,547. We store the data as a sparse matrix to optimize the computation. In this case, we use the standard hyperparameters for \gls{XGBoost}. 
Our aim in the case study is to analyze its performance and avoid overfitting.

Table~\ref{tab:xgboost_class_report} summarizes the results of the validation set per profile. 
We can appreciate that the results are excellent for most of the profiles, except for profile 6, where the classifier can not correctly label any of its points. 
In general, we obtain an \textit{accuracy} of $95.19\%$ and a \textit{weighted avg} F1-Score of $90\%$.
Overall, the results show a good capability of the trained model in predicting profiles independently from their size and starting just from a priori knowledge about a workload and its instance(s). 
This result gives us a promising path towards reproducing the proposed profiling approach, given the selected case study scenario where the granularity of information is partially insightful. 
Furthermore, the sample selected and the resulting profiles extracted from it are wide-ranged enough to constitute a complex undertaking for the model.
Finally, an additional advantage of this approach is the speed with which the model can label new workloads. We do not need any dry runs on sandboxes or runtime profiling.

\subsubsection{Results explanation}\label{subsec:metadata_analsys} 
\begin{figure}[t]
    \centering
    \includegraphics[width=0.7\columnwidth]{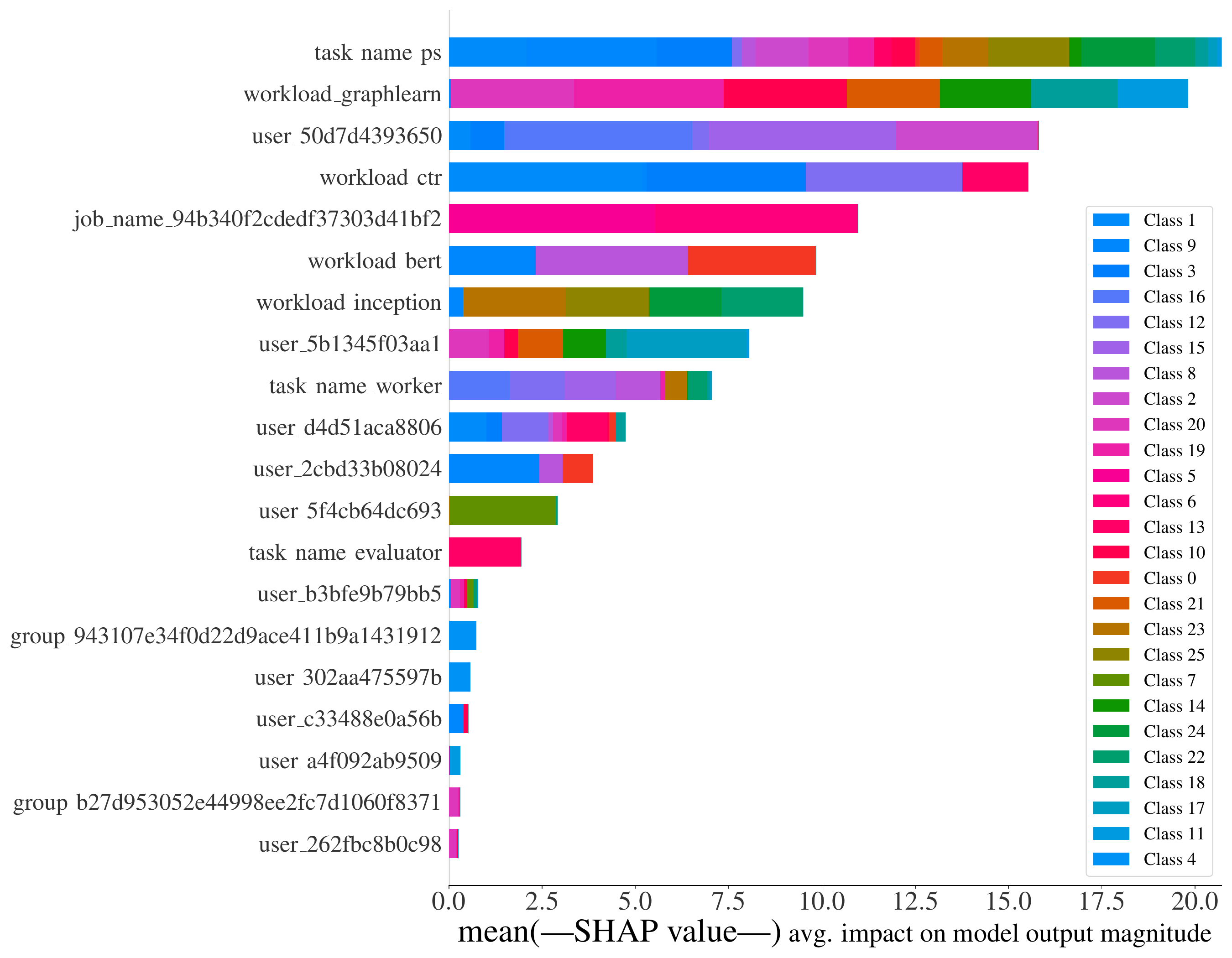}
    \caption{SHAP summary plot.}
    \label{fig:SHAP_summary_plot}
\end{figure}

A key feature is to obtain explainable results.
We achieve that using the SHAP eXplainable AI (XAI) approach~\cite{lundberg2017unified,lundberg2020local}, of the top twenty features in the \gls{XGBoost} model.~\footnote{The figure \texttt{SHAP\_summaryplot\_allclasses.pdf} is available in the repository}
Figure~\ref{fig:SHAP_summary_plot}  helps us understand how the static metadata feature influences the decision, showing the average impact.
Particularly relevant are the \textit{task name} in its ``\textit{ps}'' value, the \textit{graphlearn} \textit{workload} type, and a specific user. 
The \textit{task name}: ``\textit{ps}'' category refers to using a Parameter Server (PS) architecture for models' training. 
In this case, one or more nodes play the role of a PS, broadcasting current weights to learners before each step and aggregating gradients from them, which is an easy way to retain a global view~\cite{wang2019characterizing, li2020taming}. 
This behavior might represent a demarcation with other training architecture.
Similarly, Graph Neural Networks (GNNs) (\textit{workload}: ``\textit{graphlearn}'') have a very distinct behavior as they deal with graph data in the input. 
In particular, their distributed execution using Alibaba's developed framework can differentiate them from other workloads. 
The same goes for the NLP model labeled as (\textit{workload}: ``\textit{bert}''), which characterizes profiles 8 and 9.
Furthermore, the \textit{job name} \texttt{94b340f2cdedf37303d41bf2} is the most recurrent in our dataset, and it occurs in profiles 5 and 6. If we link this outcome with what we found in~\secref{subsec:clusters_evaluation}, we can match that these two clusters had very specific and defined resource usage values with a constantly low standard deviation. 
Therefore, it is easy to associate this metadata with a relevant decision boundary.
Overall, the use of the SHAP explainability tool reinforces the idea of our profiling approach, i.e., that the static a priori metadata represents a suitable and rich vehicle to match jobs to distinct profiles.

\subsection{Test case: predicting the workloads' behavior}\label{subsec:duration_testcase}
\begin{figure}[ht!]
  \centering
  \subfloat[a][CPU usage.]{\includegraphics[width=0.47\textwidth]{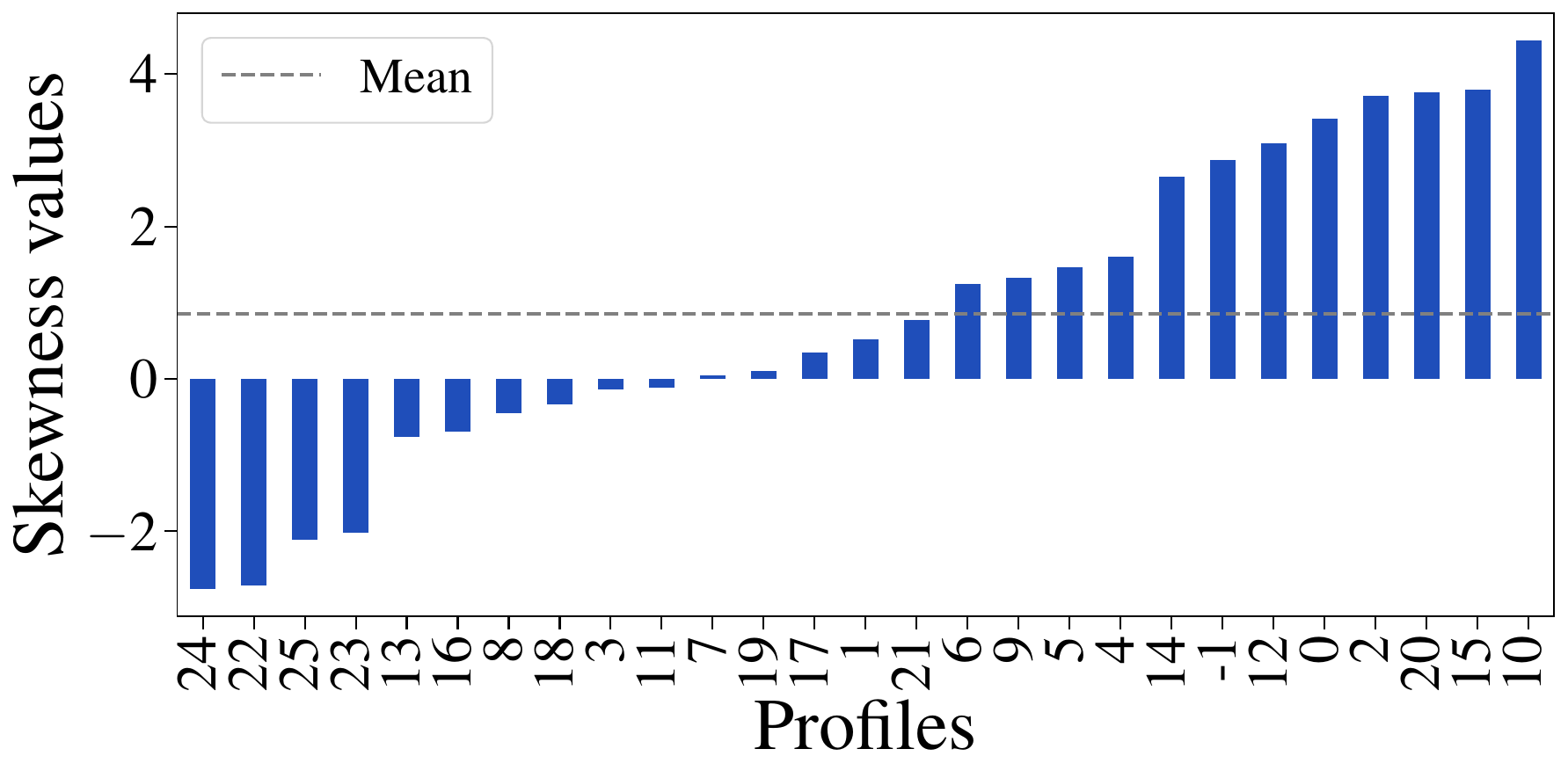} \label{fig:cpu_usage-skewness}}
  \subfloat[b][GPU working utilization.]{\includegraphics[width=0.47\textwidth]{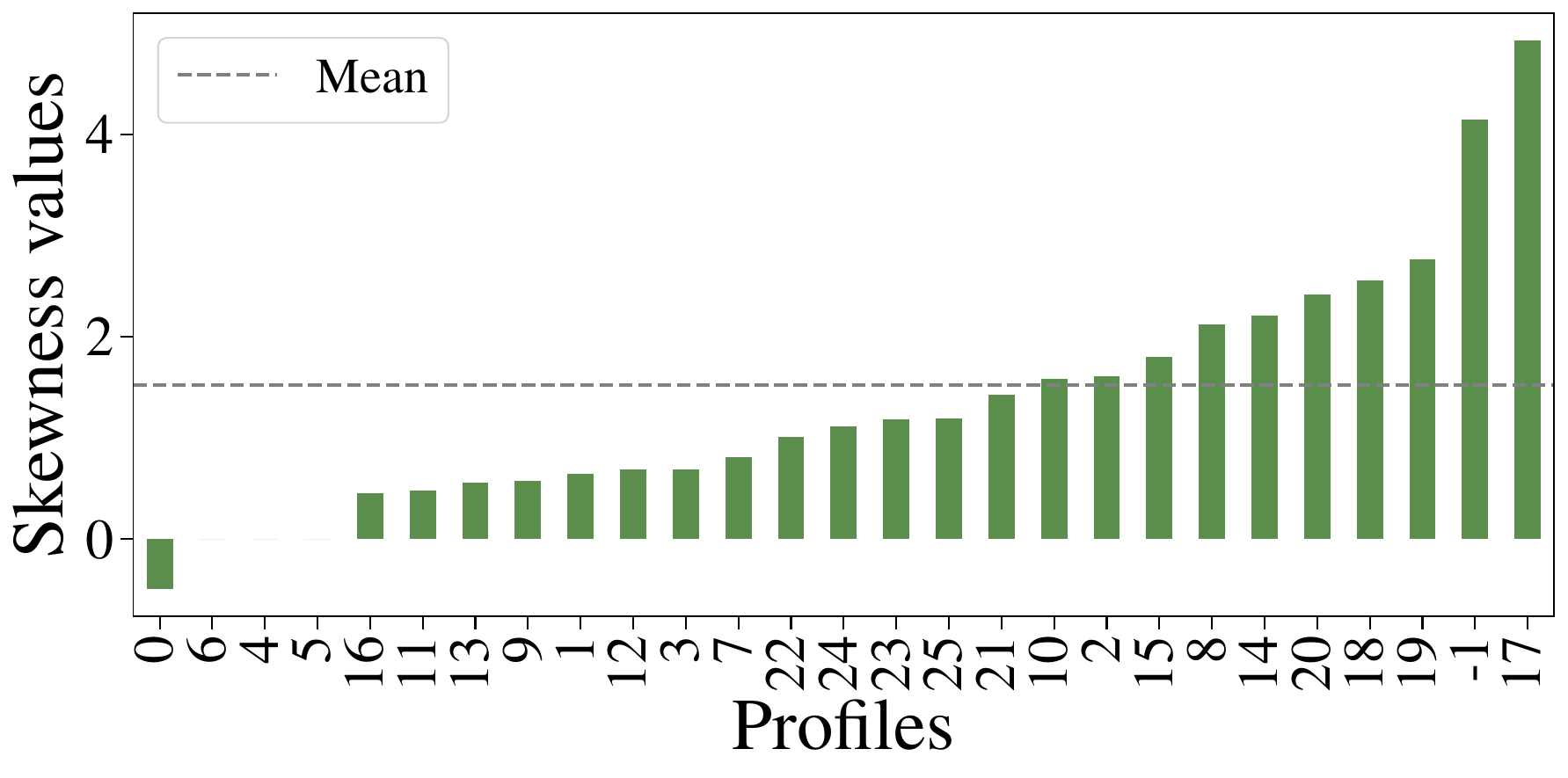} \label{fig:GPU-skewness}} \\
  \subfloat[c][Memory usage.]{\includegraphics[width=0.47\textwidth]{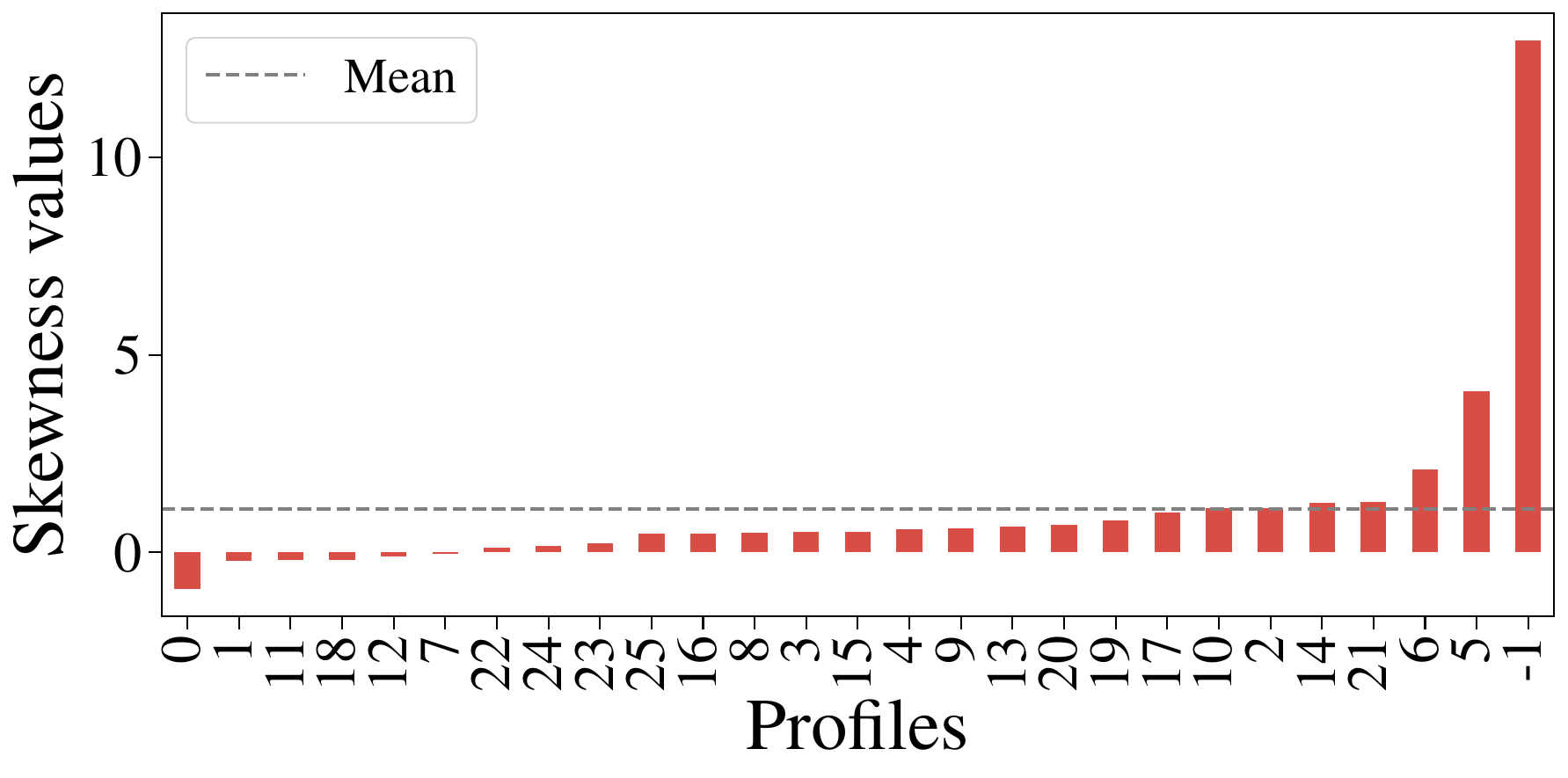} \label{fig:memory-skewness}}  
  \subfloat[d][Workload duration.]{\includegraphics[width=0.47\textwidth]{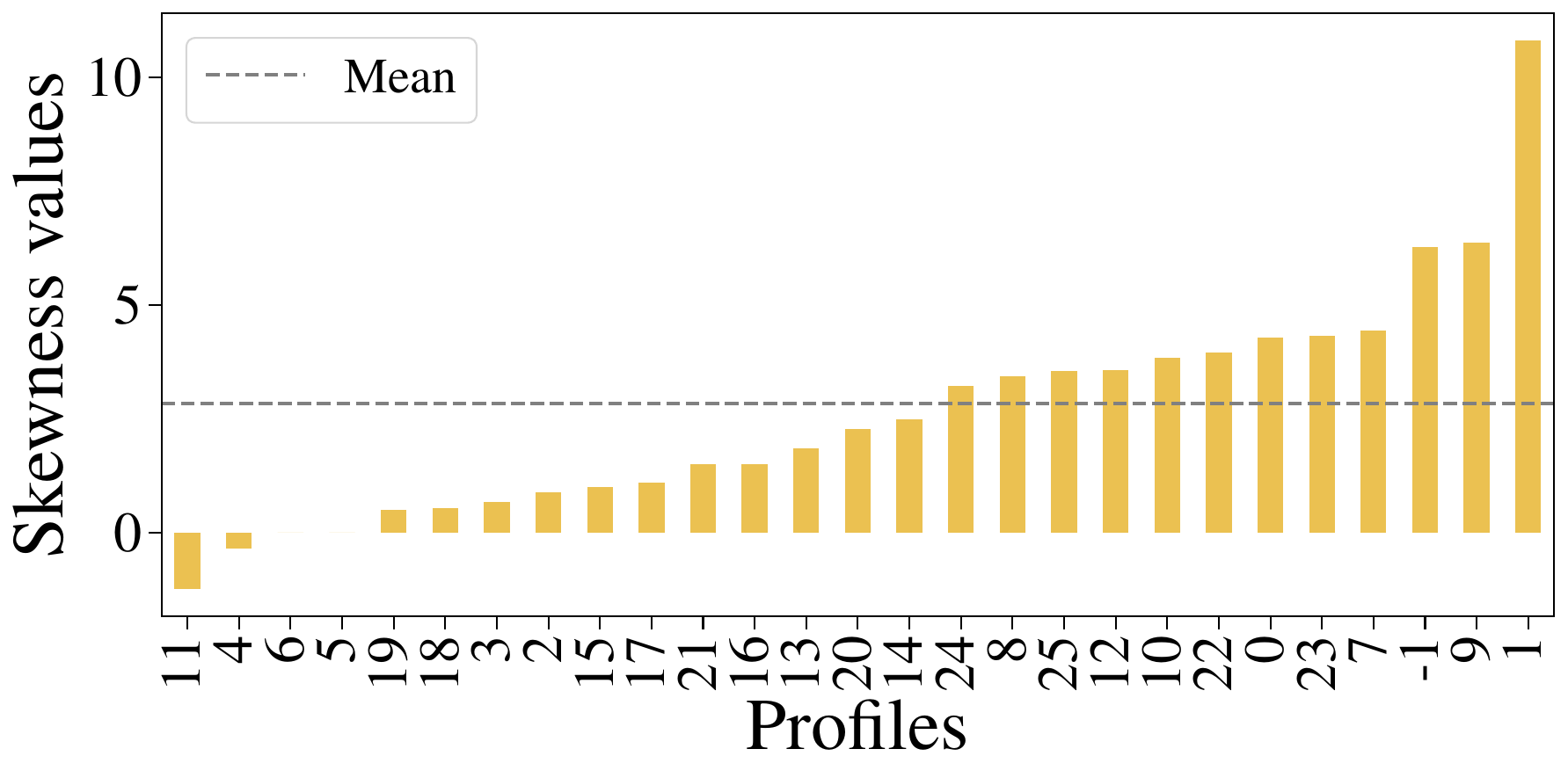} \label{fig:duration-skewness}} 
  
  \caption{Bar plots representing the skewness in the clustered profiles of CPU, GPU, Memory, and Duration.} \label{fig:skewness}
\end{figure}

Finally, we test the capability of profiles to embed relevant information. 
To do so, we extend the analysis from our previous contribution by considering 10\,000 unseen workloads. This load accounts for 10\% of the initial set, making it a realistic scenario.
In this analysis, we want to predict the workload's behavior by looking at four key resource usage indicators for ML workload, namely GPU usage, CPU usage, memory usage, and workload duration.
The first step that we follow is to analyze how to best summarize the target features for making the best prediction. Even though the development of an accurate forecasting model for resource usage is out of the scope of this work, we aim to have realistic results.
For this reason, we inspect the value distribution of the four measures, GPU usage, CPU usage, memory usage, and workload duration for each profile by computing the \textit{skewness score}. Skewness values greater than one indicate a distribution congregated towards the lower values. A skewness score lower than one instead describes distributions where there are higher values for the feature in the exam and that lower values are anomalies. A skewness value of zero indicates a normal distribution. Studying the distribution can help us understand which condensated value better represents the profiles. 
Figure~\ref{fig:skewness} represents the skewness values for every feature in each subplot. On the y-axis, we can find the skewness values, while on the x-axis, the profile labels are sorted from the lower to the higher skewness value. The grey horizontal dashed line represents the mean skewness value for the depicted feature among all the profiles, while the red one indicates the median.

As we can notice, most of the profiles, for most of the features, show skewness scores higher than one. In particular, for the memory, the mean score is  0.85 and the median is 0.52. CPU usage shows similar behavior with a mean of 1.09 and a median of 0.52. GPU and duration instead are characterized by higher skewness values, with mean and median respectively equal to 1.35 and 1.11 and 2.62 and 2.28.
These results suggest focusing on the lower spectrum of quantiles as these values show that outliers mostly lie on the higher side of the values distribution.
After experimenting with different settings, the approach that gave us the best results is to use the $5^{th}$ quantile for the prediction. Therefore, once the classifier assigns each of the sampled workloads $j \in \mathcal J_{sample}$ to a profile $p \in \mathcal P$, we use the $5^{th}$ quantile values $\hat{d}_{p}$ to assign the workload the predicted value. We do so for duration, CPU usage, and GPU usage. 

Afterward, we use the normalized Root Mean Squared Error \RMSEperc{} to compute the loss between $\hat{d}_{p}$ and the actual workload behavior $\hat{d}_{j}$. In the following, we depict the \RMSEperc{} both for each considered feature in isolation and in total, aggregating the four measures.

\paragraph{Considering features in isolation}

\begin{figure}[ht!]
  \centering
  \subfloat[a][CPU]{
    \includegraphics[width=0.47\linewidth]{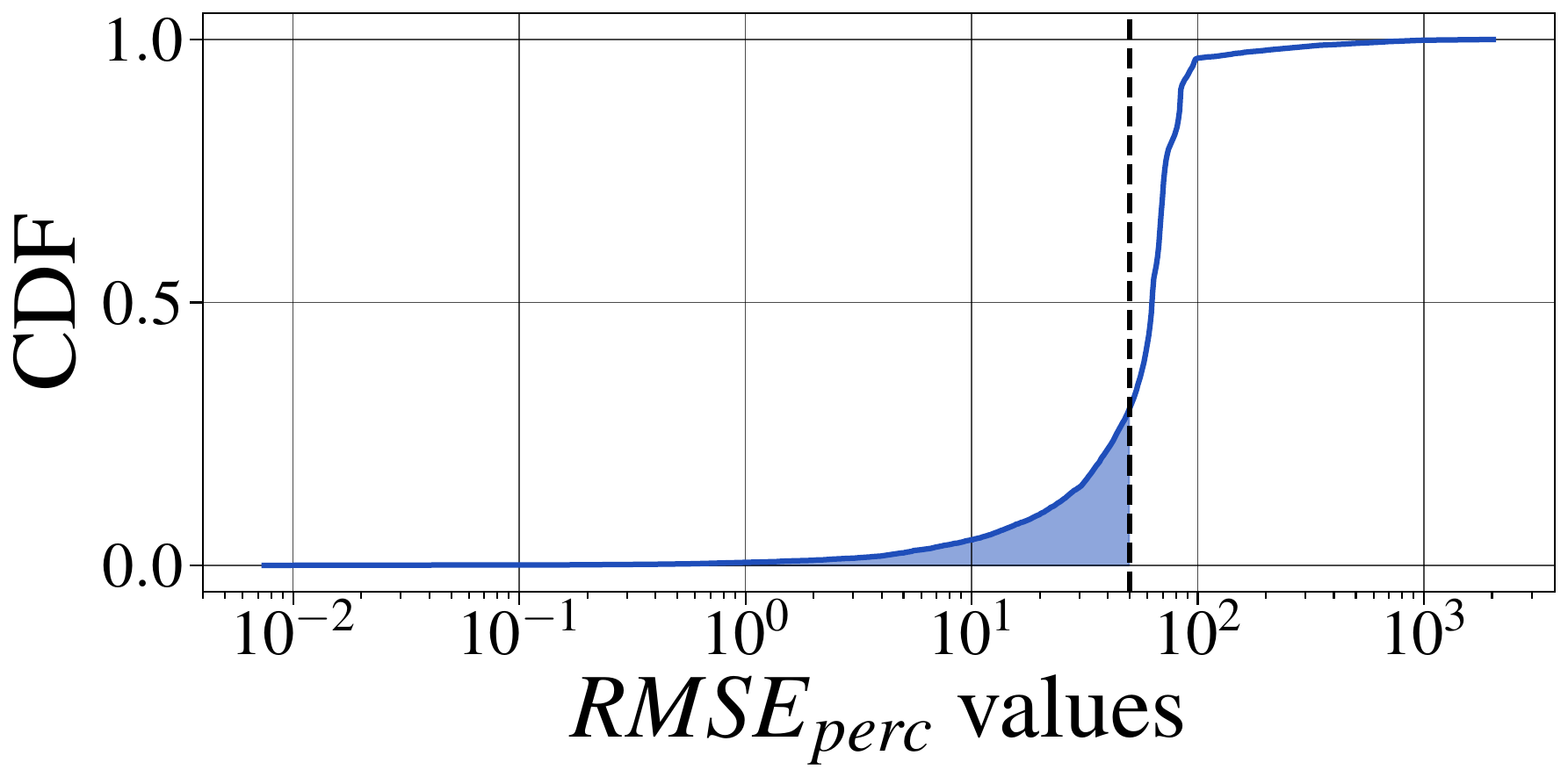} \label{fig:boxplot_cpu} }
  \hfill
  \subfloat[b][GPU]{
    \includegraphics[width=0.47\linewidth]{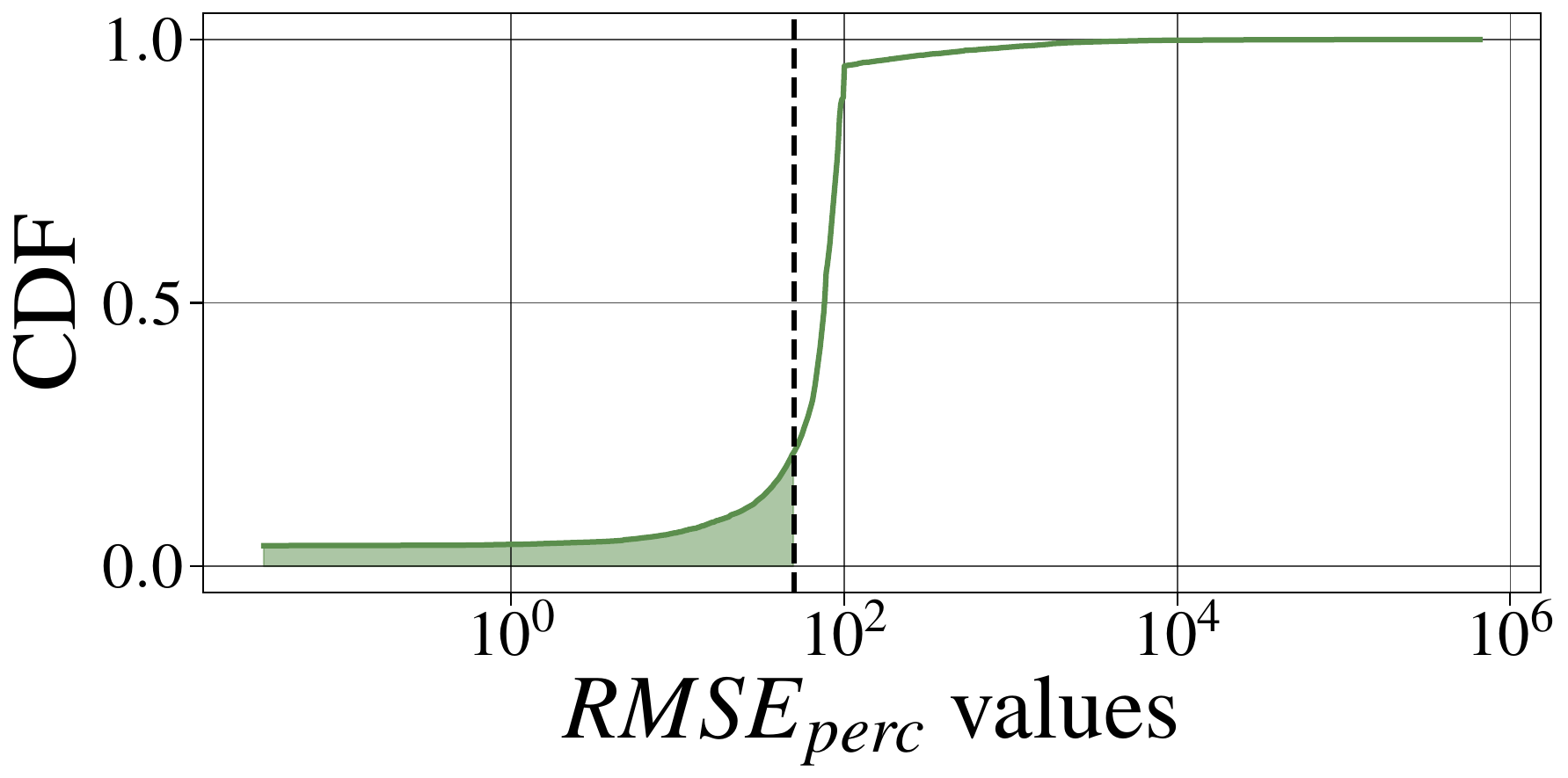} \label{fig:boxplot_gpu}}
  \\
  \subfloat[c][Memory]{
    \includegraphics[width=0.47\linewidth]{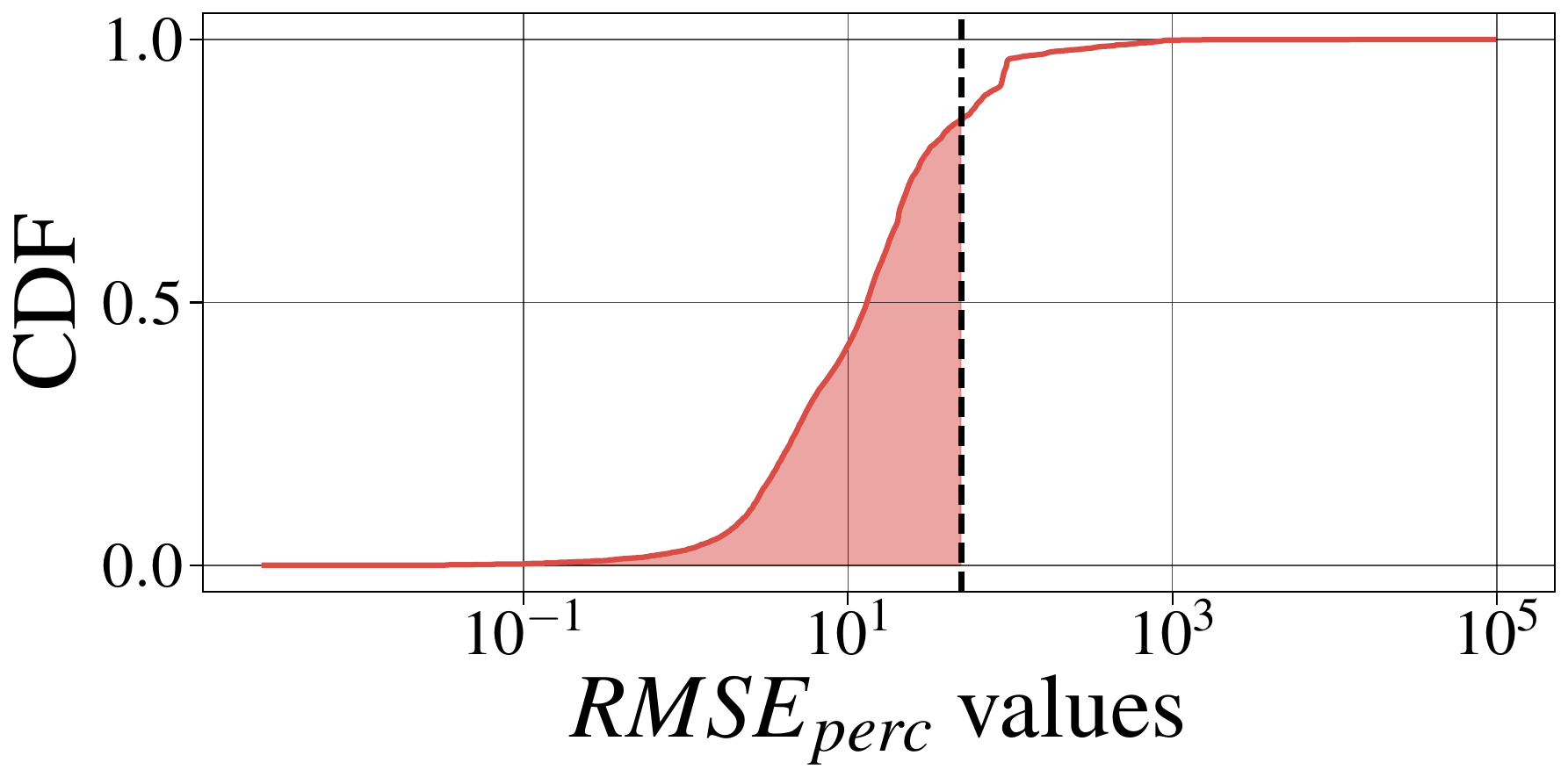} \label{fig:boxplot_memory}}
 \hfill
  \subfloat[d][Duration]{
    \includegraphics[width=0.47\linewidth]{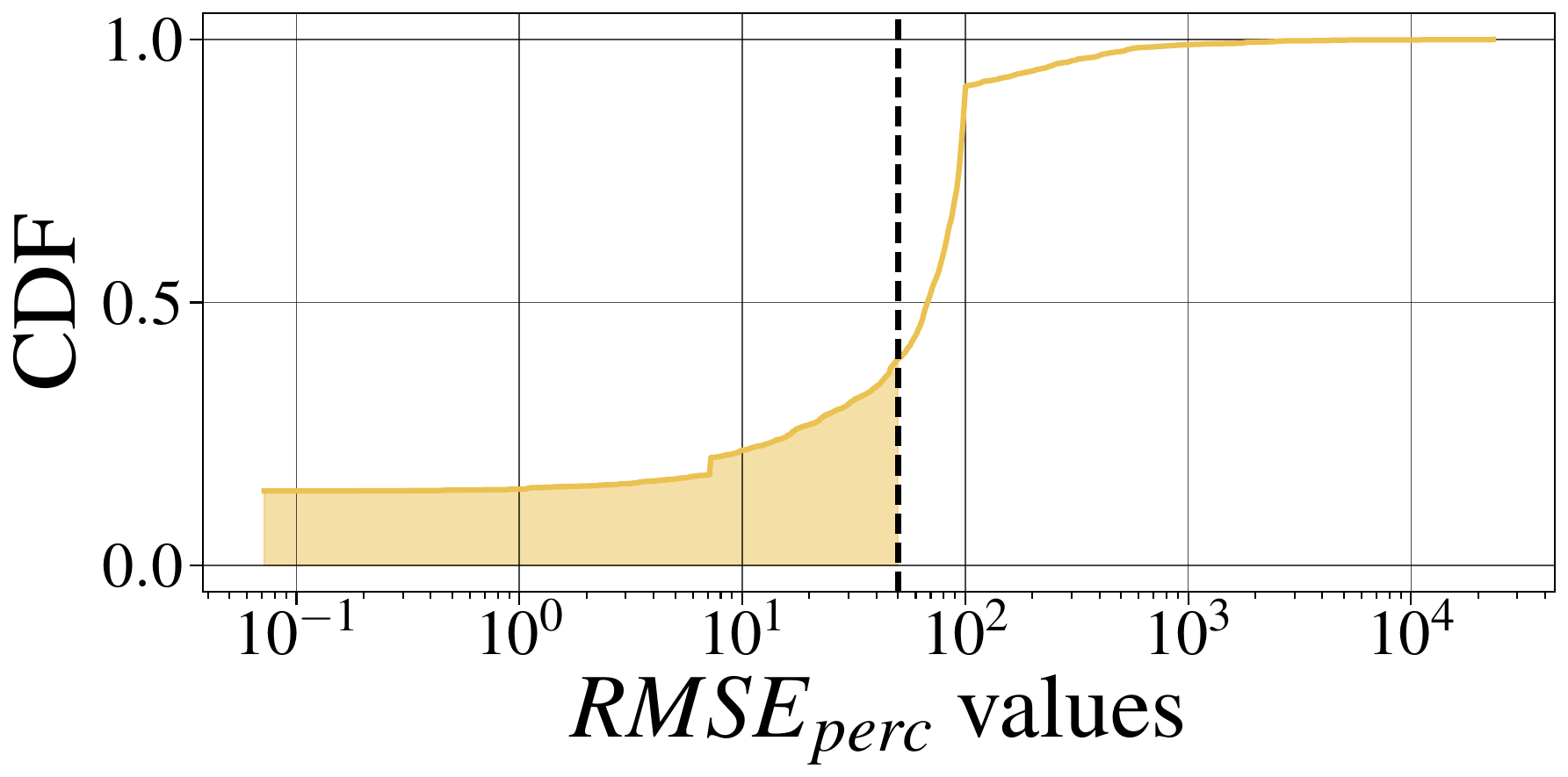} \label{fig:boxplot_duration}}
  \caption{CDFs of the $RMSE_{perc}$ values for the four considered measures: CPU, GPU, Memory, and Duration.}
  \label{fig:boxplots}
\end{figure}

Figure\ref{fig:combined_cdf_plots} represents the Empirical Cumulative Distribution Function (ECDF) of the $RMSE_{perc}$ values for the four different metrics. The x-axis shows the $RMSE_{perc}$ values in a log scale, and the y-axis the CDF.
For what concerns \textit{memory} the ECDF curve starts sharply, indicating that a significant proportion of the data has low $RMSE_{perc}$ values. More specifically, more than 80\% of the profiles have an $RMSE_{perc}$ less than 50.
Regarding \textbf{GPU} The initial rise in the GPU curve is steeper than that of Memory, highlighting that a portion of the workloads for GPU have very low error. However, the percentage of values with a $RMSE_{perc}$ lower than 50 are around 20\%.
The \textbf{CPU} curve starts off a bit slower than Memory and GPU but accelerates from $RMSE_{perc}$ values around 10; here the fraction of workloads with a \RMSEperc{} lower than 50 is around 40\%.
The \textbf{duration} curve has a shape somewhat similar to GPU, with a steep start.
A significant portion of the workloads have very small \RMSEperc{} values, as can be seen from the vast shaded region. Afterwards, the values with \RMSEperc{} below 50 settle at around 40\% of the total.
In summary, there is a good proportion of workloads with low \RMSEperc{} values, suggesting that the modeling or prediction methods are reasonably accurate for a majority of the workloads. The distribution and rate of increase in \RMSEperc{} values vary quite significantly between the metrics, with GPU and Duration showing a more pronounced initial increase, indicating that a larger percentage of their workloads have very low errors compared to Memory and CPU, but also a greater variability with outliers towards higher values.


\begin{figure}[ht!]
  \centering
  \subfloat[a][CPU]{
    \includegraphics[width=0.47\linewidth]{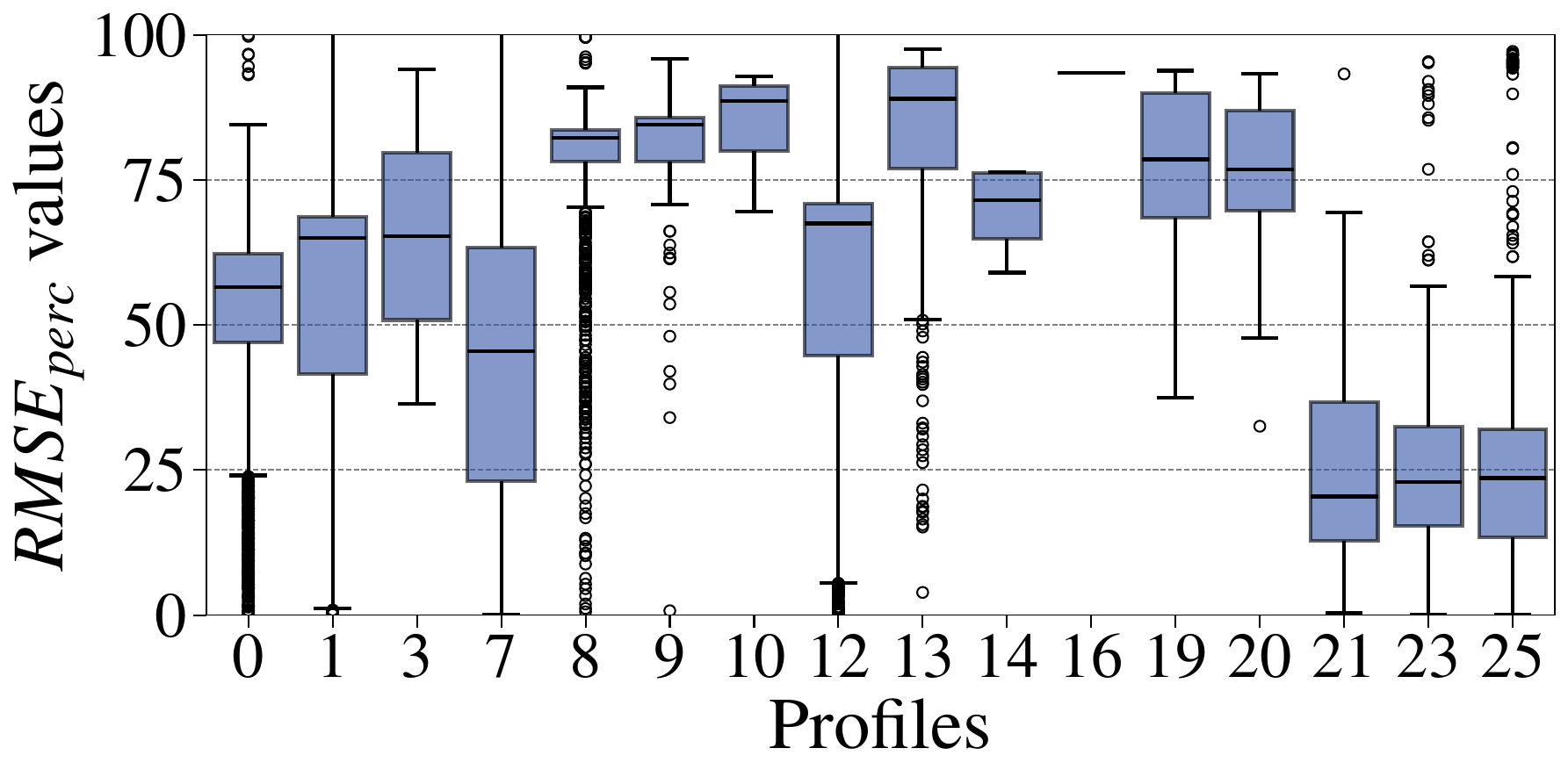} \label{fig:boxplot_cpu} }
  \hfill
  \subfloat[b][GPU]{
    \includegraphics[width=0.47\linewidth]{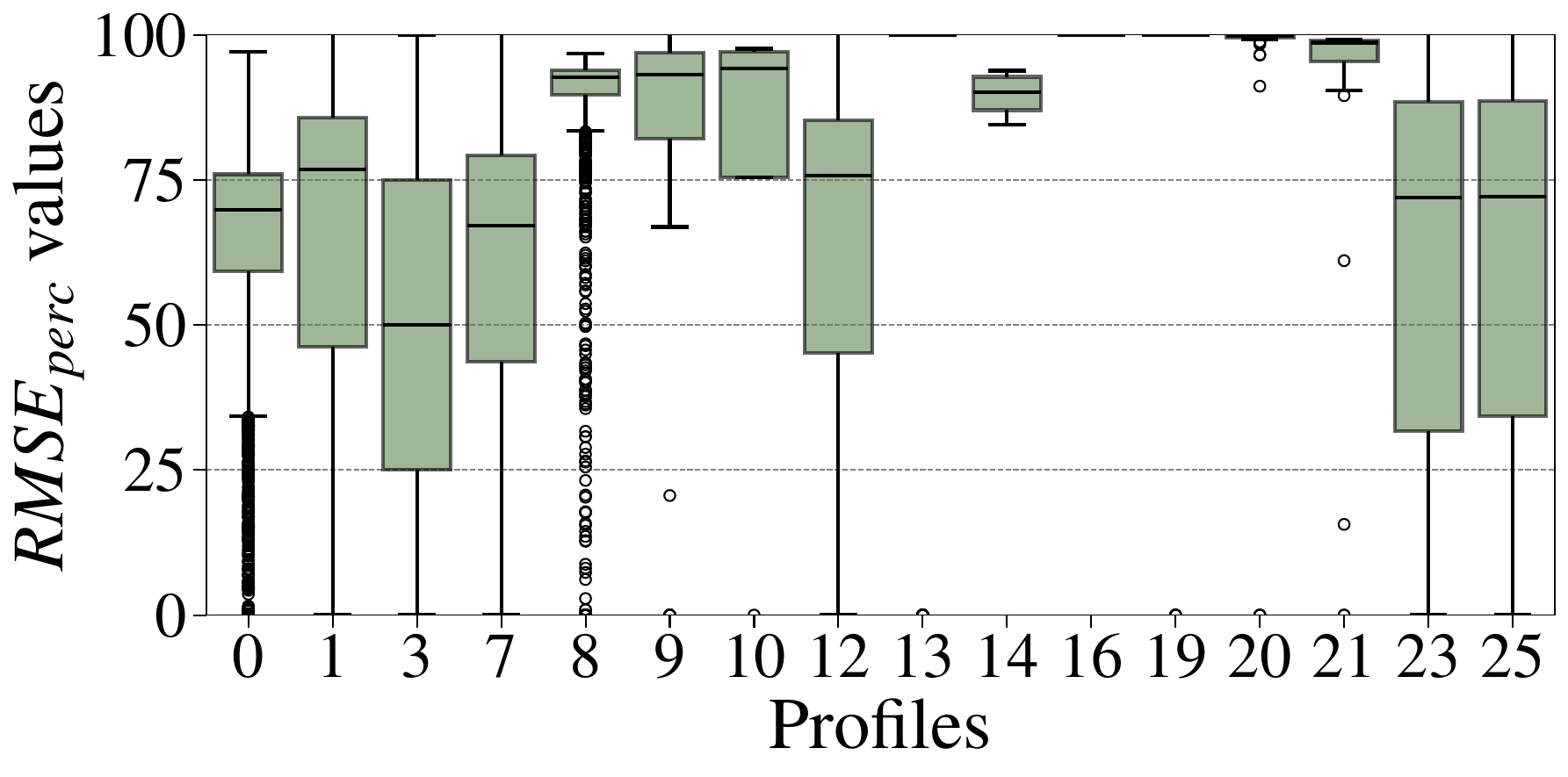} \label{fig:boxplot_gpu}}
  \\
  \subfloat[c][Memory]{
    \includegraphics[width=0.47\linewidth]{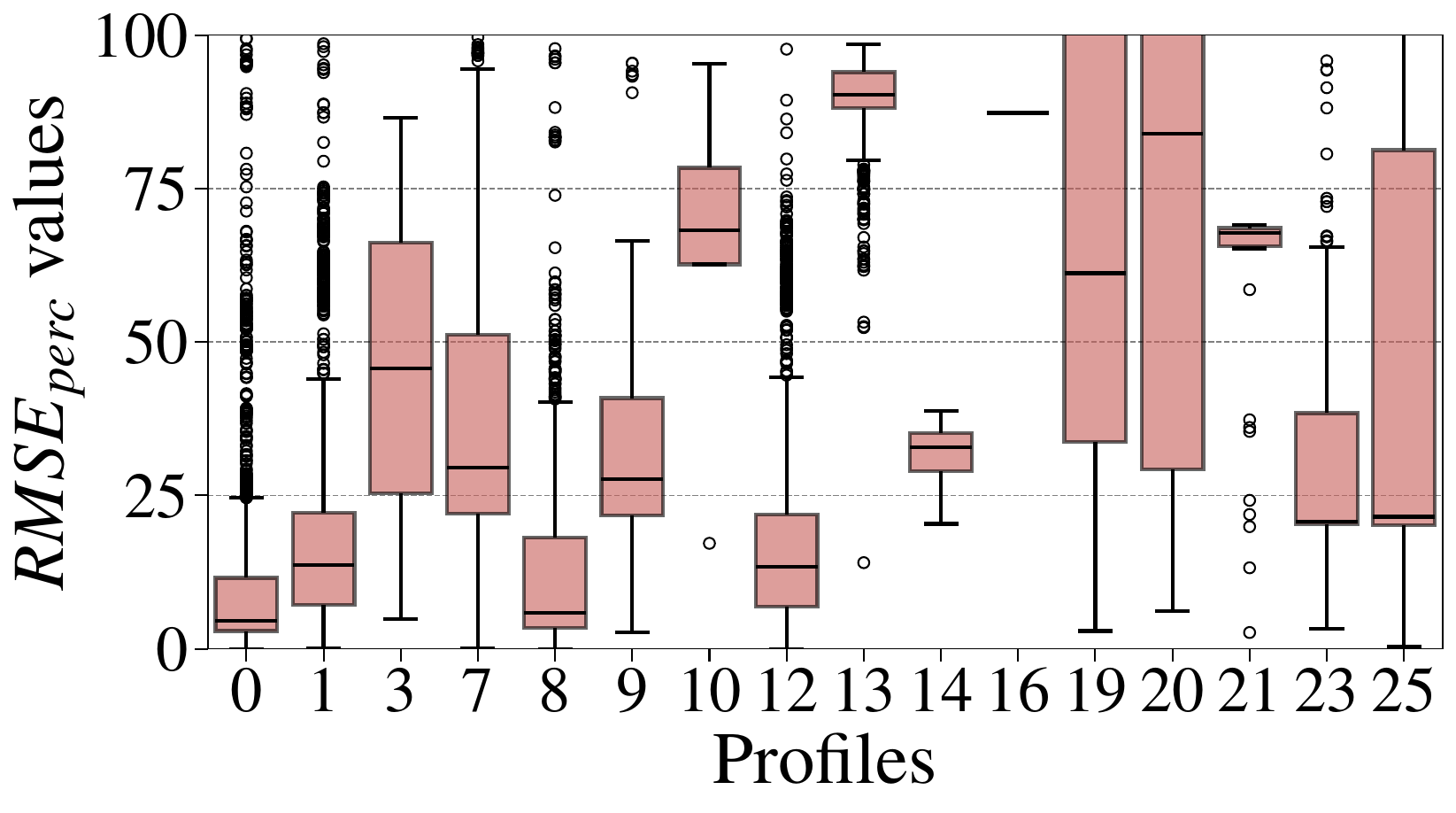} \label{fig:boxplot_memory}}
  \hfill
  \subfloat[d][Duration]{
    \includegraphics[width=0.47\linewidth]{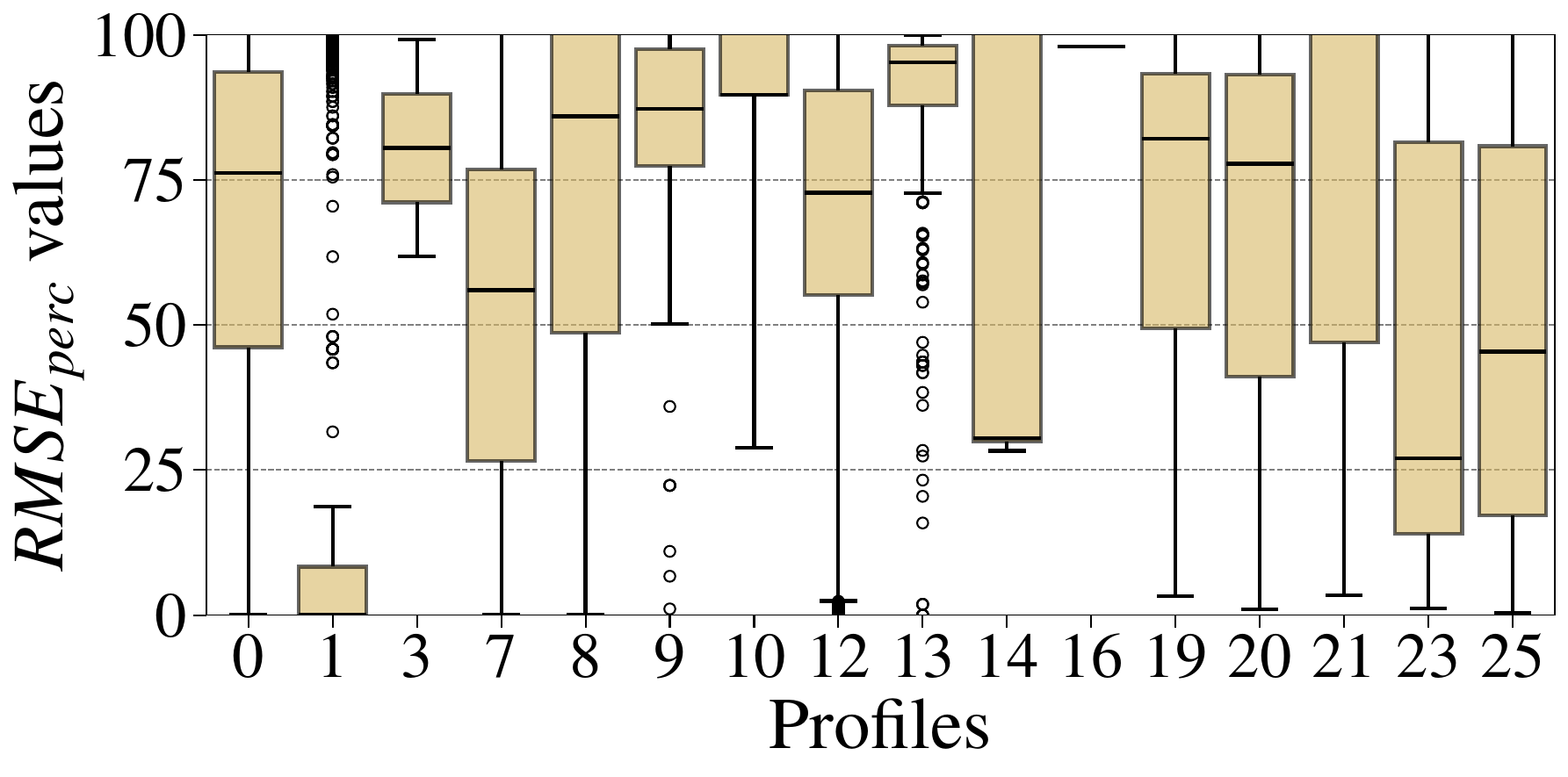} \label{fig:boxplot_duration}}
  \caption{Boxplots of the $RMSE_{perc}$ values for the four considered measures: CPU, GPU, Memory, and Duration.}
  \label{fig:combined_boxplots}
\end{figure}

Figure~\ref{fig:combined_boxplots} displays four box plots depicting the distribution of $RMSE_{perc}$ values for the four different features: CPU (Fig.~\ref{fig:boxplot_cpu}), GPU (Fig.~\ref{fig:boxplot_gpu}), Memory (Fig.~\ref{fig:boxplot_memory}), and Duration (Fig.~\ref{fig:boxplot_duration}). 
Fig.~\ref{fig:boxplot_cpu} shows that median $RMSE_{perc}$ value for CPU predictions appears to be below 50 for most of the profiles. There are quite a few outliers, especially on the higher side of the $RMSE_{perc}$ value, but overall the predictions seem accurate. 
Regarding \textbf{GPU}, Fig~\ref{fig:boxplot_gpu} shows high median $RMSE_{perc}$ values. This confirms the difficulty of fully grasping GPU using the $5_{th}$ quantile measure.
Considering \textbf{memory}, the box plot in Fig.~\ref{fig:boxplot_memory} displays a considerable variability in the data. While some profiles have their median above the 50\% of  $RMSE_{perc}$, most of them are showing lower medians, suggesting a good capability to predict this measure.
The extent in the Duration box plot (Fig.~\ref{fig:boxplot_duration}) is similar to GPU but slightly narrower. The median $RMSE_{perc}$ is generally around or slightly below the 50 mark. 
Overall, the spread, median values, and the number of outliers vary across the four features, depicting a generally good prediction capability and highlighting at the same time the complexity that comes with predicting accurately from a set of values.

\paragraph{Considering the overall result for the four features combined}
\begin{figure}[ht!]
  \centering
  \subfloat[a][CDF.]{\includegraphics[width=0.47\linewidth]{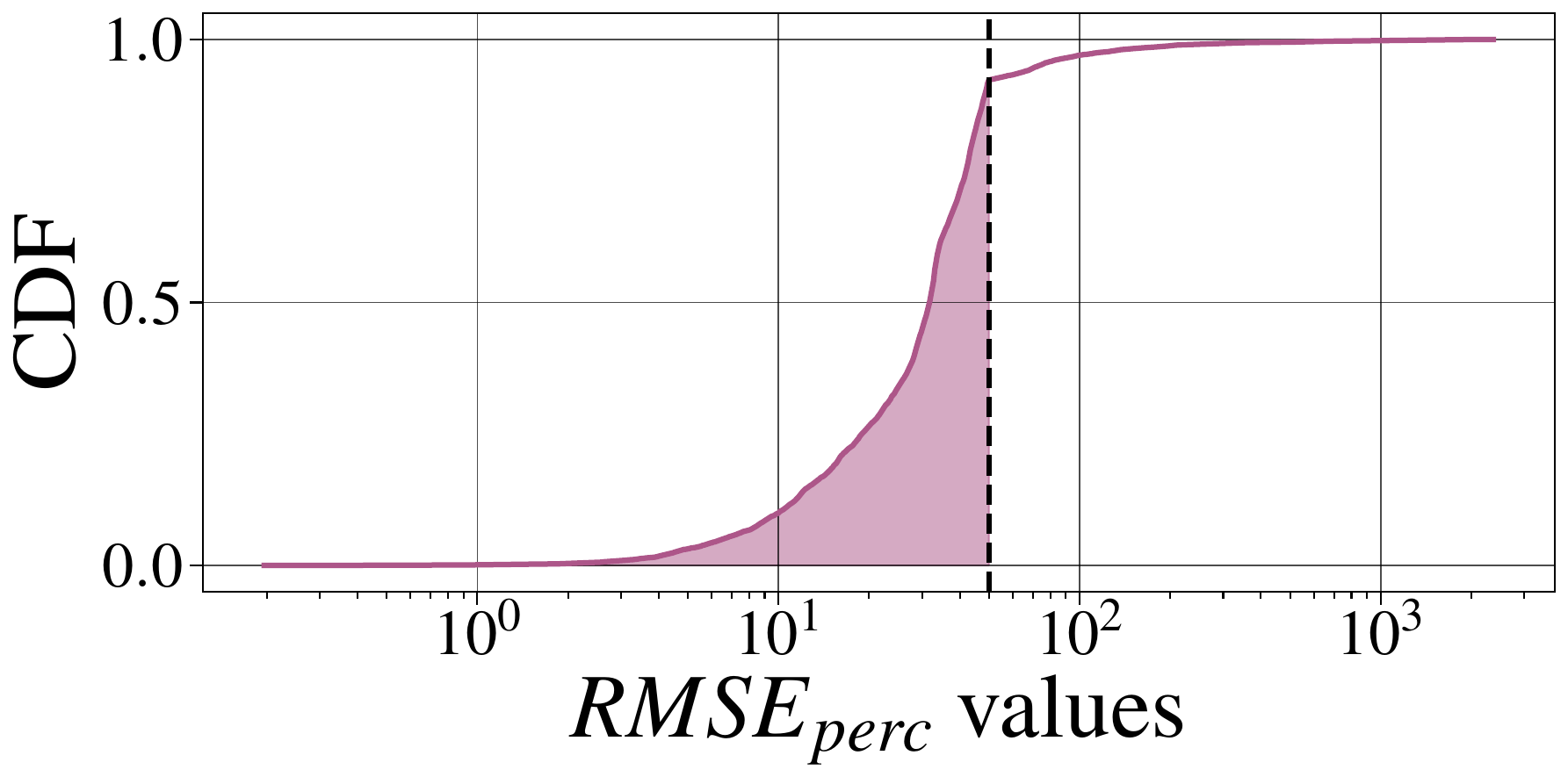} \label{fig:cdf_overall}} \hfill
  \subfloat[b][Boxplot.]{\includegraphics[width=0.47\linewidth]{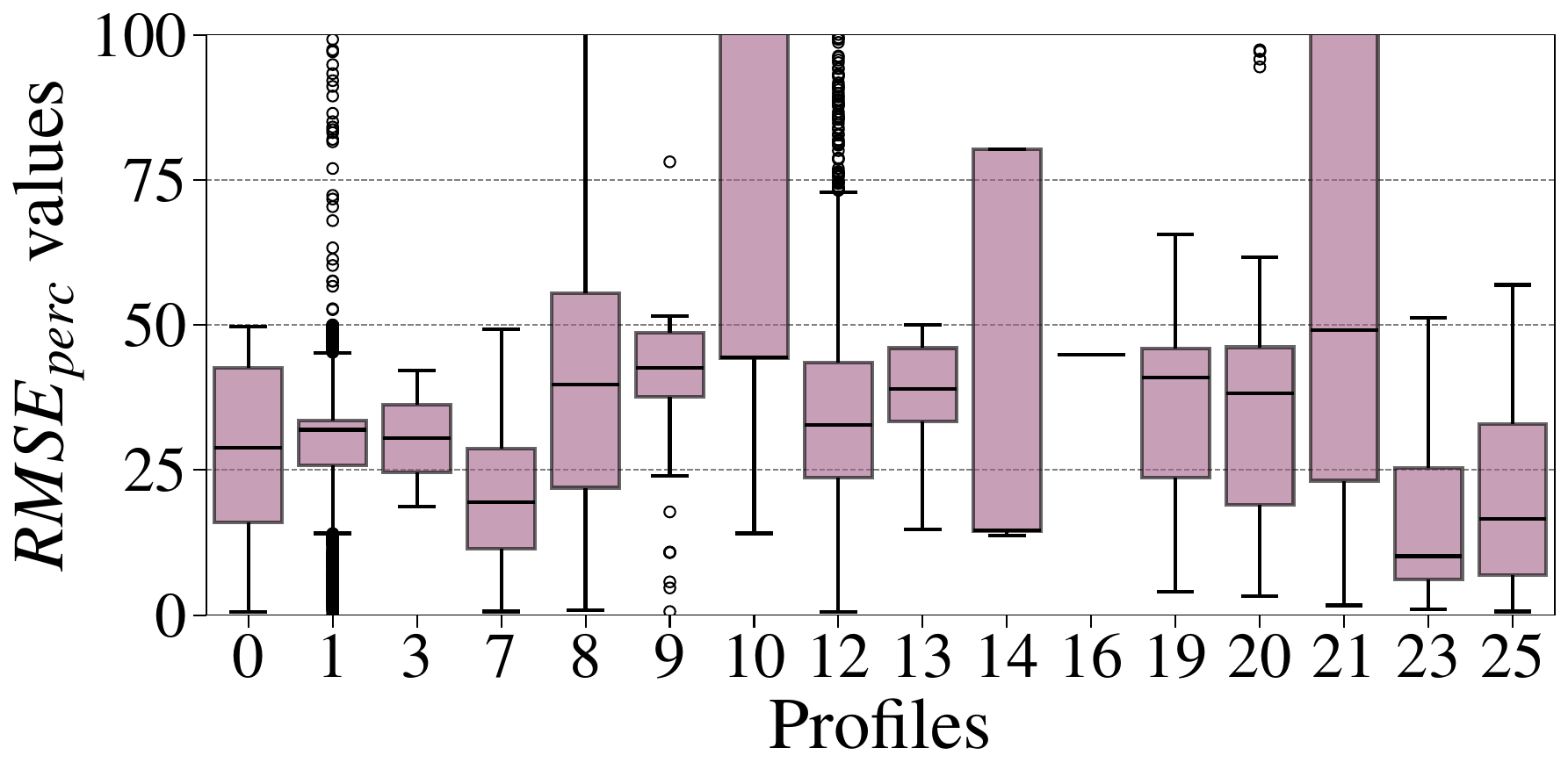} \label{fig:boxplot_overall}} 
  \caption{CDF and boxplots representing the \RMSEperc{} values by profile, considering \RMSEperc{} computed on the four main features alltogether.} \label{fig:RMSE_overall}
\end{figure}

Here, we analyze the \RMSEperc{} when considering all four measures in combination. That is the case we aim for production, where we want to know how far our prediction was considering all the selected measures.
Figure~\ref{fig:RMSE_overall} shows the CDF and the boxplots for the single measures. As we can see from Figure~\ref{fig:cdf_overall}, overall, we are able to get a very good error rate, with circa 93\% of the workloads showing a \RMSEperc{} lower than 50 and more than 40\% a \RMSEperc{} lower than 40.
These results are confirmed when looking at the boxplot in Figure~\ref{fig:boxplot_overall}. Here, we can see how, for most of the profiles, the median is lower than 50. Some profiles still show many outliers or a greater distribution, but this is to be expected given that we use the same prediction metric for all of them.

\noindent \textbf{\textit{Takeaway}} The results show us how, in combination, we are able to obtain good predictions only using an overarching, simple statistical metric. This result is promising as we can already define a good approximation for most of our workloads. The boxplots show us how, however, not all the profiles can be easily summarized by a single metric. Therefore, this behavior calls for future improvement in the prediction mechanisms by considering more sophisticated solutions.

\subsection{Feedback loop}
\label{sec:evaluation-feedback_loop}
The last evaluation involves checking how the proposed profiling can improve over time. 
Subsequent application of the feedback loop on the 10,000 new unseen workloads provided encouraging results. A mechanism is instituted to trigger re-clustering if there's a misjudgment rate of 1\% for the entire workload set, which in this case equates to 1\,000 violations. 
Since using the $5^{th}$ quantile would not produce enough violations, we tweak the prediction using another setting we previously tested; namely, we consider the $5^{th}$ quantile for predicting the feature value if the skewness value of the profile, for that feature, is greater than one; otherwise, we rely on the \textit{median} value. In this instance, out of the 10,000 workloads, 1\,013 are originally flagged for violations. This experiment thus triggers the re-clustering when the system records 1\,000 violations. In this case, the system detects a reduction of the ACQUIRES values and triggers a re-clustering. The re-clustering strategy proves to be effective, successfully avoiding the 13 following violations.

\section{Extended analysis on Google cluster data}
\label{sec:google}

\begin{table*}[ht!]
\centering
\caption{Test Set performance of XGBoost over the Google cluster data traces.}
\label{tab:google_test_performance_xgboost}
\resizebox{\textwidth}{!}{
\begin{tabular}{|c|c|c|c|c|c|c|c|c|c|c|c|c|c|c|c|}
\hline
\textbf{Metric}    & \textbf{0}    & \textbf{1}    & \textbf{2}    & \textbf{3}    & \textbf{4}    & \textbf{5}    & \textbf{6}    & \textbf{7}    & \textbf{8}    & \textbf{9}    & \textbf{10}   & \textbf{11}   & \textbf{12}   & \textbf{Macro Avg} & \textbf{Weighted Avg} \\ \hline
\textbf{Precision} & 0.97          & 1.00          & 0.93          & 1.00          & 0.96          & 0.97          & 0.94          & 1.00          & 0.80          & 1.00          & 0.94          & 0.89          & 1.00          & 0.95                & 0.99                 \\ \hline
\textbf{Recall}    & 1.00          & 1.00          & 0.96          & 1.00          & 0.77          & 0.99          & 0.91          & 0.92          & 0.55          & 0.94          & 0.73          & 0.65          & 0.95          & 0.87                & 0.99                 \\ \hline
\textbf{F1-Score}  & 0.99          & 1.00          & 0.94          & 1.00          & 0.85          & 0.98          & 0.93          & 0.96          & 0.65          & 0.97          & 0.82          & 0.75          & 0.97          & 0.91                & 0.99                 \\ \hline
\textbf{Support}   & 1268          & 7189          & 384           & 210           & 57            & 1095          & 347           & 139           & 22            & 65            & 45            & 48            & 59            & 10928              & 10928               \\ \hline
\end{tabular}}
\end{table*}

\begin{figure}[ht!]
  \centering
  \subfloat[a][CPU]{
    \includegraphics[width=0.47\linewidth]{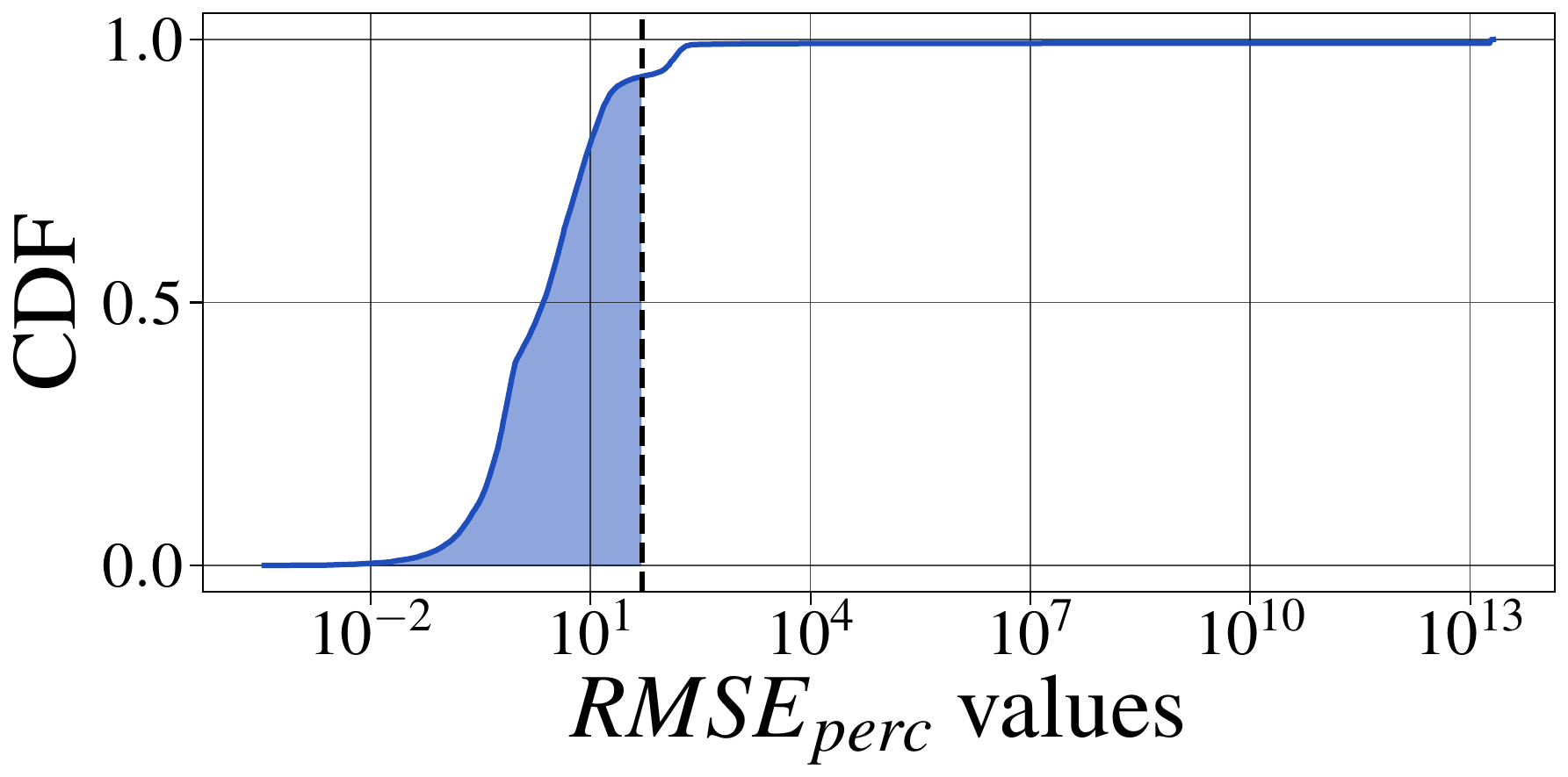} \label{fig:cdf_google_cpu} }
  \hfill
  \subfloat[b][CPU boxplot]{
    \includegraphics[width=0.47\linewidth]{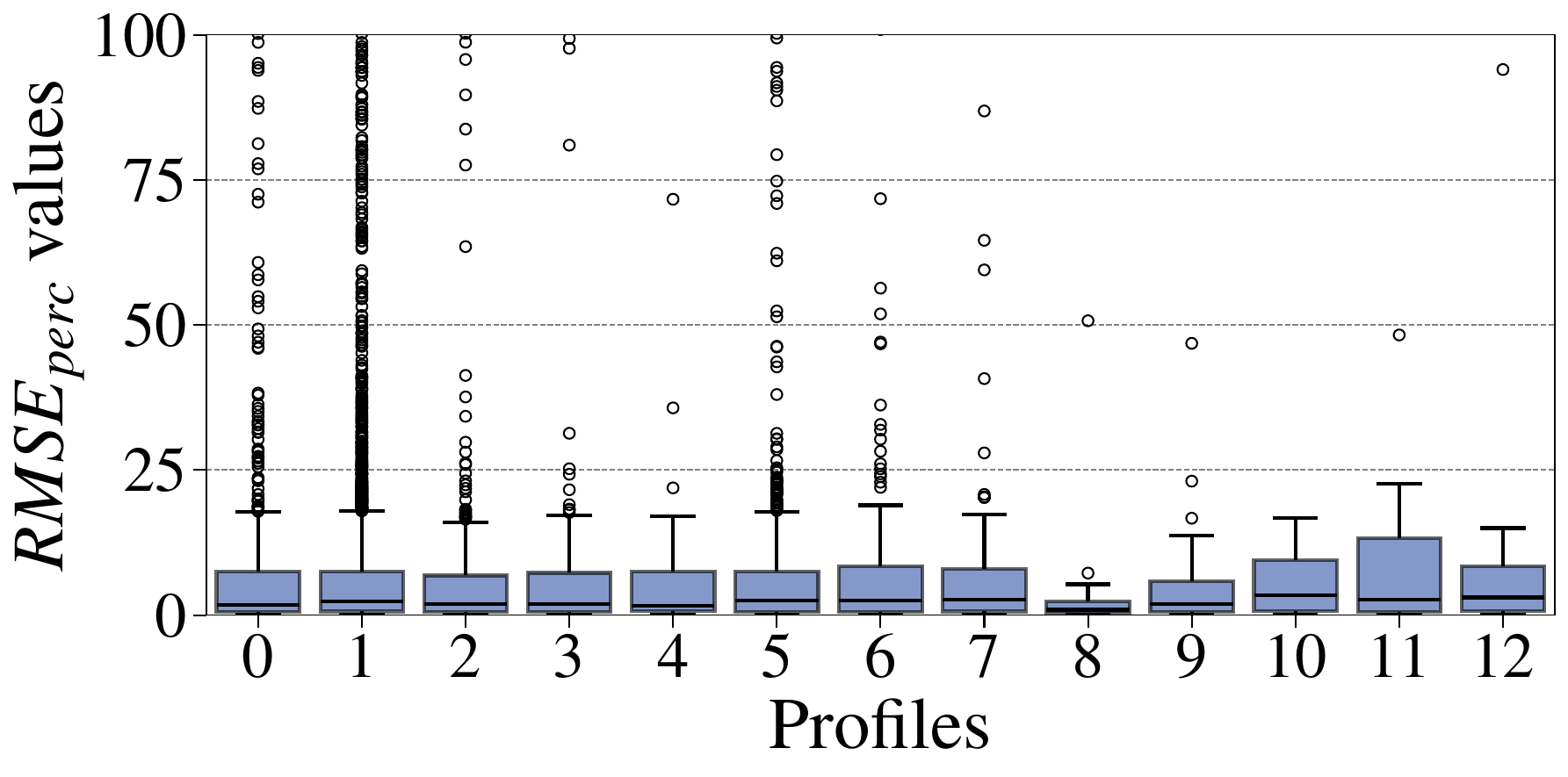} \label{fig:bxplt_google_cpu}}
  \\
  \subfloat[c][Memory]{
    \includegraphics[width=0.47\linewidth]{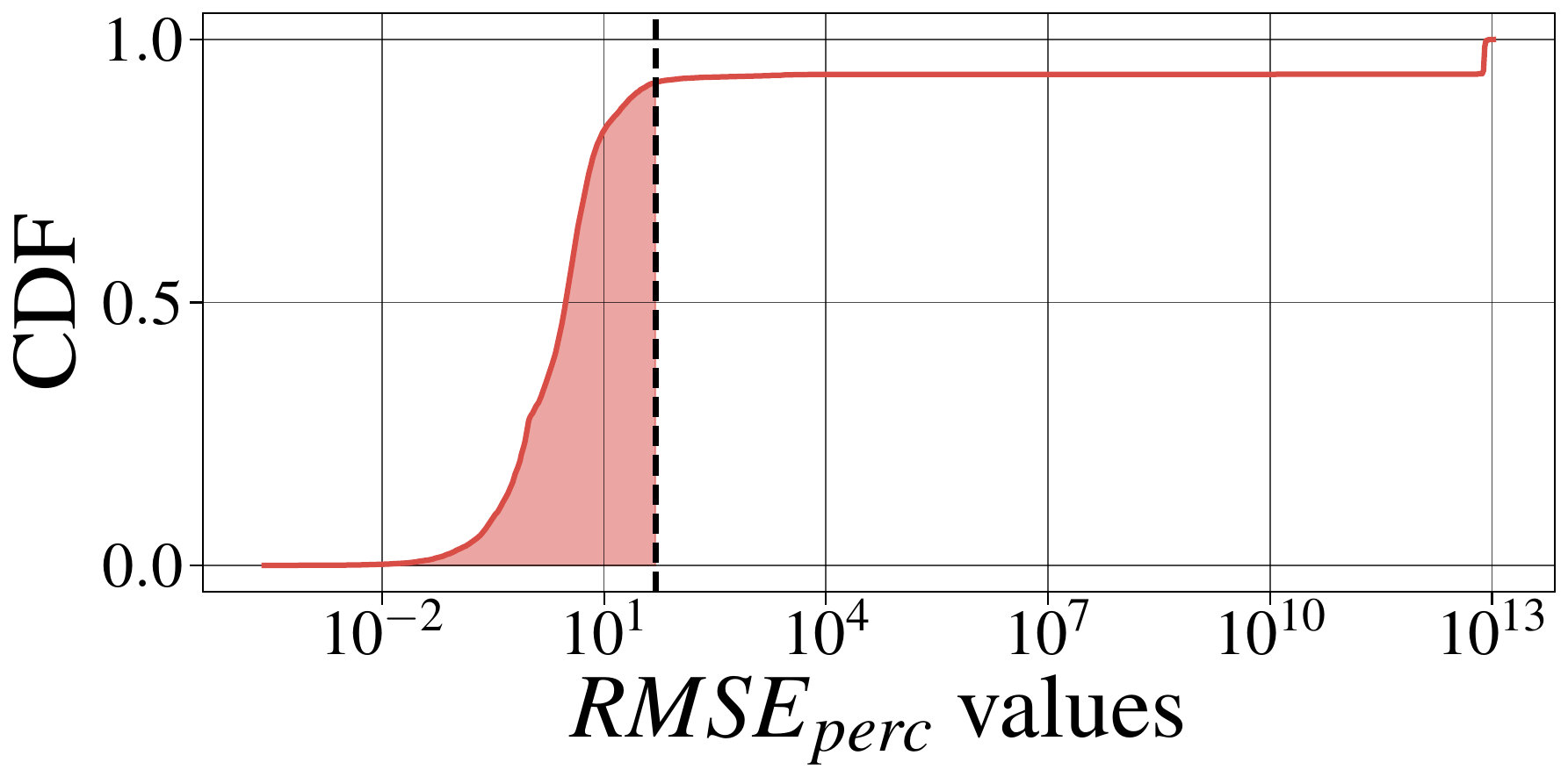} \label{fig:cdf_google_mem} }
  \hfill
  \subfloat[d][Memory boxplot]{
    \includegraphics[width=0.47\linewidth]{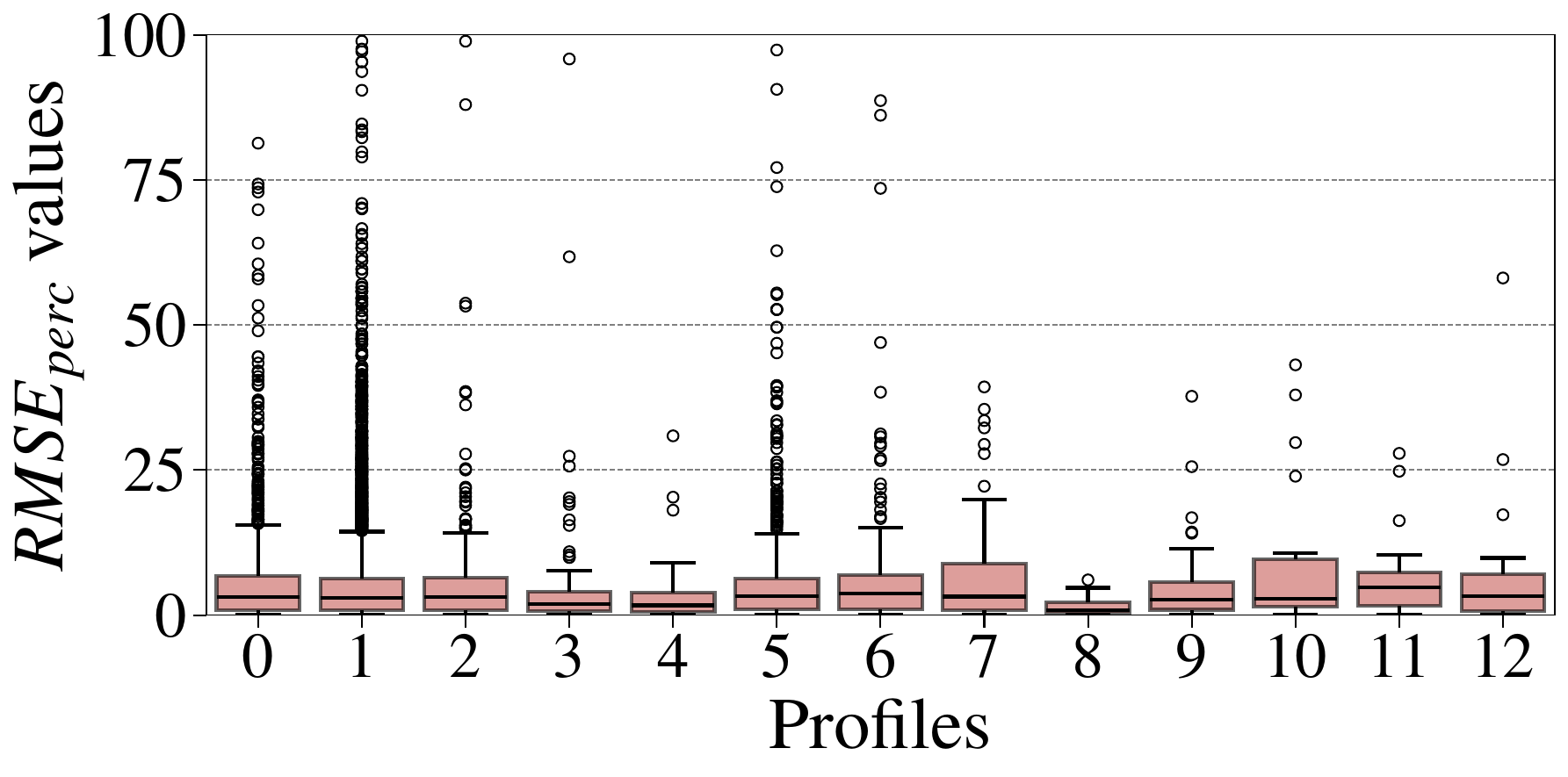} \label{fig:bxplt_google_mem}}
  \\
    \subfloat[e][Duration]{
    \includegraphics[width=0.47\linewidth]{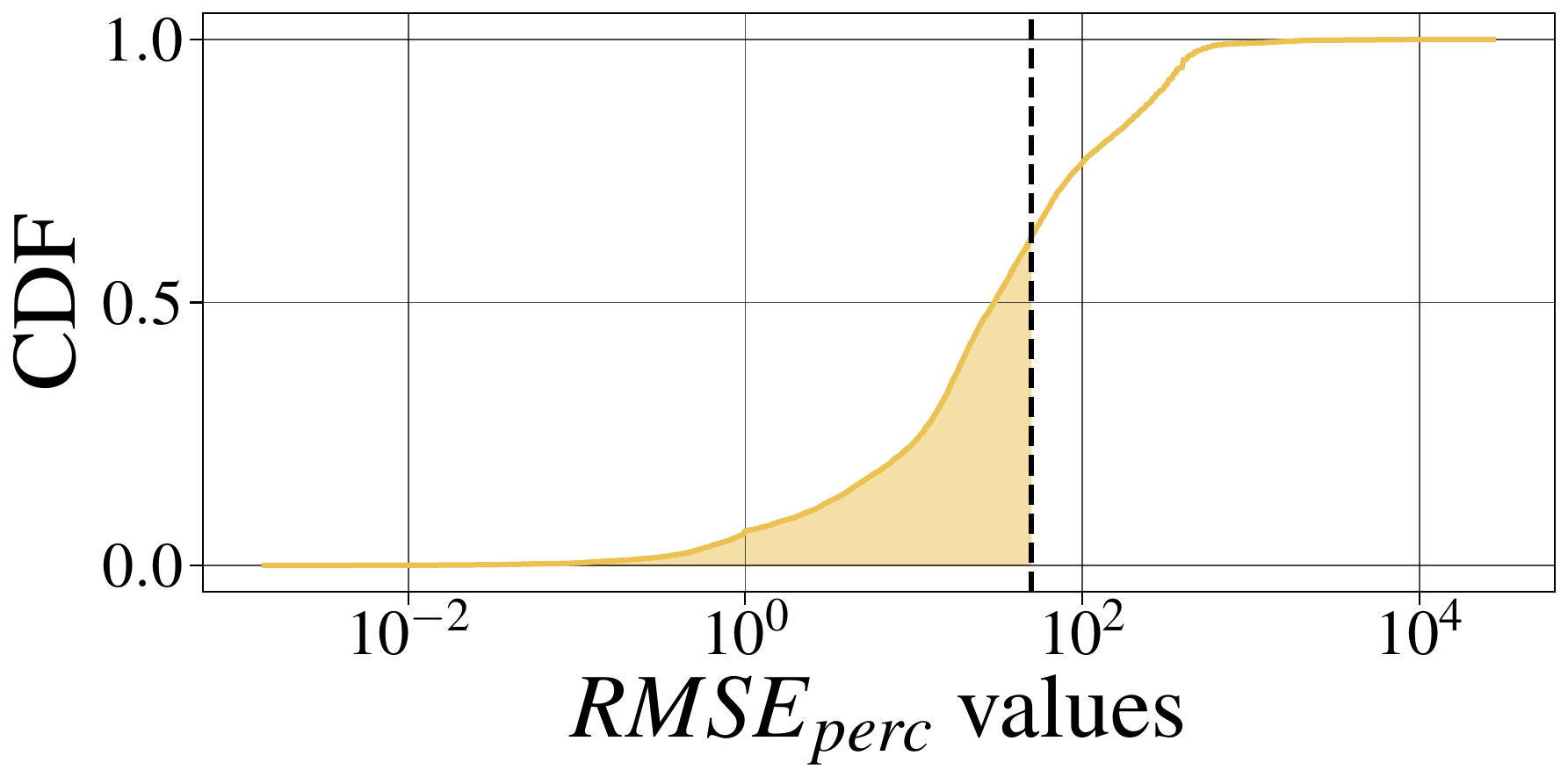} \label{fig:cdf_google_duration} }
  \hfill
  \subfloat[f][GPU]{
    \includegraphics[width=0.47\linewidth]{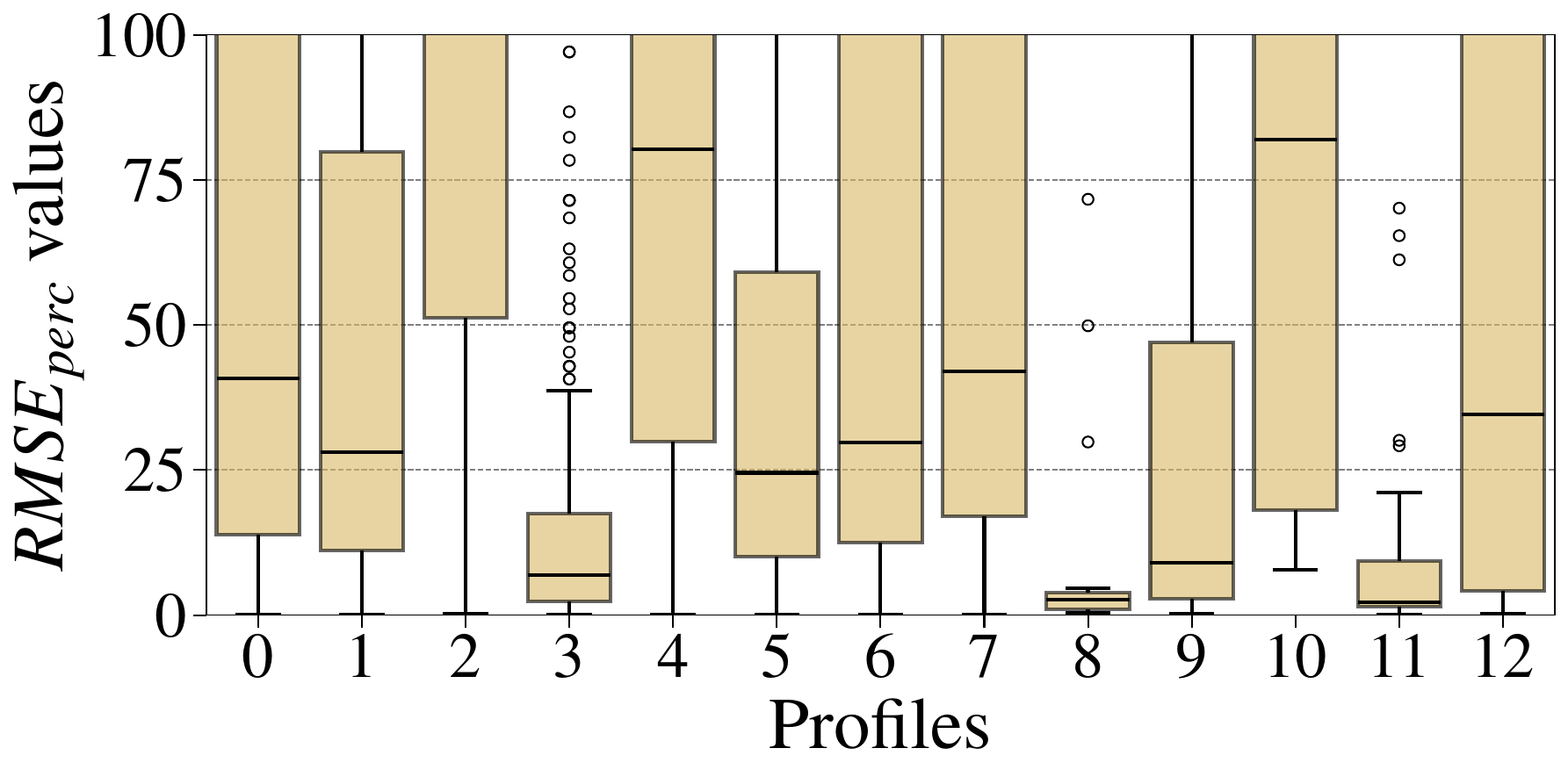} \label{fig:bxplt_google_duration}}

  \caption{Plots of the $RMSE_{perc}$ values on the Google cluster data for CPU, Memory, and Duration.}
  \label{fig:Google_combined_plots}
\end{figure}

We extend our analysis to estimate the capability of the proposed approach to work in different scenarios. To this extent, we rely on the well-known and established Google cluster data traces~\cite{verma2015large}. The dataset represents each workload (or job) scheduled on a node as a set of tasks. All tasks within a job execute the same binary, sharing the same options and requests. Hence, different task categories run as separate jobs. Further, each task runs within its container and has some runtime metrics associated with it and collected through \texttt{cgroup}, from CPU rate to memory usage to disc I/O time.
As the Google cluster data trace is too large to be completely analyzed, we perform a thorough but lean analysis for this evaluation. We extract circa 65000 jobs by stratified sampling over a relevant but unbalanced feature, \textit{scheduling class}, that separates workloads according to their latency sensitiveness (0 less sensitive, three most sensitive).
For the runtime metrics, we use the \texttt{cgroup} telemetry shared by Google.
The metadata features are less than in the Alibaba use case, making the test more challenging. In our case, we use workload-related information as the priority label and the scheduling class, plus information on the required amount of resources expressed in quartiles; namely, we consider disk space, memory, and CPU request.

For this experiment, we keep the formal structure of our PolarisProfiler, changing some mechanisms to prove how it might work with different algorithms. First, for the Profile Generator, we rely on the flat DBSCAN instead of the Hierarchical one. We use  \PowerTransform{} to transform the runtime metrics. Furthermore, as the high dimensionality is a problem for clustering methods, before giving our data in input to DBSCAN, we project it in a three-dimensional space using an autoencoder.
The Profile Classifier still relies on XGBoost, as in the Alibaba use case. However, here we replace one-hot-encoding of the categorical features with an autoencoder mechanism.~\footnote{https://github.com/AlliedToasters/dfencoder}

After the profile groups generation, we divide the data in training and test set for the Profile Classifier and the behavior prediction.
The results on the test set (circa 11\,000 workloads) are promising. First, the XGBoost classifier offers a strong performance over the thirteen clusters identified by DBSCAN, as shown in Table~\ref{tab:google_test_performance_xgboost}. In particular, it demonstrates resilience to different data and clusters with a cardinality imbalance.
Figure~\ref{fig:Google_combined_plots} reports the results of our analysis, showing the capability of our approach to work with different data sets and algorithms. The left column shows the Cumulative Distribution Functions (CDFs) of $RMSE_{perc}$ values for the four considered metrics: CPU, Memory, Duration, and GPU. These highlight the overall distribution of errors. In particular, the $RMSE_{perc}$ values for CPU and Duration show a wide range and some significant outliers, suggesting high variability. Conversely, the Memory metric exhibits a more contained distribution, indicating more consistent predictions across the clusters. GPU errors exhibit a larger spread, as evidenced by both the CDF and boxplot visualizations.
The right column provides boxplots of $RMSE_{perc}$ values across the 13 clusters identified by DBSCAN. These visualizations show the variation in prediction errors across the clusters. Notably, the approach demonstrates robustness, with the XGBoost classifier handling the cardinality imbalance of the clusters effectively. While some clusters, such as Profiles 1 and 5, exhibit more pronounced error variations, the overall performance confirms the resilience of our profiling mechanism, especially when modern techniques like autoencoders and flat DBSCAN are employed.
These results confirm the adaptability of our approach to varying data distributions and clustering outcomes, providing a promising basis for further evaluations.

\section{Related work}
~\label{sec:related}

Here, we highlight the state of the art in workload profiling. We start by describing runtime solutions; we then present static solutions underlying the approaches that resemble our model more. Finally, we shed light on specific methods for machine learning workload profiling, a recent trend that shows the relevance of the proposed approach.

\subsection{Runtime profiling}
Many authors focused to improve runtime workload profiling. Kairos~\cite{delgado2018kairos} does not require any a priori knowledge of task runtime.
Instead, Kairos employs preemption to estimate the predicted remaining runtime of tasks from when they have already been completed. Similarly, Jajoo et al.~\cite{jajoo2022case} propose a learning-in-space approach (\textit{SLearn}). They select and schedule only a portion of each workload's tasks. This method takes advantage of the similarities between the runtime characteristics of the tasks inside a single workload. Still, these and similar online approaches are subject to ``environment inconsistency.''
PARTIES~\cite{chen2019parties} offers online profiling. As Kairos and SLearn approach, it uses runtime information, discarding a priori knowledge, highlighting the difficulties of having information from user-submitted workloads. On the same line, Kaushik et al. ~\cite{kaushik2022study} profile application at runtime for improving vertical scaling. Inagaki et al. ~\cite{inagaki2022detecting} worked on profiling microservices to detect runtime bottlenecks. Gibilisco et al.~\cite{gibilisco2016stage} also focus on runtime profile sampling for Spark performance. Manner et al.~\cite{manner2021optimizing} perform dynamic profiling through simulations.
Rao et al.~\cite{rao2021soda} combine static and dynamic profiling with a focus on Spark. On the contrary, we use available static metadata as any workload is submitted, making the a priori matching trouble-free.
Other works~\cite{narayanan2020heterogeneity,xiao2018gandiva, mahajan2020themis} collect ``offline'' runtime information, running the workloads in exclusive mode. This approach, though, suffers from the environment's inconsistency.


\subsection{Offline profiling}
Building statistics and extracting patterns from metadata has seen diverse applications across domains. Gupta et al.\cite{gupta2013profile} proposed a profile-based network intrusion detection system for cloud environments, leveraging network behavior profiles of virtual machines (VMs) to identify threats. Bartzas et al.\cite{bartzas2010software} focused on profiling software metadata in embedded systems, with an emphasis on runtime memory behavior. In the context of structured datasets, WebLens~\cite{khan2020weblens} offers metadata profiling for large-scale data integration. Similarly, Calzarossa et al.~\cite{calzarossa2016workload} surveyed workload characterization techniques, showing how most approaches rely heavily on runtime metrics and execution traces. 
Previous work used offline-based approaches to estimate the duration of workloads~\cite{ferguson2012jockey,
karanasos2015mercury} 
These works estimate the duration by using assumptions on specific features, e.g., task type and dataset size. Instead, our work relies on a generic approach that uses old dynamic information to infer the specific static and a priori metadata features to detect homogeneous profiles. Other approaches, like 3Sigma~\cite{park20183sigma}, rely on the total historical workload duration distributions to predict how long the new workloads will start. Similarly, Weng et al.~\cite{weng2022mlaas} use the Alibaba dataset past estimation and a set of fixed parameters, i.e., \textit{group} and \textit{user}, to estimate the workload completion time. Conversely, we create specific profiles to address such challenges.

Similar to our method, Hu et al.~\cite{hu2021characterization} rank workloads using GPU time, correlating it to attributes, such as workload name, user, and submission time. They leverage these attributes to predict the workloads' priority in scheduling. This approach follows a similar methodology. However, we aim to provide a more generic approach to automatically extract these correlations and patterns.  
InfaaS~\cite{romero2021infaas} proposes using statically-profiled metadata, plus the tracking of dynamic state for high-level-requirement-based distributed inference serving.
On a close path, Kattepur et al.~\cite{kattepur2017priori} have a methodology in principle similar to our approach, but, in practice, it is runtime based and focusing on robotics through fog networks.



\subsection{Profiling machine learning workload}
Finally, we focus on recent research that aims at characterizing machine learning workloads, as it has been the main focus of our case study. Some works~\cite{yeung2020towards, geoffrey2021habitat} approach the profiling using historical execution traces containing hardware attributes and runtime data to forecast the duration of a DNN's training iteration. Aryl~\cite{li2022aryl} leveraging the former approach to estimate the DNN workload duration, using the history of the runs of the same workload. SCHEDTUNE~\cite{albahar2022schedtune} leverages historical execution traces to build profiles to predict resource usage. Our case study differentiates from that as we follow a more generic approach, i.e., we do not focus solely on training and do not consider hardware assumptions. Loki~\cite{ahmad2024loki} addresses hardware and accuracy scaling in inference serving pipelines by dynamically optimizing resource allocation using metadata, although it centers on runtime adaptation. Themis~\cite{razavi2024tale} extends autoscaling strategies, combining horizontal and vertical scaling for deep learning inference but remains tied to dynamic profiling methods. FlexLLM~\cite{miao2024flexllm} incorporates metadata in parameter-efficient fine-tuning for large language models, focusing on LLM-specific scenarios. FaaSwap~\cite{yu2023faaswap} applies metadata-driven scheduling policies to GPU-efficient serverless inference but emphasizes runtime execution metrics. Based on Habitat, EOP~\cite{xu2022eop} aims at characterizing deep learning inference tasks by looking at three main characteristics of the DNN, such as the \textit{batch size}, \textit{Height-weight-weight}, and \textit{Height-weight-weight}. Again, this approach targets a narrow problem and makes strong a priori assumptions on the features that can better represent the workloads. Shin et al.~\cite{shin2022xonar} developed an approach to profile the workload of AI applications.

\subsection{Takeaways}
In contrast to these works, we utilize static, a priori metadata for profiling workloads, offering a generalizable and environment-resilient solution that avoids the complexities of runtime dependency. This approach is particularly suited to distributed systems, where static metadata aids in efficient and robust workload profiling.

\section{Discussions and future work}

\paragraph{Long-time evolution analysis} An essential part of the profiles is to be dynamic and adapt to changing workloads and environments. Despite our accurate analysis, it would be key to evaluate how the profiling method would work in deployment over long periods. This aspect is especially relevant in an open infrastructure such as the computing continuum, where the workload types can drastically change over time.
\paragraph{Integrating the metadata-based profiling with runtime solutions} Both history-based and online approaches have advantages and disadvantages.  History-based approaches benefit from rich metadata but may struggle with real-time variability. The latter case can respond better to the current infrastructure and environment state. However, it can suffer from time constraints, making the analyses naturally less precise; plus, it can go too far, overestimating current variations by not looking at past patterns. The former does not have any of these problems, but it could scarcely tolerate significant variations typical of systems' evolution. Therefore, a promising future direction involves enriching our history-based approach with an online component. In particular, we plan it to integrate it with predictive monitoring tools~\cite{morichetta2023demystifying,lackinger2024timeseries}, which help in estimating workloads' runtime properties.
\paragraph{Improving the profiling feedback} There are limits to using a single statistical values for predictions, as we perform in our case study for the job duration. Deploying more elaborate solutions, e.g., Bayesian approaches, could help the prediction accuracy. We can see this for the test samples in which \RMSEperc value is more than 100\%. This behavior calls for further research and investigations, which can be tailored to the specific target.
 
\paragraph{Testing the approach on other case studies} To consolidate the results shown in this case study, it is essential to extend the approach to other scenarios. Some examples might be creating profiles explicitly related to certain SLOs. In addition, we aim at building more comprehensive testbeds. An important aspect is to check the whole orchestration pipeline, where the workload can be scheduled~\cite{nastic2021polaris} on specific nodes~\cite{VPujol2023Intelligent} based on its profile. Being able to check its impact on the infrastructure, the capability to fulfill its SLOs, would help us further improve the methodology.

\section{Conclusions} \label{sec:conclusion}

This paper formalized a methodology for profiling workload in virtualized and shared computing infrastructures, leveraging static, a priori metadata. Our goal is to create a generalizable, fast, and precise workload profiler capable of estimating runtime characteristics before execution.
We then validated our approach through two use cases. First, we comprehensively analyzed real ML workload traces, leveraging the Alibaba dataset. We outlined practical methods and algorithms to implement the previously defined conceptual approaches and showed how the specific technologies, together with the general approach, can lead to good results. In particular, we presented how, on 10\,000 unseen data, our system can improve and correct the divergent behaviors that can naturally appear after adding many new workloads to the existing profiles.
Finally, we presented how the PolarisProfiler can generalize across various workload types by testing our approach on the Google cluster traces. The results are promising, highlighting PolarisProfiler as a reliable mechanism for workload profiling and pushing toward further improvement of the model.

\bibliographystyle{acm}
\bibliography{references}

\end{document}